\documentclass[superscriptaddress,a4paper,preprint,amsmath,amssymb,aps,prstab,]{revtex4-1}
\usepackage[colorlinks=true,linkcolor=black,citecolor=blue,urlcolor=blue,]{hyperref}
\usepackage[utf8]{inputenc}
\usepackage[english]{babel}\addto\captionsenglish{}
\usepackage[T1]{fontenc}
\usepackage[misc]{ifsym}
\usepackage{yfonts}
\usepackage{graphicx}
\usepackage{amsmath}
\usepackage{color}
\usepackage{ulem}
\usepackage{booktabs}
\usepackage{float}
\usepackage[top=2cm,bottom=2cm,left=2cm,right=2cm]{geometry}

\begin{document}
  \title{\textbf{Simulation studies for upgrading a high-intensity surface muon beamline at Paul Scherrer Institute}}

  \author{Lu-Ping Zhou}
  \email{zhoulp@ihep.ac.cn (L.P. Zhou)}
  \affiliation{Institute of High Energy Physics, CAS, Beijing 100049, China}
  \affiliation{University of Chinese Academy of Sciences, CAS, Beijing 100049, China}
  \affiliation{Spallation Neutron Source Science Center, Dongguan 523803, China}
  \affiliation{Paul Scherrer Institute, Laboratory for Muon Spin Spectroscopy, CH-5232 Villigen PSI, Switzerland}
  \author{Xiao-Jie Ni}
  \affiliation{Paul Scherrer Institute, Laboratory for Muon Spin Spectroscopy, CH-5232 Villigen PSI, Switzerland}
  \author{Zaher Salman}
  \affiliation{Paul Scherrer Institute, Laboratory for Muon Spin Spectroscopy, CH-5232 Villigen PSI, Switzerland}
  \author{Andreas Suter}
  \affiliation{Paul Scherrer Institute, Laboratory for Muon Spin Spectroscopy, CH-5232 Villigen PSI, Switzerland}
  \author{Jing-Yu Tang}
  \affiliation{Institute of High Energy Physics, CAS, Beijing 100049, China}
  \affiliation{University of Chinese Academy of Sciences, CAS, Beijing 100049, China}
  \affiliation{Spallation Neutron Source Science Center, Dongguan 523803, China}
  \author{Vjeran Vrankovic}
  \affiliation{Paul Scherrer Institute, Division of Large Research Facilities, CH-5232 Villigen PSI, Switzerland}
  \author{Thomas Prokscha}
  \email{thomas.prokscha@psi.ch (T. Prokscha)}
  \affiliation{Paul Scherrer Institute, Laboratory for Muon Spin Spectroscopy, CH-5232 Villigen PSI, Switzerland}

  \date{\today}

  \begin{abstract}
    	The $\mu$E4-LEM beamline at Paul Scherrer Institute (PSI, Switzerland) is a special muon beamline combining the hyprid type surface muon beamline $\mu$E4 with the low energy muon facility (LEM) and delivers  ${\mu}^{+}$ with tunable energy up to 30 keV for low-energy muon spin rotation  experiments (LE-$\mu$SR). We investigate a possible upgrade scenario for the surface muon beamline $\mu$E4 by replacing the last set of quadrupole triplet with a special solenoid to obtain 1.4 times original beam intensity on the LEM muon moderator target. In order to avoid the muon beam intensity loss at the LEM spectrometer due to the stray magnetic field of the solenoid, three kinds of solenoid models have been explored and the stray field of the solenoid at the LEM facility is finally reduced to the magnitude of the geomagnetic field. A more radical design, "Super-$\mu$E4", has also been investigated for further increasing the brightness of the low energy muon beam, where we make use of the current $\mu$E4 channel and all sets of quadrupole triplets are replaced by large aperture solenoids. Together with the new slanted muon target E, at least 2.9 times the original muon beam intensity can be expected in the Super-$\mu$E4 beamline. Our work demonstrates the feasibility of upgrading surface muon beamlines by replacing quadrupole magnets with normal-conducting solenoids, resulting in higher muon rates and smaller beam spot sizes.
  \end{abstract}
  
  \maketitle

  \begin{center}\section*{I. Introduction}\end{center}
  	
\begin{figure}[tbp]
		\centering
		\includegraphics[width=12cm]{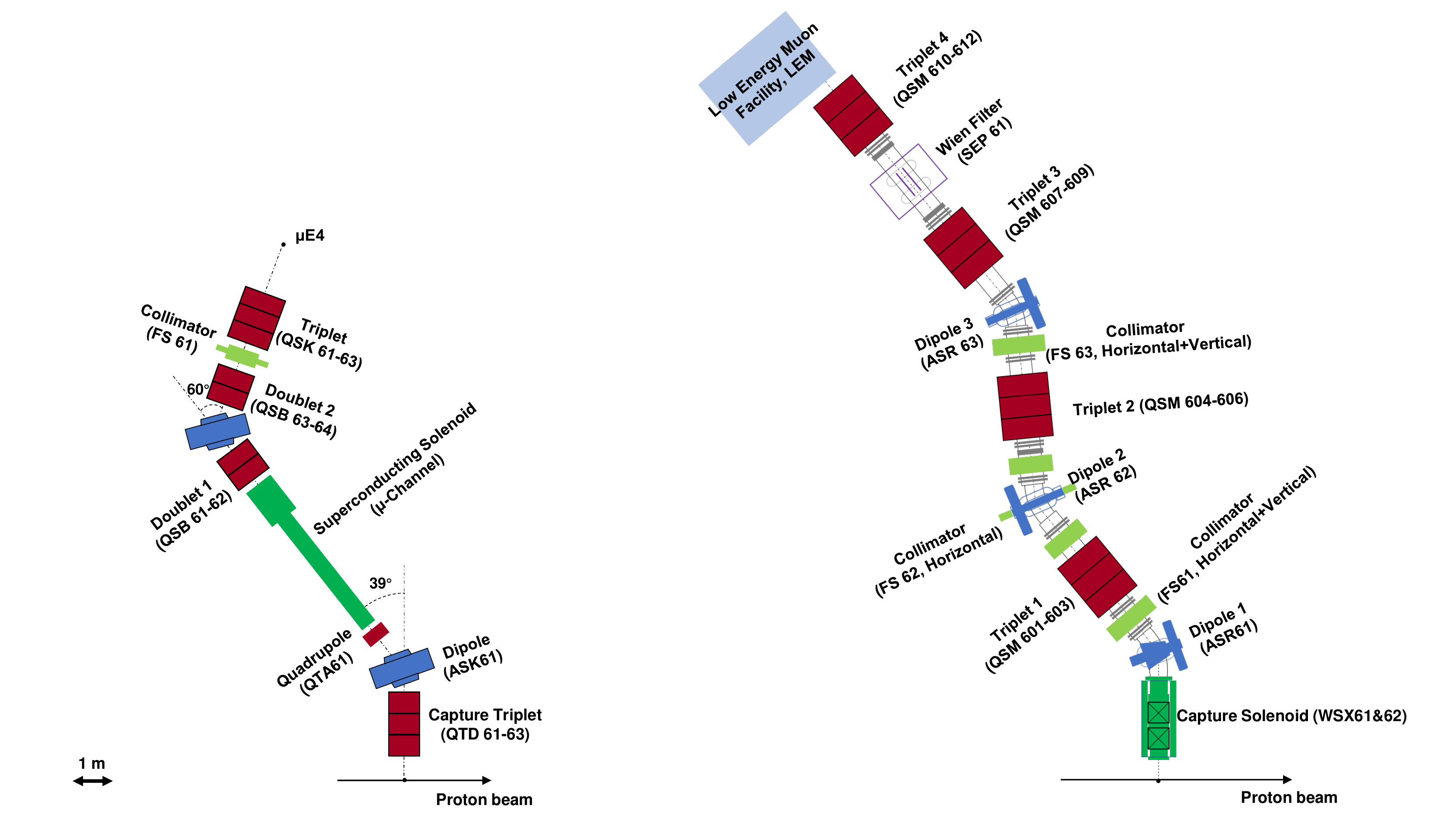}
		\caption{Left: the old $\mu$E4 decay channel; Right: the hybrid type $\mu$E4 beamline rebuilt in 2003/2004 as a surface muon injector for the LEM facility}
		\label{FIG1} 
	\end{figure}

\begin{figure}[tbp]
		\centering
		\includegraphics[width=8cm]{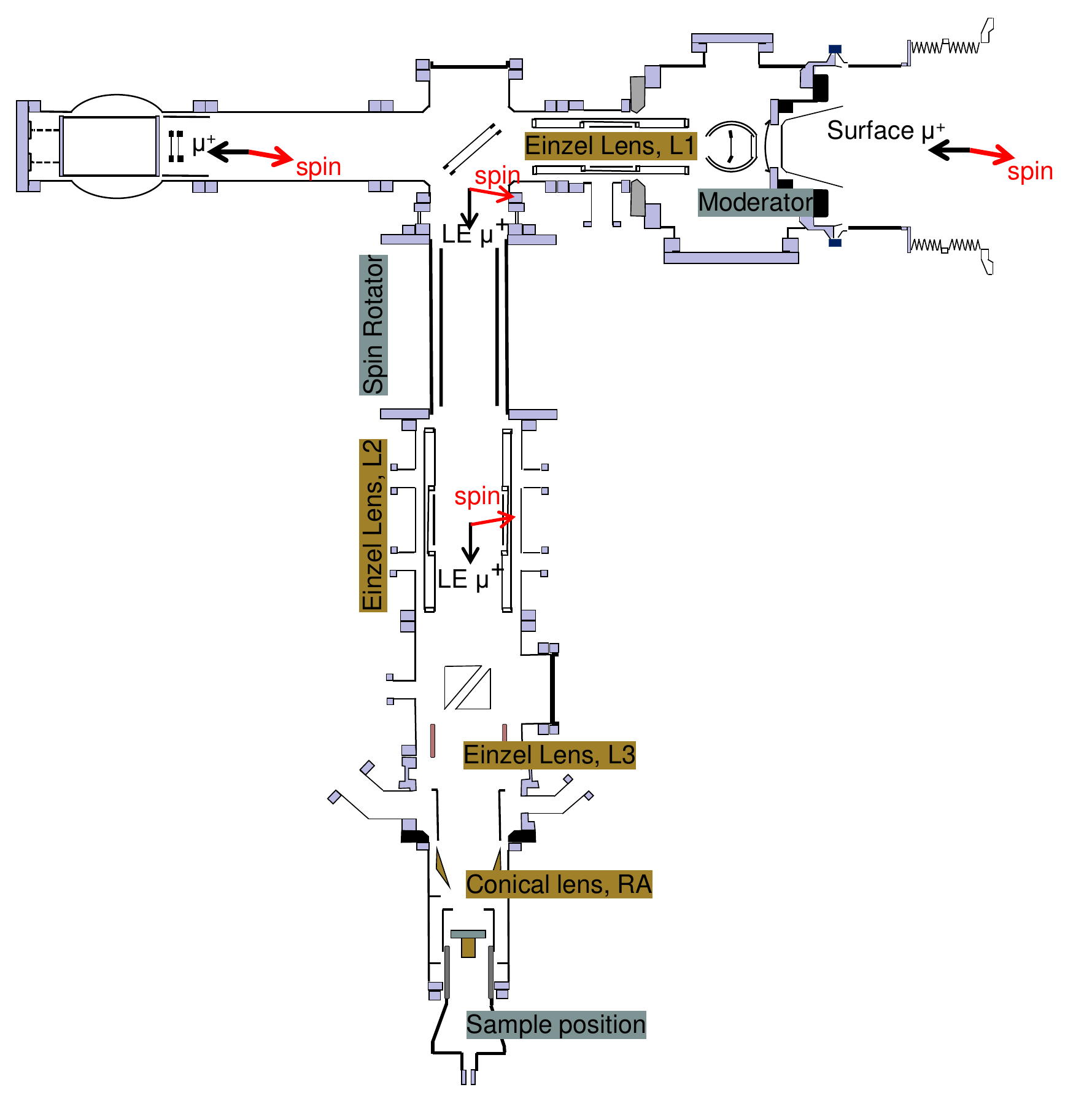}
		\caption{The layout of LEM facility}
		\label{FIG2} 
	\end{figure}

	The hybrid-type surface muon beamline $\mu$E4 at Paul Scherrer Institute (PSI, Switzerland) has already been in operation for more than a decade. The $\mu$E4 beamline was first built in the 1970s as a decay muon channel, then rebuilt in 2003/2004 as a high intensity surface muon beamline together with the low energy muon (LEM) facility \cite{Foroughi_Upgrading_2001,Prokscha_NewMB_2004,Prokscha_NewMuE4_2008,Morenzoni_LEmuPSI_2000,Morenzoni_LEmu_2003,Prokscha_Thinfilm_2004,Morenzoni_NanoFilm_2003}. Figure~\ref{FIG1} shows the layout of the $\mu$E4 beamline. Although originally designed to have the ability of transmitting both surface and decay muon beams, the new $\mu$E4 beamline has been operating in the surface muon mode and the structure remained unchanged since then. Two normal-conducting solenoids WSX61 and WSX62 close to the muon production target E serve to increase the acceptance, and the following large-aperture magnetic quadrupoles and dipoles transport the surface muon beam to the LEM moderator. The overall acceptance is limited to around 135 mSr due to the employment of quadrupoles and dipoles. The beam momentum is selected by dipoles and the contamination of cloud muons is 5\% - 10\% \cite{Prokscha_NewMuE4_2008}, causing a reduction of beam polarization to about 95\%. The Wien Filter SEP61 rotates the muon spin by about 10$^\circ$ before being focused onto the muon moderator target in the LEM apparatus. The muon beam is moderated subsequently to a mean energy of about 15~eV in a cryogenic moderator layer of solid argon (or solid neon), deposited on a 125-$\mu$m-thick Ag foil \cite{Prokscha_ModforMup_2001}, with an efficiency of $\sim 6\times 10^{-5}$($\sim 1\times 10^{-4}$ for solid neon). The produced low energy muons are re-accelerated up to 20~keV, and then transported and focused by three electrostatic Einzel lenses and a conical lens \cite{Xiao_Segmented_2017} to the final sample position, where the beam energy can be tuned in the range from 1 to 30 keV by applying an electrostatic bias to the sample  to conduct  low energy $\mu$SR (LE-$\mu$SR) experiments. Figure~\ref{FIG2} shows the layout of the LEM facility \cite{Khaw_SimuLEM_2015}. During the beam transport, a spin rotator (Wien Filter) is used to rotate the muon spin and to separate protons from the muon beam. These protons originate from the muon moderator by impact ionization of water or hydrogen molecules sticking at the moderator surface.
		
	
	In recent years, most of muon beamlines around the world have been further upgraded to obtain better performances. In UK, the ISIS Neutron and Muon Source (ISIS) has developed its muon beamline at the EC muon facility by replacing doublets with large-aperture triplets, and got about a factor of two  increase in muon rate and a beam spot with  0.6 cm RMS radius in the simulation \cite{HillierAD_ISISmuon_2014, HillierAD_ISISmuon_2019}. In Japan, the RCNP of Osaka university has built the MuSIC muon beamline for the verification of a beamline fully based on the solenoid capture and transport for muon physics experiments such as COMET at J-PARC \cite{Comet_collaboration_comet_2020}, and later transformed for $\mu$SR applications \cite{Cook_MuSIC_2017, Tomono_MuSIC_2018}. Newly designed surface muon beamlines are more likely to use large aperture solenoids to increase the beam intensity on the sample. For instance, J-PARC in Japan has succeeded to build the second slow muon beamline in the world, the Ultra Slow Muon beamline (USM). It uses the so-called “Super Omega” beamline, which is mainly composed of solenoids \cite{Strasser_ULM_2014,Nakamura_ULM_2014,Ikedo_SuperOmega_2011,Nakahara_SuperOmega_2009}, to supply a high intensity surface muon beam for USM generation. On the opposite side of the muon target station, J-PARC prepares to use large aperture capture solenoids to increase the acceptance of its H-line and then transport the high intensity muon beam by using a pair of two solenoids which have reverse currents \cite{Miyake_MuSE_2018,kawamura_new_2018}. In Switzerland, PSI is investigating new high intensity muon beams (HIMB) with a surface muon rate up to $10^{10}$/s based on its new target M (so-called “target H”) \cite{Apapa_HiMB_2019,KSKirch_HiMB_2020,Aiba:2021bxe}. In China, the first muon source EMuS is going to use solenoids to transport high intensity surface muon and pion beams in the baseline scenario \cite{Tang_EMuS_2018}.
	
	At PSI, the existing $\mu$E4 beamline, using large aperture quadrupole triplets for beam transport, can deliver about 40\% of the surface muon beam onto the rectangular moderator target with a size of $30\times 30$~mm$^2$. The causes for low transmission efficiency are i) the large transverse emittances of $\varepsilon_x = 550$~$\pi\cdot$cm$\cdot$mrad and $\varepsilon_y = 1000$~$\pi\cdot$cm$\cdot$mrad, and ii) the asymmetric focusing in $x$- and $y$-directions of the last quadrupole triplet, causing an asymmetric beam spot on the moderator with respective $x$- and $y$-RMS values of $\sigma_x = 2.9$~cm and $\sigma_y = 1.4$~cm \cite{Prokscha_NewMuE4_2008}. These two beam characteristics limit the rate of moderated muons to $1.2\times10^4$/s (solid argon) and $2.0\times10^4$/s (solid neon) at a proton beam current of 2~mA. An increase of these rates will have a large positive impact on the entire user program at the LEM facility. We therefore started a study to increase the muon beam intensity and to reduce the beam spot size due to the large transverse emittances and asymmetric focusing. As a first step, replacing the last quadrupole by a solenoid can already significantly improve the beam spot on moderator. However, a solenoid generates extended stray magnetic fields, which will seriously deteriorate the transport of the low-energy muon beam after moderation. Thus, the solenoid has to be designed with a sufficiently small residual stray field at the LEM facility to avoid significantly distorting the transmission of the low-energy muon beam to the LE-$\mu$SR spectrometer, while keeping the ability of focusing the muon beam on the moderator. In a second step, the replacement of all quadrupole triplets by solenoids is investigated ("Super-$\mu$E4"), which gives the largest possible improvement of the beam transport of $\mu$E4. 
	
	This paper is organized as follows: Sec.~II introduces the $\mu$E4 beam simulations and the comparison between utilizing a solenoid or a quadrupole triplet for the final focusing, followed by Sec.~III, where we introduce the exploration of the possible solenoid models with low stray field and their impact on the muon beam transport to the LE-$\mu$SR spectrometer. In Sec.~IV, we introduce the high intensity muon beamline "Super-$\mu$E4" study based on large aperture normal-conducting solenoids.
	\\
	\\
	
	\begin{center}{\section*{II. Simulations for different designs of the muon targets and beamlines}}\end{center}
	
\begin{figure}[tbp]
		\centering
		\includegraphics[width=12cm]{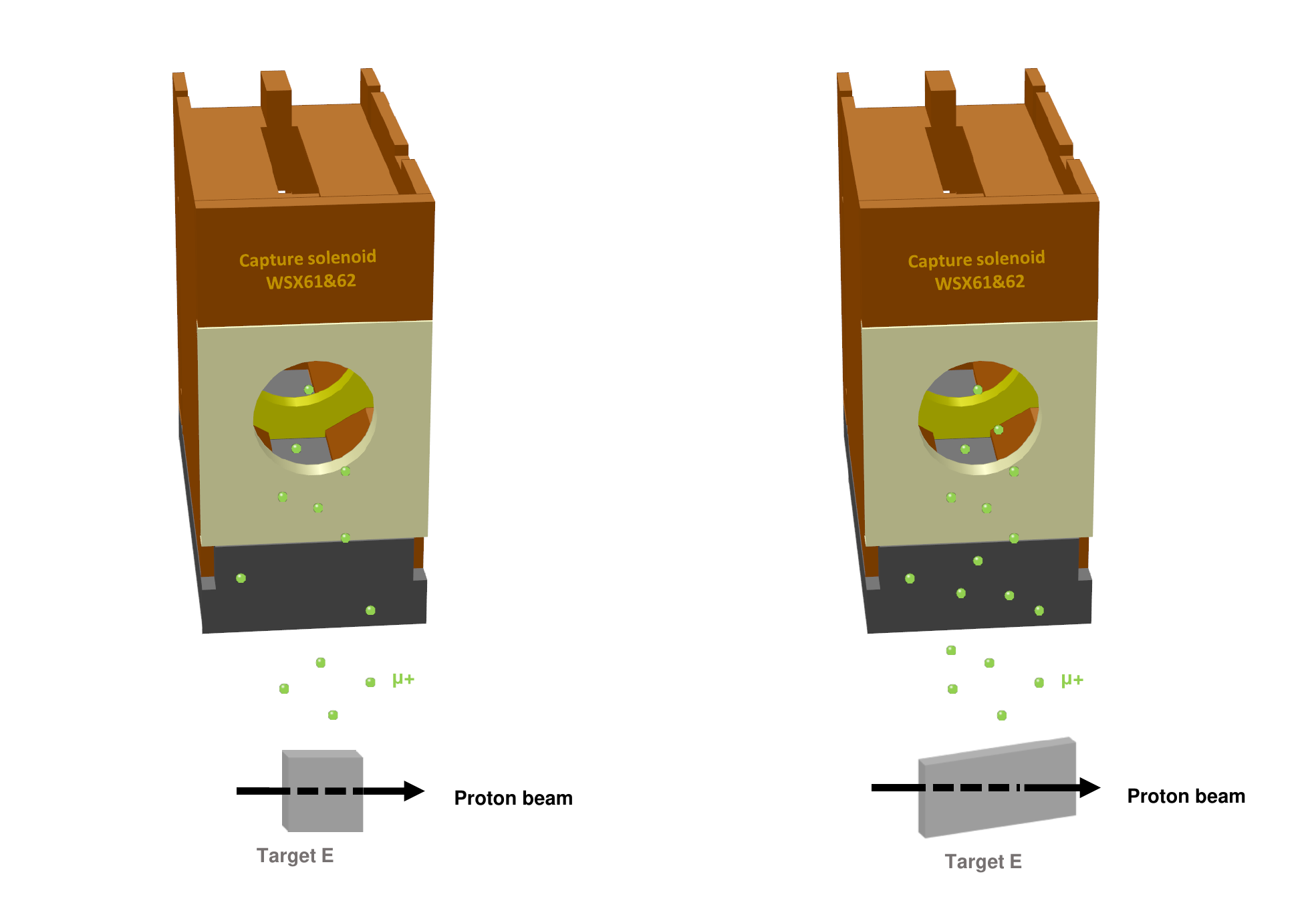}
		\caption{The OTE, as shown on the left side, has been replaced by the NTE in November 2019, shown in the right figure. The two targets are designed to have the same effective length for the proton beam to ensure that energy loss, scattering, and the deposited power of the beam in the target remains unchanged while the new target has a larger surface producing more surface muons. In order to see the difference more clearly, both targets have been magnified proportionally.}
		\label{FIG3} 
	\end{figure}

\begin{figure}[tbp]
		\centering
		\includegraphics[width=16cm]{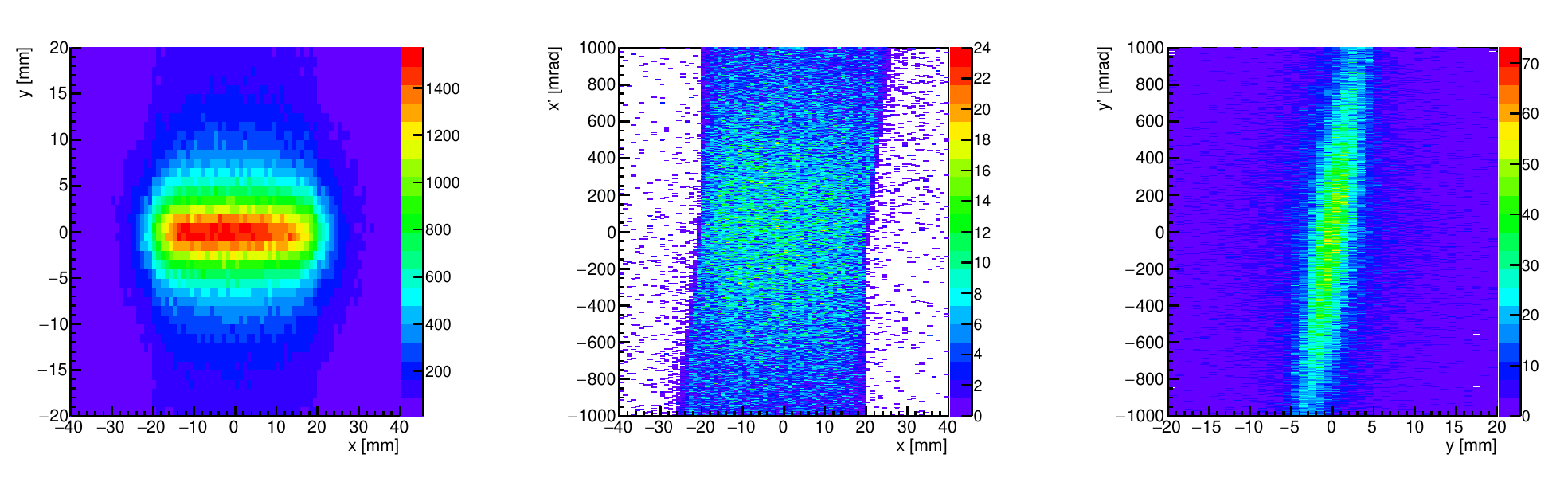}
		\\
		(a)
		\\
		\includegraphics[width=16cm]{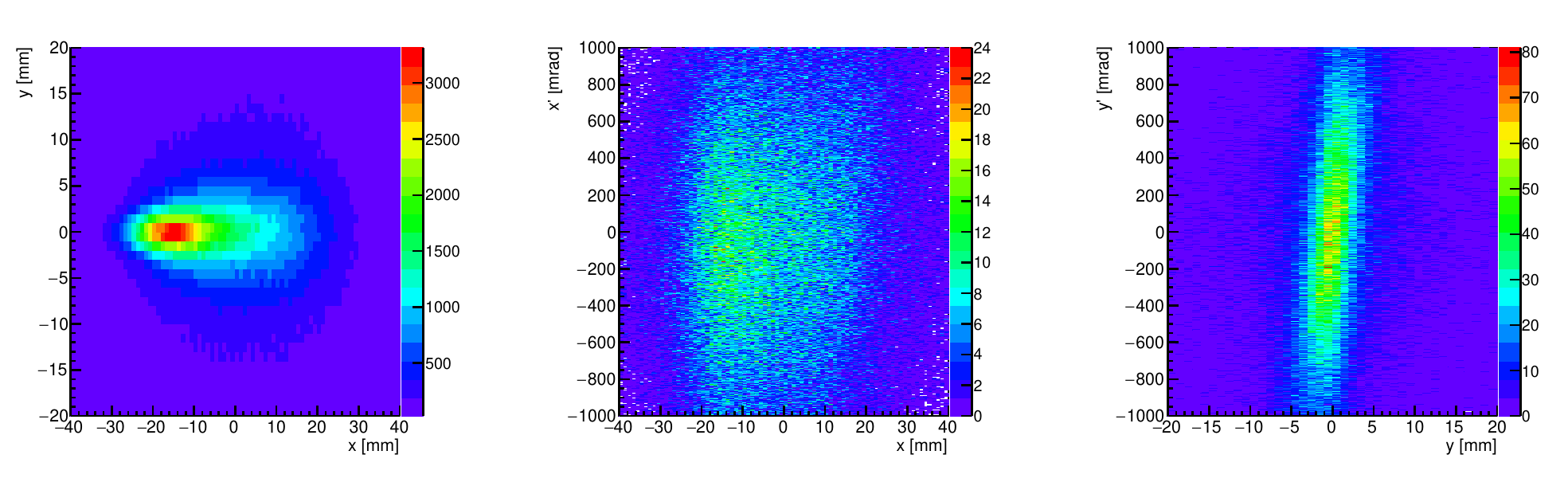}
		\\
		(b)
		\\
		\caption{The figures (a) show the real space (x-y) distribution and phase spaces (x-x’ and y-y’) of the OTE, and the figures (b) show the corresponding distributions of the NTE. Virtual detectors are attached to the surface of the target to record surface muons in the simulation. $10^{11}$ protons are simulated for each target and we use "QGSP\_BIC" as the physical model for the interaction of the proton beam for both target simulations in G4beamline.}
		\label{FIG4} 
	\end{figure}

    The two pion/muon production targets "M" and "E" of the PSI High Intensity Proton Accelerator facility HIPA are slowly rotating graphite wheels, with a straight pion/muon production volume in the proton beam direction and lengths of 5~mm and 40~mm, respectively \cite{Berg_SMtarget_2016}. Recent studies of the muon production target E at PSI have revealed that a slanted target E, with same effective length for the proton beam as the original straight target, can yield a 30$\%$ -- 40$\%$ increase of surface muon beam rate by sideway collection depending on different target rotation angles \cite{KSKirch_HiMB_2020,Berg_SMtarget_2016}. This increase is mainly caused by the effectively larger surface and production volume of this target geometry. In the studies of $\mu$E4, G4beamline \cite{Kaplan_G4beamline_2007} is used to simulate both the production of pions and muons in target E, and the transport of the surface muons through the beamline. G4beamline uses Geant4 cross section models for pion production, yielding relatively larger or smaller rates of surface muons depending on different physics models (reference \cite{Berg_SMtarget_2016} offers the corrected cross section). The detailed simulation of pion/muon production in the target, however, exceeds the scope of this paper since we are mainly interested in studying the relative increase of surface muon beam rate with a simplified generation of the source particles for the beamline simulation. The parameters of the target and proton beam are taken from reference \cite{Berg_SMtarget_2016}. The RMS values of the incident proton beam at the target are $\sigma{_x}$ = 0.75~mm and $\sigma{_y}$ = 1.25~mm, and Fig.~\ref{FIG3} shows the two target geometries used in the simulation. The original, 40-mm-long straight target E (OTE) is simulated as a rectangular graphite box of 40~mm length, 6~mm width and 100~mm height. The parameters of the new slanted target E (NTE) is simulated as a rectangular graphite box of 83.5~mm length, 5.56~mm width and 100~mm height, with an 8 degrees rotation around the vertical axis y. Only muons originating in the region -20~mm < y < 20~mm are taken into consideration for both targets because of their wheel-like shape \cite{Berg_SMtarget_2016}. The initial phase spaces of the muon beam produced by the two targets are shown in Fig.~\ref{FIG4}. Using in G4beamline the model QGSP\_BIC for pion production in target E, the experimentally observed muon rates at the end of the $\mu$E4 beamline are fairly well reproduced, see Appendix.  
	
	
	Due to the higher order effects, both focusing by the large aperture quadrupole triplets and the solenoid will introduce a filamentation of transverse phase spaces for the beam with both transverse emittances being of the order of 100~$\pi\cdot$cm$\cdot$mrad. When using the solenoid for transporting and focusing of the large emittance beam with a wide momentum spread of a few percent, the higher order effects of the solenoid and the coupling of different planes will probably increase the 4-D RMS emittance \cite{Eshraqi_Solenoid_2009}. However, the beam transport benefits from the more symmetric focusing in the two transverse directions, resulting in a smaller beam spot size.
	
\begin{figure}[tbp]
		\centering
		\includegraphics[width=16cm]{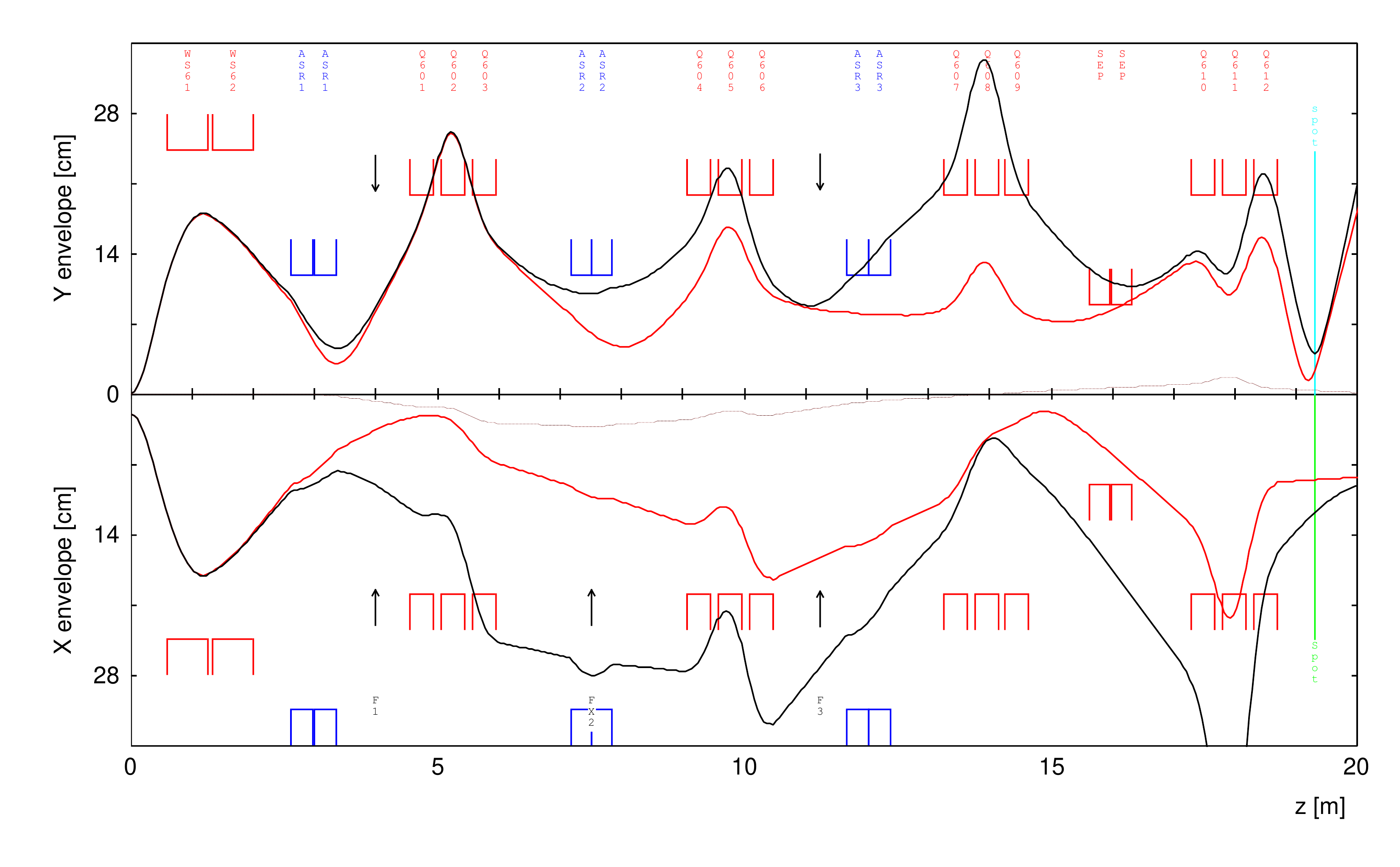}
		\\(a)
		\\
		\includegraphics[width=16cm]{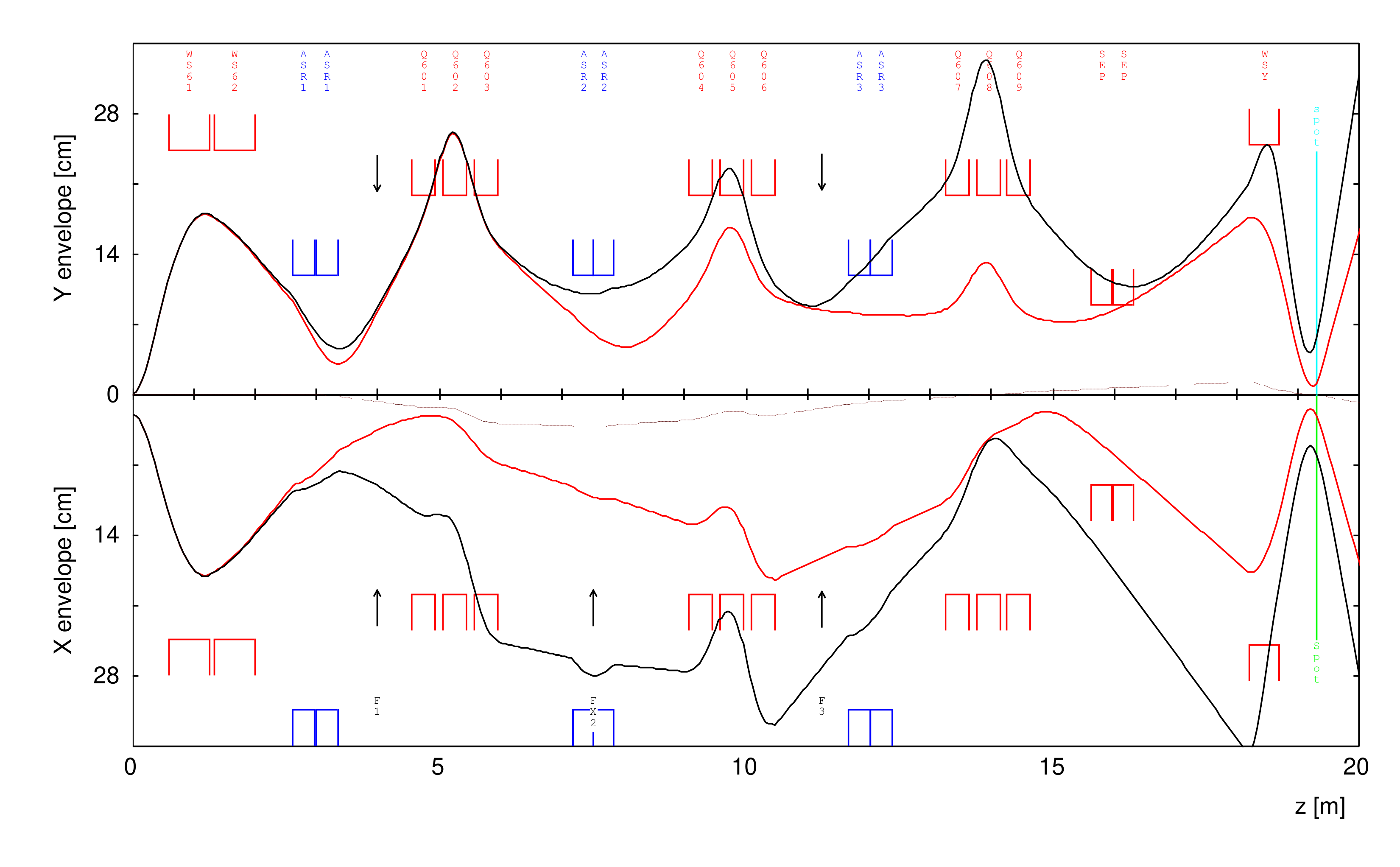}
		\\(b)
		\\
		\caption{Beam envelopes calculations of the $\mu$E4 beamline using {\tt TRANSPORT}: (a) the current $\mu$E4 beamline; (b) the upgraded $\mu$E4 beamline with a solenoid replacing the last quadrupole triplet. Initial parameters used in {\tt TRANSPORT} are (half-width): x = 20 mm, x’ = 200 mrad, y = 2.5 mm, y’ = 200 mrad, dp/p = 0 (red lines) and 4.5$\%$ (black lines). The red lines indicate the first order calculations and the black lines are the second order calculations, respectively. The arrows represent the three slit systems with maximum jaw opening.}
		\label{FIG5} 
	\end{figure}

	Figure~\ref{FIG5} shows the $\mu$E4 beam envelope optical calculations using the program {\tt TRANSPORT} \cite{Rohrer_TRANSPORT}, which is based on the calculation of beam transfer matrices. In contrast to the existing $\mu$E4, the beam waist at the end of the beamline at the moderator position can be made symmetrical in $x$ and $y$ using a solenoid instead of the last quadrupole triplet. This indicates that a higher beam intensity on the moderator area can be achieved by symmetric focusing. In {\tt TRANSPORT}, the transfer matrix of solenoids is defined for "long" coils, where the length is much larger than the aperture. In the $\mu$E4 case with "short" solenoids, the length is about the same as the aperture. Also, the iron housing of the $\mu$E4 solenoids affects the fringe fields of the solenoids, which is not taken into account by the standard transfer matrix for a solenoid in {\tt TRANSPORT}. Nevertheless, as has been shown in \cite{Prokscha_NewMuE4_2008}, the transfer matrix formalism gives a sufficiently well description of the beam optics with the short solenoids, which is beneficial for efficient beamline optimization and for a quick and instructive evaluation of the main characteristics of the beamline.
	
\begin{figure}[tbp]
		\centering
		\includegraphics[width=12cm]{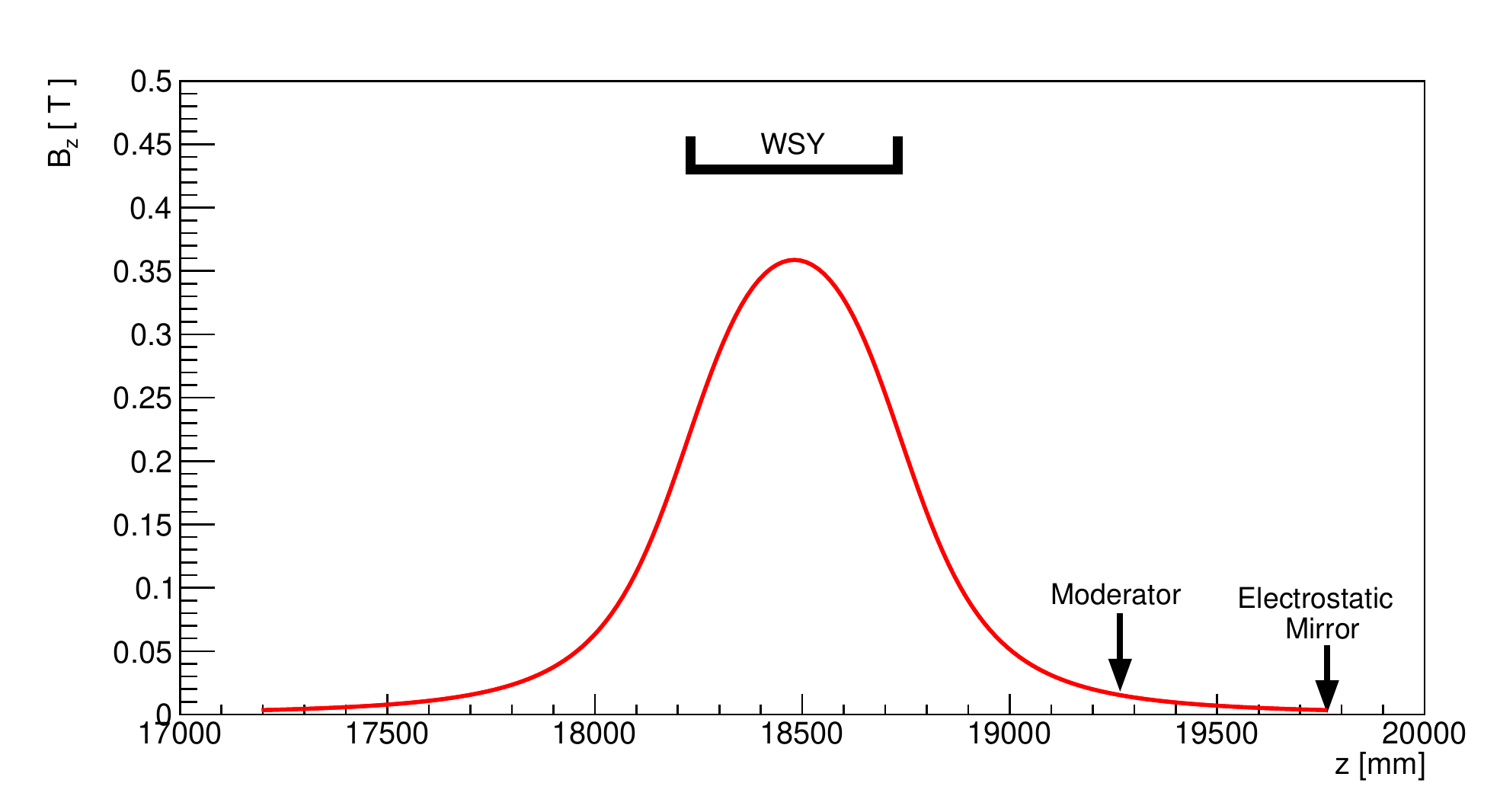}
		\\(a)
		\\
		\includegraphics[width=12cm]{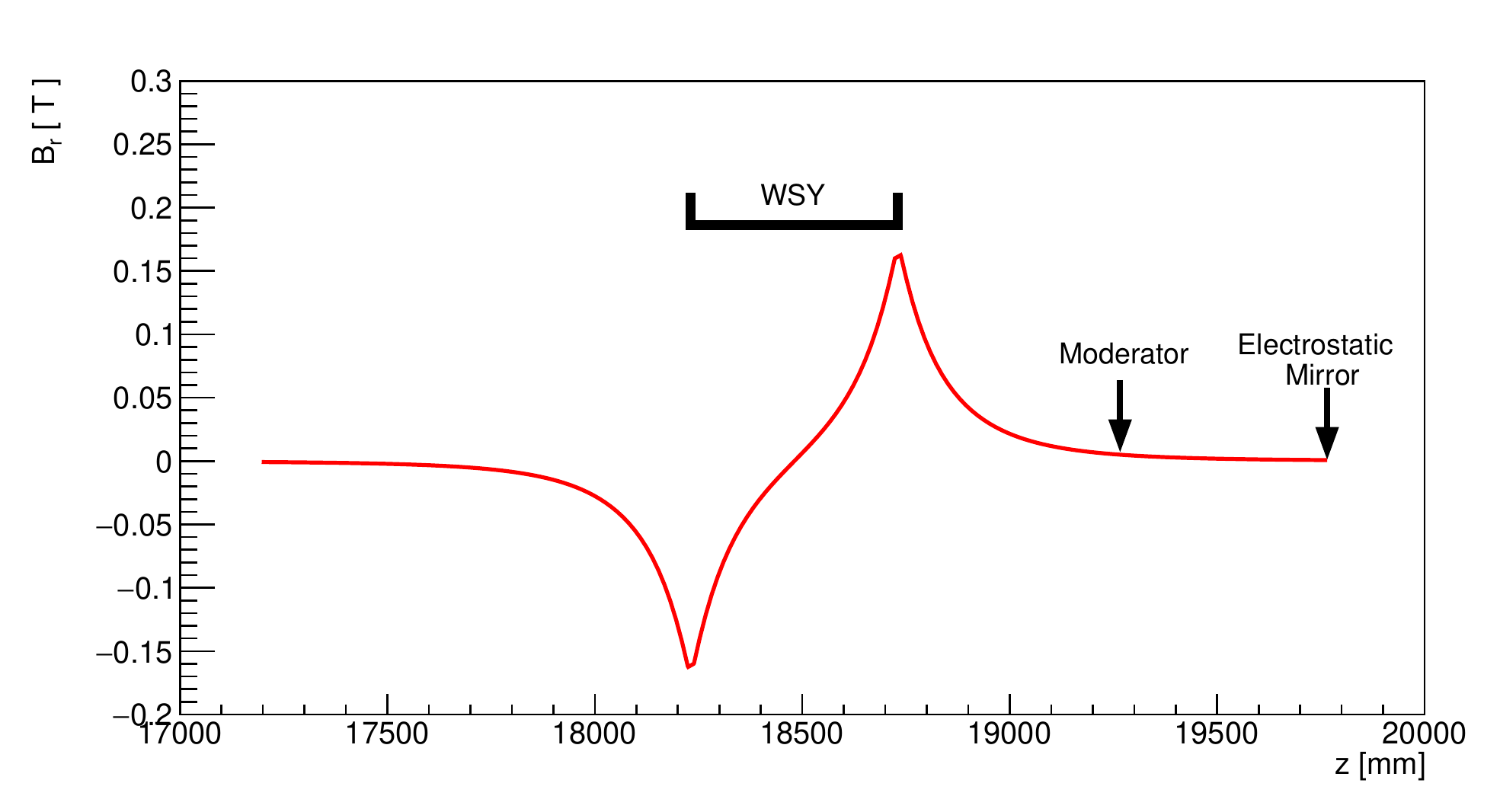}
		\\(b)
		\\
		\caption{The magnetic field of the solenoid used in G4beamline: (a) the $B_z$ component along the center line; (b) the $B_r$ component at maximum aperture r = 20 cm. The field has been tuned to get the maximum intensity on moderator. The horizontal axis shows the position in the beamline. The center of the solenoid WSY is located at z = 18481 mm and moderator is located at z = 19266 mm.}
		\label{FIG6} 
	\end{figure}

\begin{figure}[tbp]
		\centering
		\includegraphics[width=12cm]{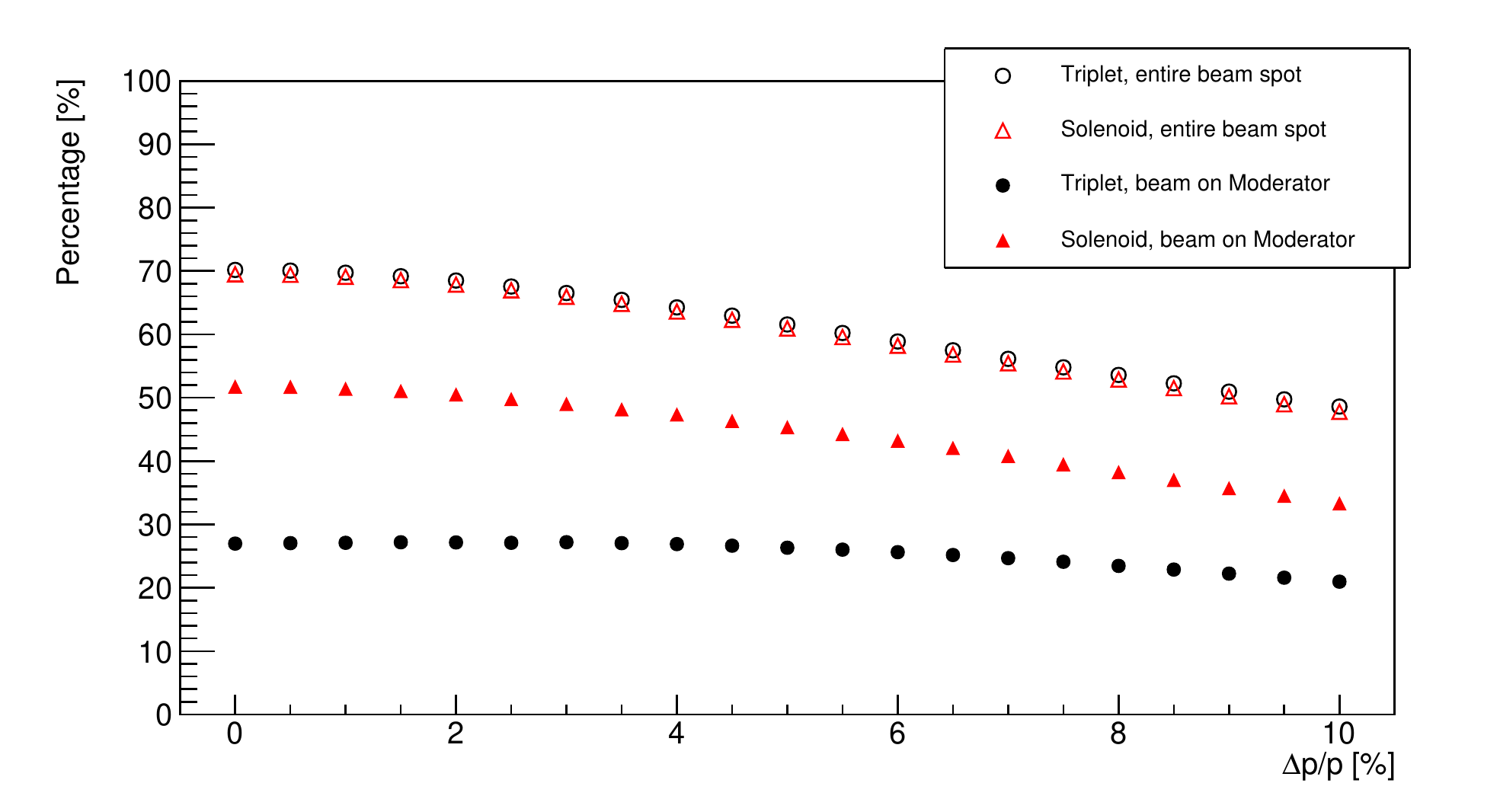}
		\caption{The fractions of muons with the initial phase spaces used in {\tt TRANSPORT} that are transported to the position of the moderator in G4beamline for the current $\mu$E4 beamline and the upgraded $\mu$E4 beamline. 100000 muons for each momentum bite settings are generated in the simulation at the muon target E. Statistical errors of simulations are less than 1\%. } 
		\label{FIG7} 
	\end{figure}

	The beam optics is only calculated to 2nd order in {\tt TRANSPORT}. 
	For the transmission studies of the large emittance beams, including a correct transport through the solenoids, a multi-particle Monte-Carlo simulation is therefore indispensable. Therefore in a next step, G4beamline is employed for the simulation of beam transport in which the uniform distribution of the initial phase spaces in {\tt TRANSPORT} is used as the input beam source. In the Monte-Carlo simulation, the fractions of muons transported onto the moderator target of the existing $\mu$E4 beamline for different momentum bites are compared with those of the upgraded $\mu$E4 beamline, where the last quadrupole triplet is replaced by a simple solenoid named "WSY" with an aperture of 0.5~m and a length of 0.5~m. The field of the G4beamline solenoid model, which is shown in Fig.~\ref{FIG6}, is calculated based on the Biot-Savart law and represents a real solenoid without any iron housing. Figure~\ref{FIG7} shows the simulation results by using the G4beamline solenoid WSY. In the simulation, the existing $\mu$E4 beamline can focus 38\% -- 44\% of the whole beam on the moderator, depending on the momentum bite. The G4beamline solenoid WSY can focus more muons on the moderator regardless of different momentum bite, where the total transmission efficiency remains almost unchanged since the WSY has the same aperture as the quadrupole magnets. For example, using this simplified initial phase space, $\sim$~70\% of the beam is focused on the moderator at the full momentum bite of 10\% of the $\mu$E4 beamline by using WSY.


\begin{figure}[tbp]
		\begin{center}
			\includegraphics[width=16cm]{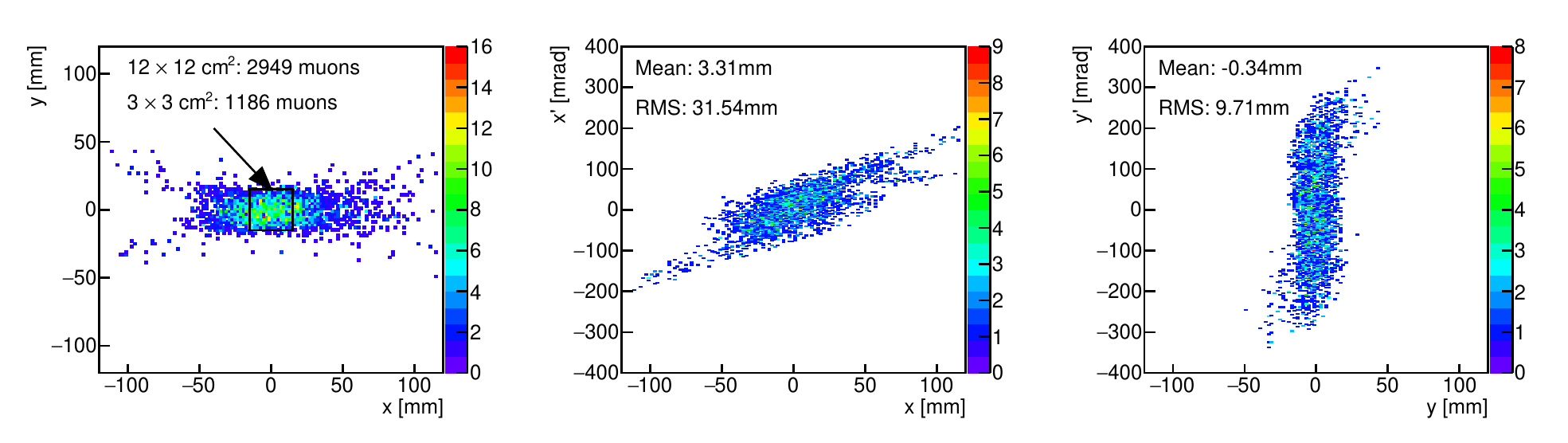}
			\\(a)
			\\\includegraphics[width=16cm]{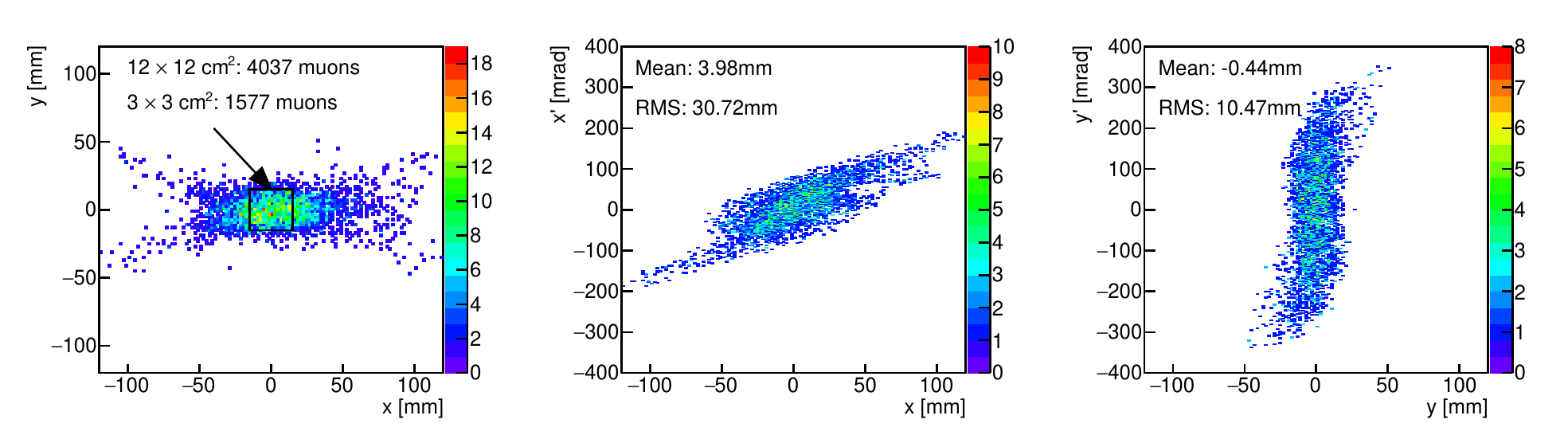}
			\\(b)
			\\\includegraphics[width=16cm]{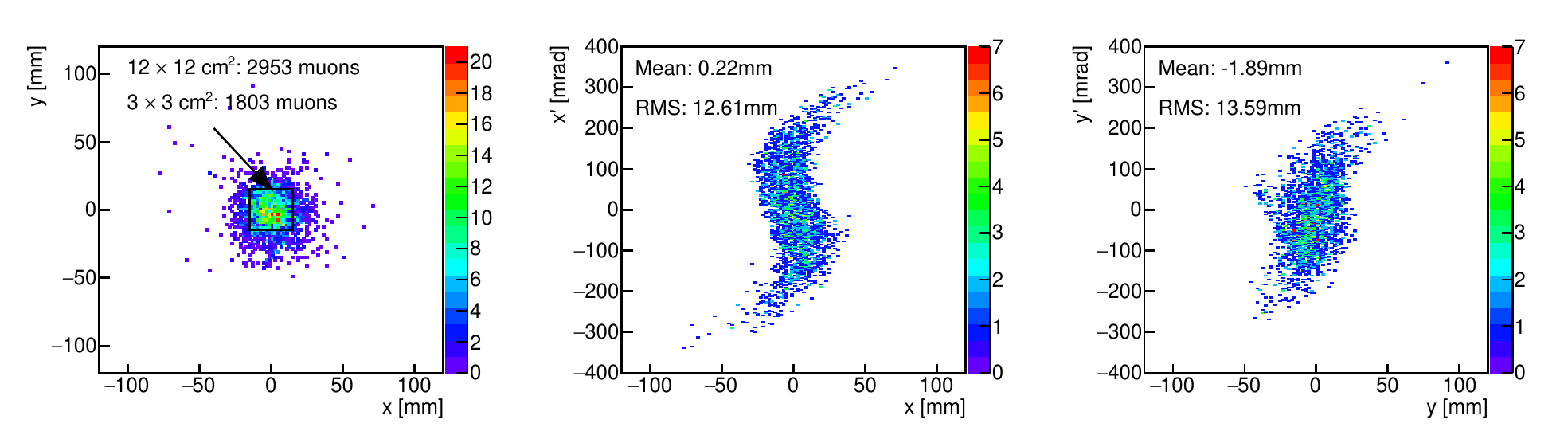}
			\\(c)
			\\\includegraphics[width=16cm]{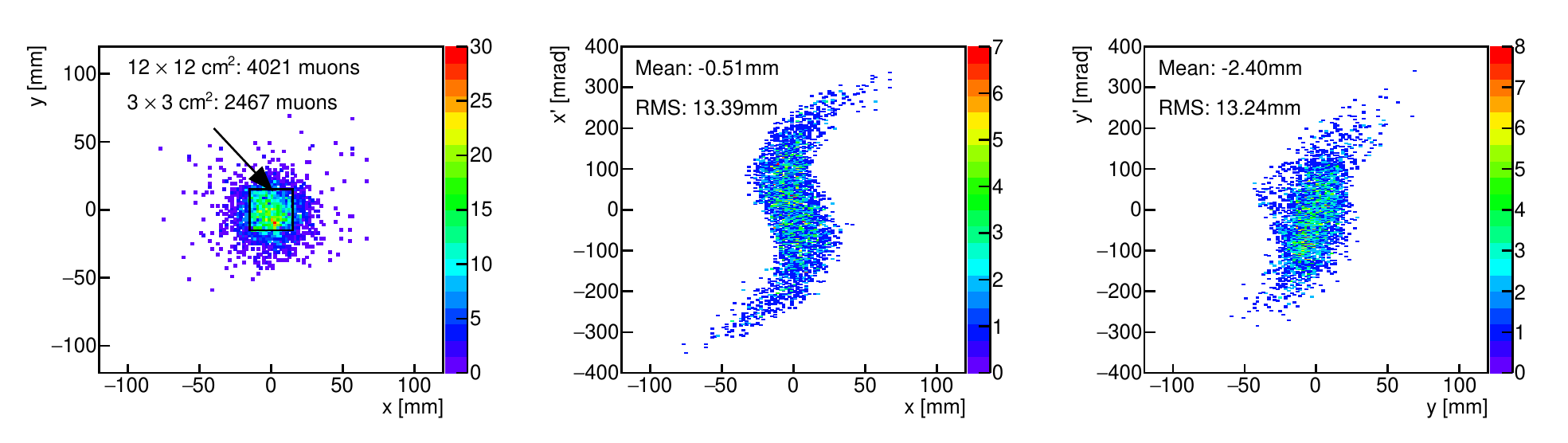}
			\\(d)
		\end{center}
		\caption{Simulated beam spots and phase spaces at the LEM moderator position for different muon target E geometries and beamline settings: (a) OTE + quadrupole triplet(i.e., the original setting of the $\mu$E4); (b) NTE + quadrupole triplet; (c) OTE + G4beamline solenoid WSY; (d) NTE + G4beamline solenoid WSY. For the NTE simulation, the center of the $\mu$E4 beamline is aligned to the center of the muon distribution ($x$ = -8~mm in FIG.~4b).}
		\label{FIG8} 
	\end{figure}

\begin{figure}[tbp]
		\centering
		\includegraphics[width=12cm]{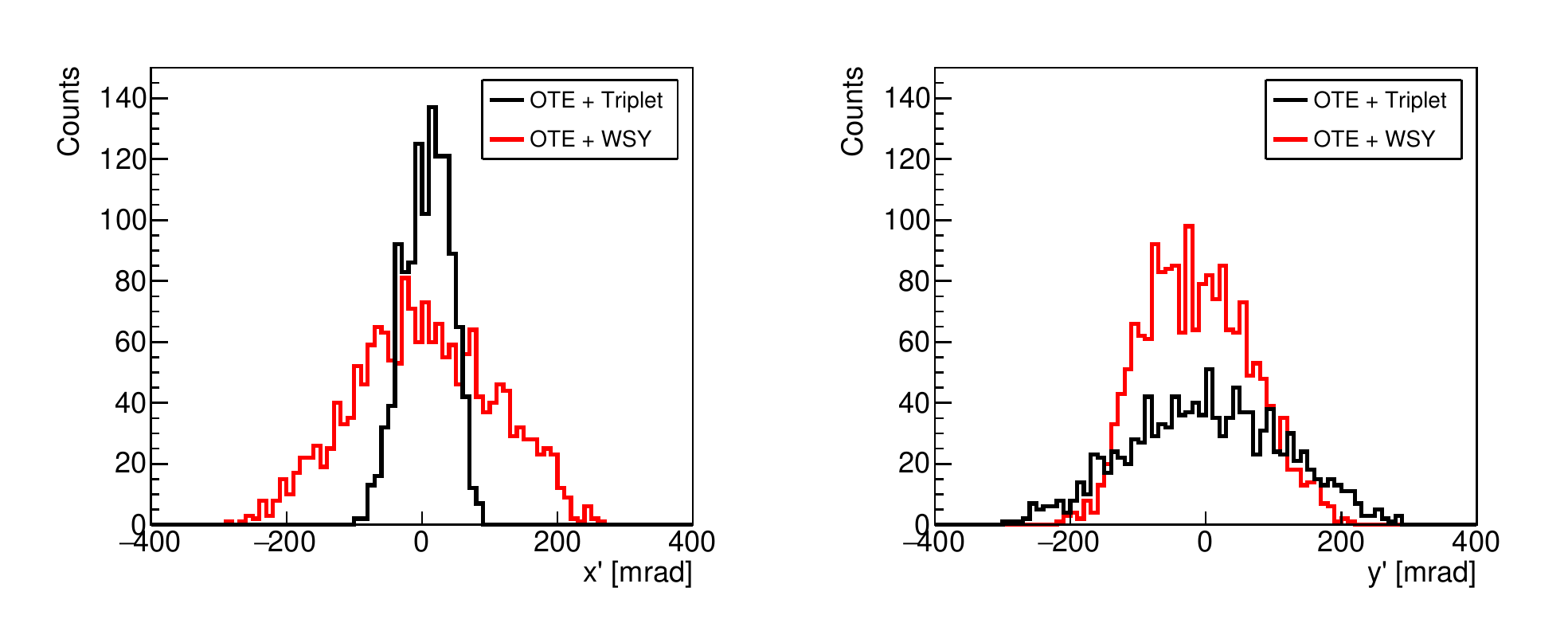}		
		\caption{Comparisons of the transverse divergences for different beam spots. Left: horizontal direction; right: vertical direction.}
		\label{FIG9} 
	\end{figure}

	The simulated beam spots and phase spaces at the moderator position using the initial phase spaces of Fig.~\ref{FIG4} for the OTE and the NTE are compared and shown in Fig.~\ref{FIG8}. To optimize the fraction of muon beam focused on the moderator with a fixed area size ($30\times 30$~mm$^2$), the divergences of the focused beam $x'$ and $y'$ have to be maximized. For obtaining the highest intensity on moderator, the triplet can generate a beam waist with large divergence in only one direction ($y$), while the divergence in the other direction ($x$) is limited to $\sim 100$~mrad. In the solenoid case, both divergences $x'$ and $y'$ can reach large values up to 200 mrad (see Fig.~\ref{FIG9}), and the original rectangle-like shape of the beam spot becomes more like a circle symmetric about $x$ and $y$, as shown in Fig.~\ref{FIG8}. The gain in muon rate on the moderator by using the G4beamline solenoid WSY compared with the quadrupole triplet is 52\% for the OTE, which is less than the 70\% using the simplified initial phase space of the {\tt TRANSPORT} simulation.
	
	Compared to the OTE, the employment of the NTE in the simulation results in additional increases of muon rate of 33\% for the current $\mu$E4 beamline by comparing Fig.~\ref{FIG8}(b) with Fig.~\ref{FIG8}(a), and 37\% for the upgraded $\mu$E4 beamline by comparing Fig.~\ref{FIG8}(d) with Fig.~\ref{FIG8}(c). The overall increase of muon rate on moderator by using the NTE with larger muon production volume/surface, and by using a solenoid at the end of the beamline amounts to 108\% by comparing Fig.~\ref{FIG8}(d) with Fig.~\ref{FIG8}(a).
	

\begin{figure}[tbp]
		\centering
		\includegraphics[width=12cm]{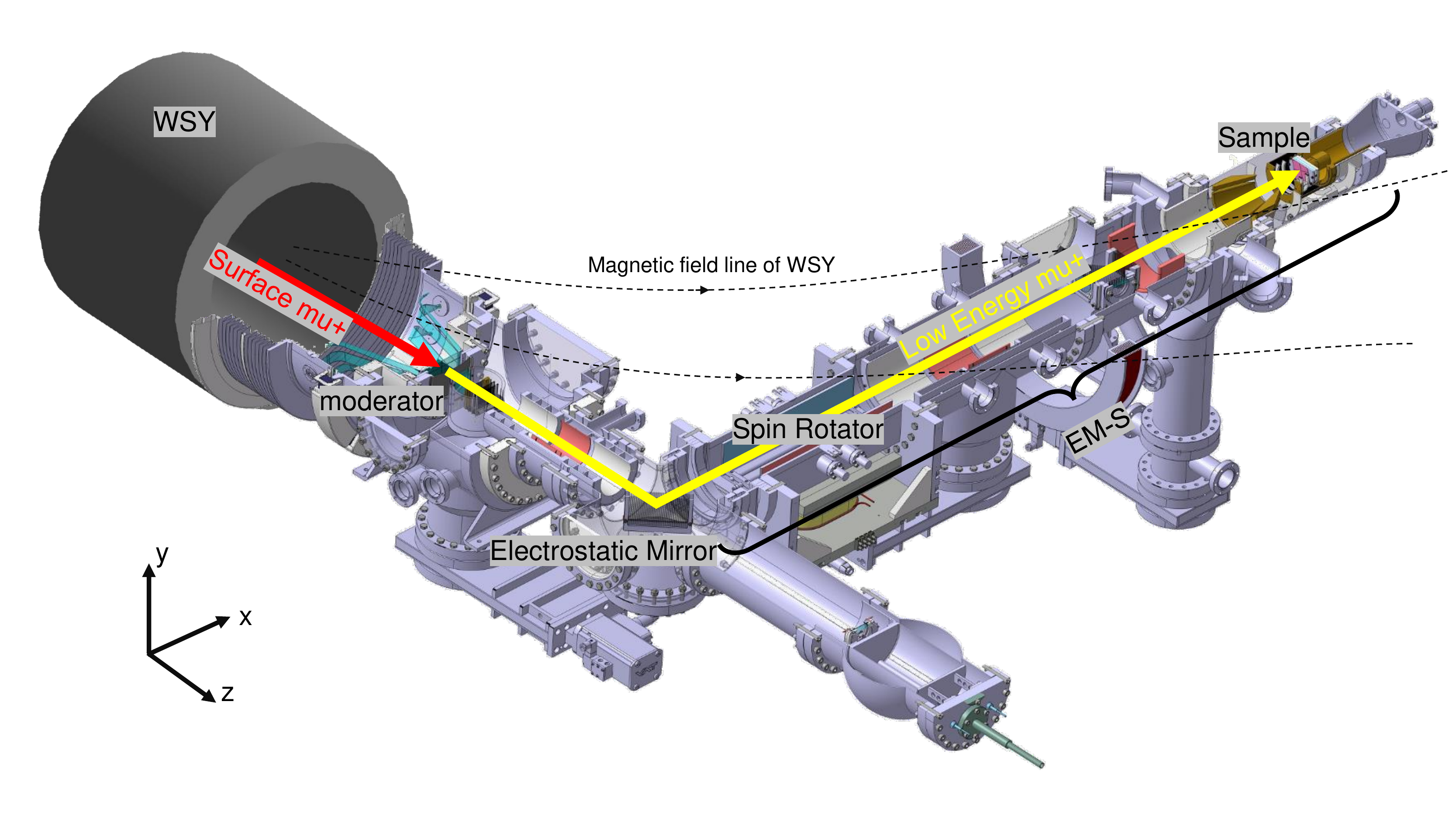}
		\caption{The profile of the LEM facility, showing the section EM-S between the Electrostatic Mirror (EM) and the position of the Sample (S), where the LE-$\mu$SR experiments are carried out.}
		\label{FIG10} 
	\end{figure}

\begin{figure}[tbp]
		\centering
		\includegraphics[width=12cm]{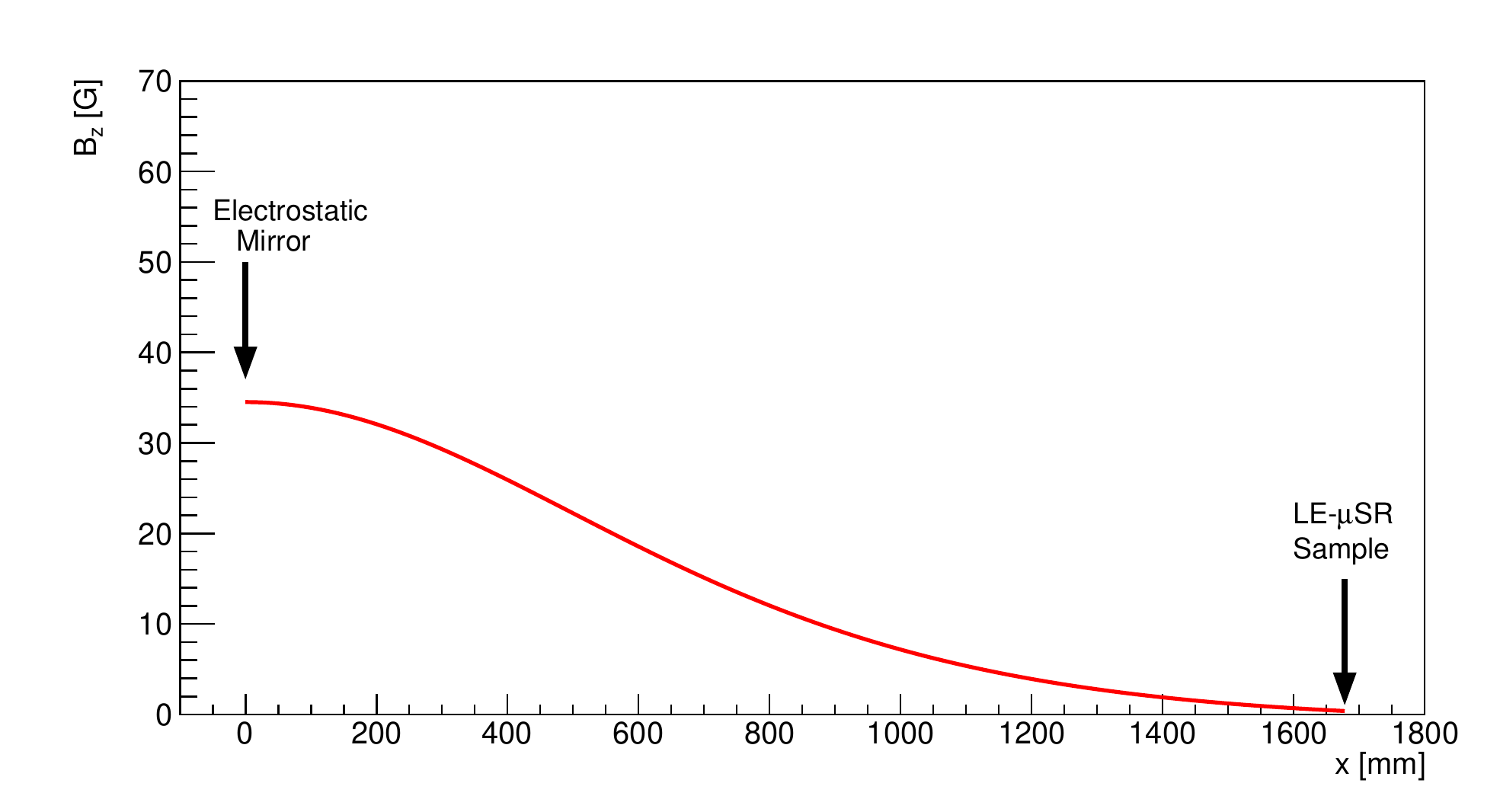}
		\caption{The stray field $B_z$ of the G4beamline solenoid WSY along the section EM-S of the LEM facility.}
		\label{FIG11} 
	\end{figure}

    Although the use of the G4beamline solenoid WSY instead of the last quadrupole triplet (QSM610-612) demonstrates the significantly better achievable intensity
    at the final beam focus, the stray field of this solenoid has a serious influence on any experiment downstream of the beam focus. We will use the LEM facility to study
    exemplarily the serious influence of these stray fields on a low energy muon beam. The deflection of the muon in a homogeneous transverse magnetic field can be calculated by the following formula \cite{Prokscha_NewMuE4_2008}
	\begin{equation}
		d(mm) = \int_{0}^{l}\varphi{_B}\textit{d}l = 15\cdot l_{B}^{2}(m)\cdot\frac{B(G)}{p(MeV/c)}
	\end{equation}
	where $l_B$ is the length of the muon trajectory and $\varphi{_B}$ is the deflection angle of the muon. The section EM-S of the LEM beamline from the Electrostatic Mirror to the position of sample is 1.678~m, as shown in Fig.~\ref{FIG10}. The deflection will be $\sim$24~mm for muons with a typical kinetic energy of 15~keV (p = 1.78~MeV/c) under 1~G homogeneous field. The overall deflection of the beam spot would be smaller due to the focusing effect of the Einzel lenses and the conical lens. However, the stray field still needs to be shielded below 1~G to keep the muon on the correct trajectory and to minimize rate loss. The stray field component $B_z$ of the G4beamline solenoid WSY without any iron housing along the EM-S beam section is shown in Fig.~\ref{FIG11}. The maximum value is 35 G when the center field B${_0}$ reaches 3588~G inside the solenoid. From Eq.~(1), the low energy muon beam would be quickly deflected and lost in the EM-S beam section. Therefore, it is necessary to design a special solenoid WSY which does not only satisfy the focusing demands of the $\mu$E4 beamline to increase the muon rate on the moderator, but which also generates sufficiently small stray fields in an acceptable range for the LEM facility, not only on-axis but also far off-axis.
	

	\begin{center}\section*{III. Investigation of possible models for the solenoid WSY}\end{center}
	
\begin{figure}[tbp]
		\centering
		\includegraphics[width=12cm]{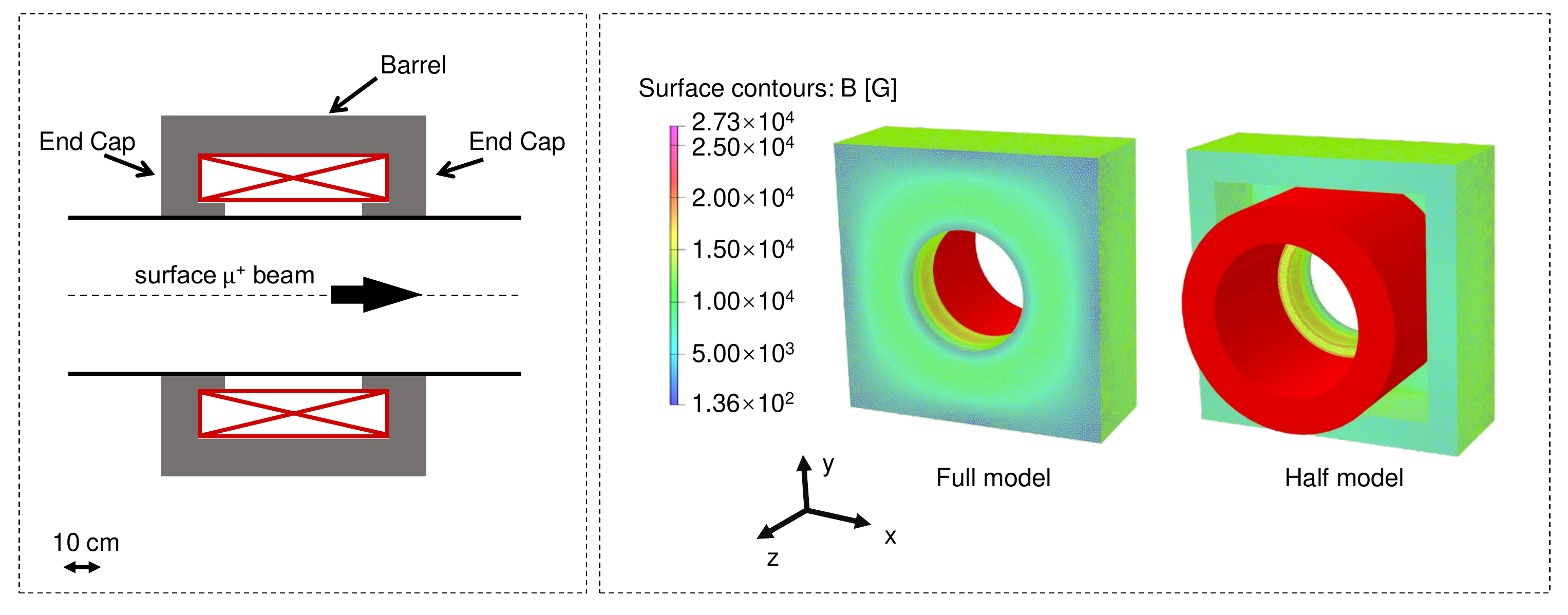}
		\caption{Model~1: the typical iron housing solenoid in Opera.}
		\label{FIG12} 
	\end{figure}

\begin{figure}[tbp]
		\centering
		\includegraphics[width=12cm]{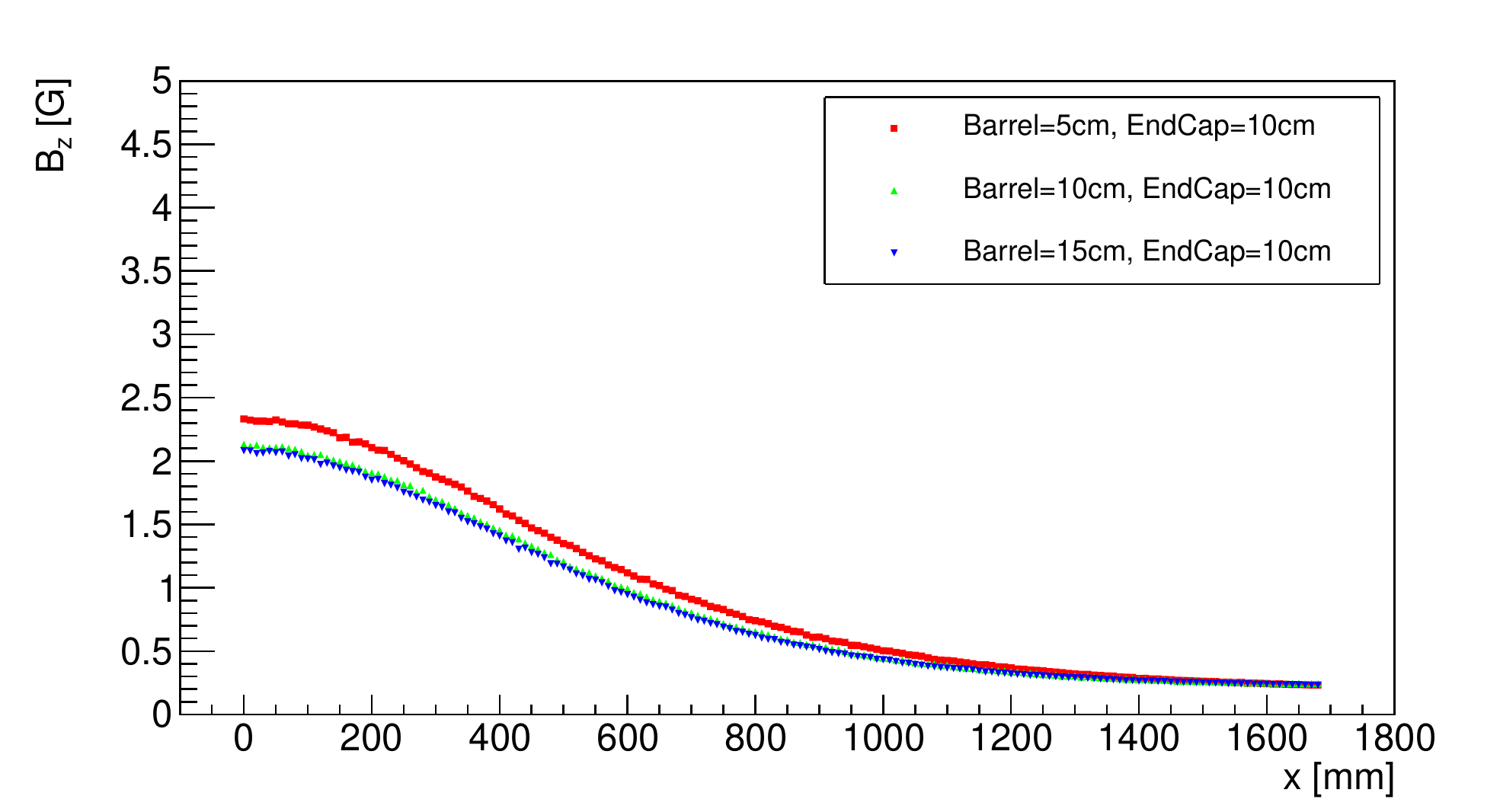}
		\\
		(a)
		\\
		\includegraphics[width=12cm]{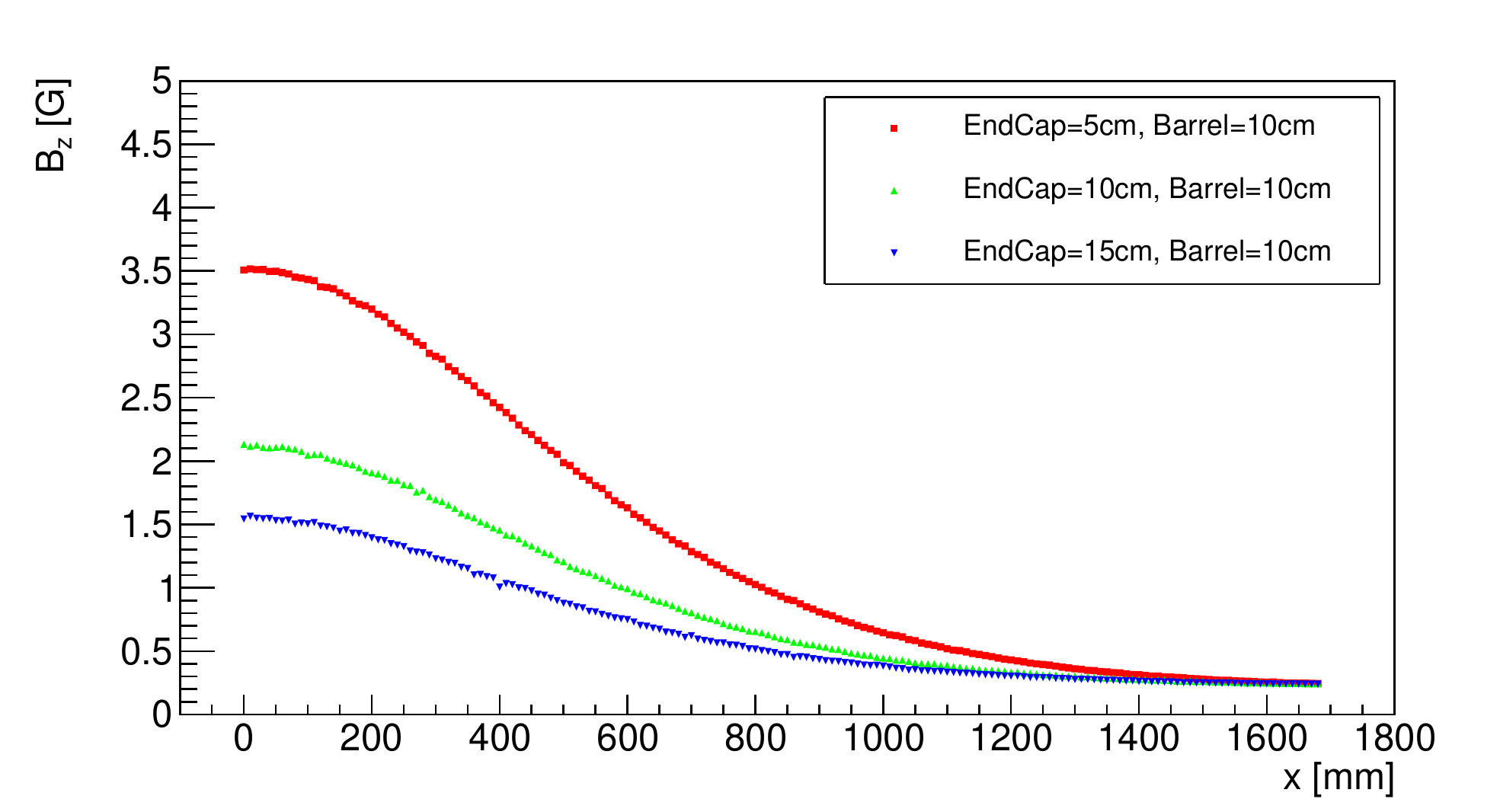}
		\\
		(b)
		\\
		\caption{Model~1: the stray field component $B_z$ of the solenoid WSY along the EM-S beam section for different thicknesses of the iron housing. (a) the WSY with different barrel thicknesses and end cap thicknesses fixed at 10~cm; (b) the WSY with different end cap thicknesses and barrel thicknesses fixed at 10~cm.}
		\label{FIG13} 
	\end{figure}

	Three kinds of models have been investigated in attempting to reduce the stray field in the LEM region. The idea is to either use the iron housing or add compensation coils with reverse current direction on the outside, or combine them together. Solenoid models are built and analyzed with the finite element analysis software Opera \cite{OPERA-3D_web}. Figure~\ref{FIG12} shows the traditional iron housing method that was used in the existing solenoids WSX61/62 of the $\mu$E4 beamline. The relatively small moderator area requires accurate focusing of the beam. Any small offset of the solenoid in its position could lead to unacceptable beam losses. The compensation iron inside the solenoid helps ensuring the precise installation and enables a more efficient operation. For the strict requirement of the LEM facility, different acceptable sizes of iron housing have been investigated for reducing the stray field component $B_z$ along the EM-S beam section. The analysis results are shown in Fig.~\ref{FIG13} and all their focusing powers $\frac{1}{f}$ \cite{Kumar_Solenoid_2009} have been normalized to what is used in the G4beamline simulation as mentioned in the Fig.~\ref{FIG7}. The outer iron housing is important to shield the stray field leaking into LEM, as long as it remains unsaturated. In general, the use of the iron housing can reduce the stray field to a few Gauss with acceptable iron thicknesses. Increasing the barrel thickness beyond a certain value does not further reduce the overall stray field at the EM-S beam section. Increasing the thickness of the end cap is effective for reducing the stray field at the position closer to the end cap, that is, it can reduce the stray field strength at the Electrostatic Mirror. As the distance from the end cap increases, the shielding ability of the end cap decreases.

	
\begin{figure}[tbp]
		\centering
		\includegraphics[width=12cm]{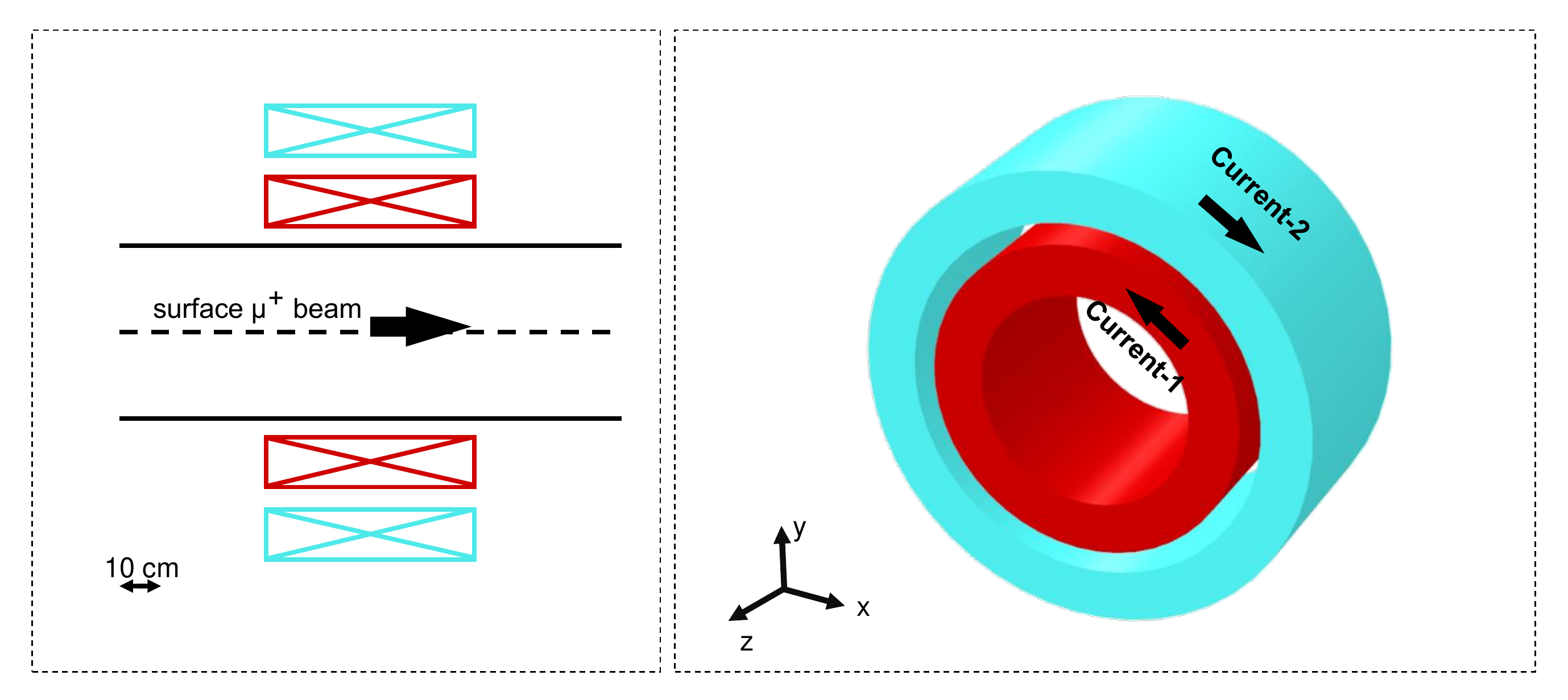}
		\caption{Model~2: the dual-layer solenoid. The outer coil has the direction of current opposite to the inner coil (blue: the outside coil, red: the inner coil).}
		\label{FIG14} 
	\end{figure}

\begin{figure}[tbp]
		\centering
		\includegraphics[width=12cm]{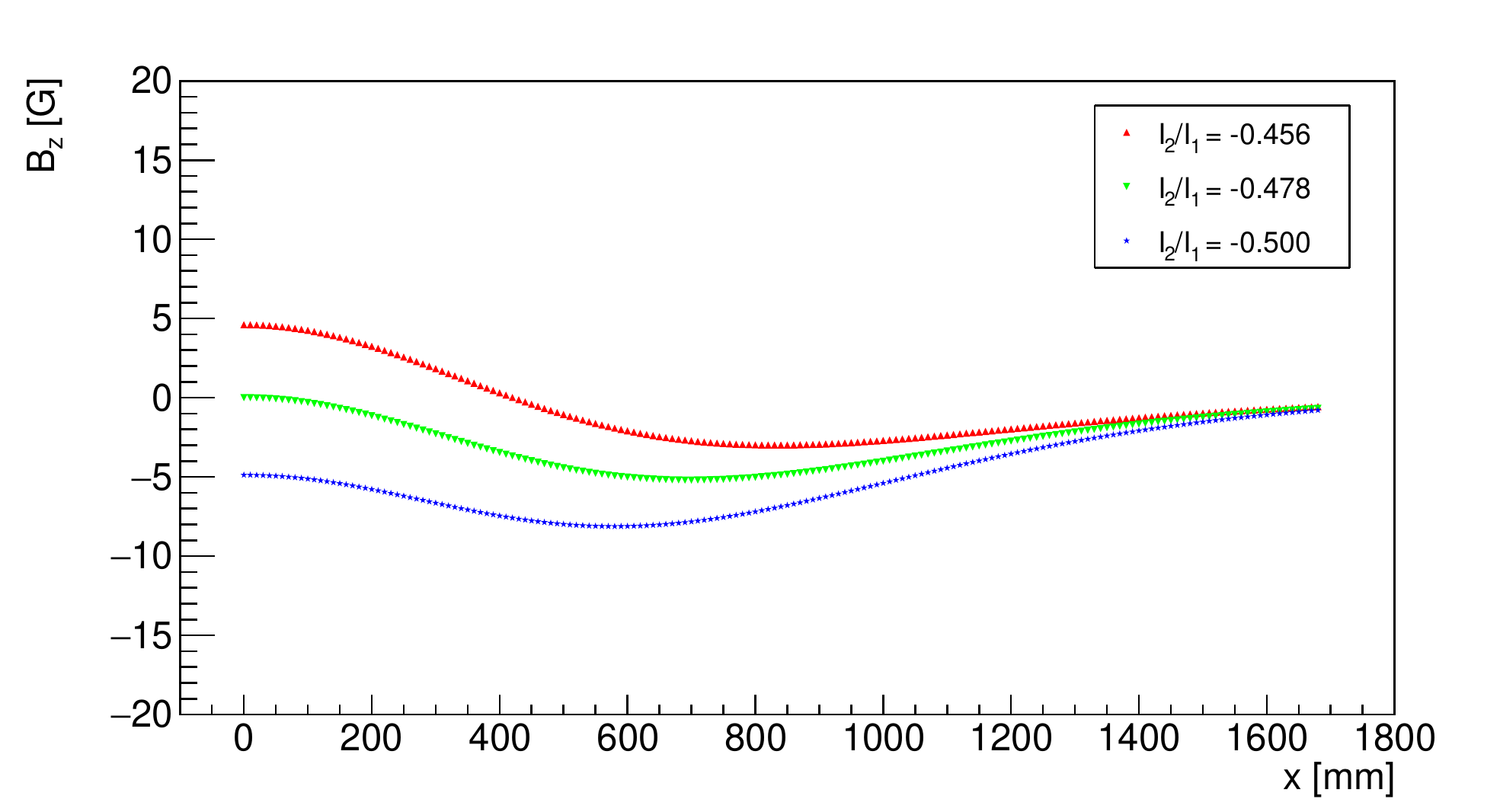}
		\caption{Model~2: stray field $B_z$ along the EM-S beam section for the dual-layer solenoid with different current ratios.}
		\label{FIG15} 
	\end{figure}

	The second model investigated is the dual-layer solenoid, which has a main solenoid and an outer solenoid with current in the opposite direction \cite{Laxdal_Solshielding_2013,Takeuchi_Dualsol_1997}, see Fig.~\ref{FIG14}. By varying the ratio of the currents of the two solenoids, the field inside the solenoid and the stray field at the EM-S beam section can be changed. Stray fields with different ratios of currents are shown in Fig.~\ref{FIG15}. All of their focusing powers have been normalized to what is used in the G4beamline simulation as mentioned in the first model. To achieve optimum focusing on the moderator and minimum stray fields, the inner coil needs to be operated  at a high current density exceeding 5 A/mm${^2}$ to compensate the negative effect on the focusing power by the outer coils. This would probably require the use of superconducting technology, since $\sim 5$~A/mm$^2$ is the limit for normal-conducting coils.

	
\begin{figure}[tbp]
		\centering
		\includegraphics[width=12cm]{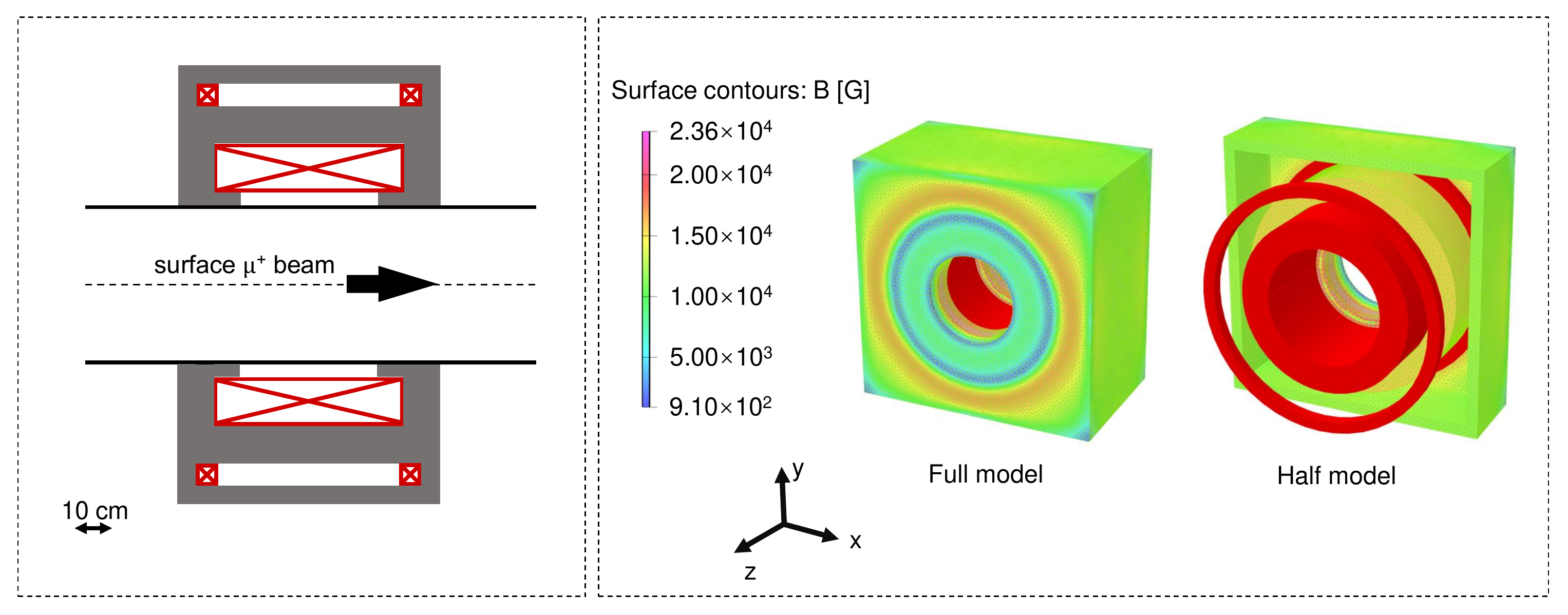}
		\caption{Model~3: the iron housing solenoid with compensation coils.}
		\label{FIG16} 
	\end{figure}

\begin{figure}[tbp]
		\centering
		\includegraphics[width=12cm]{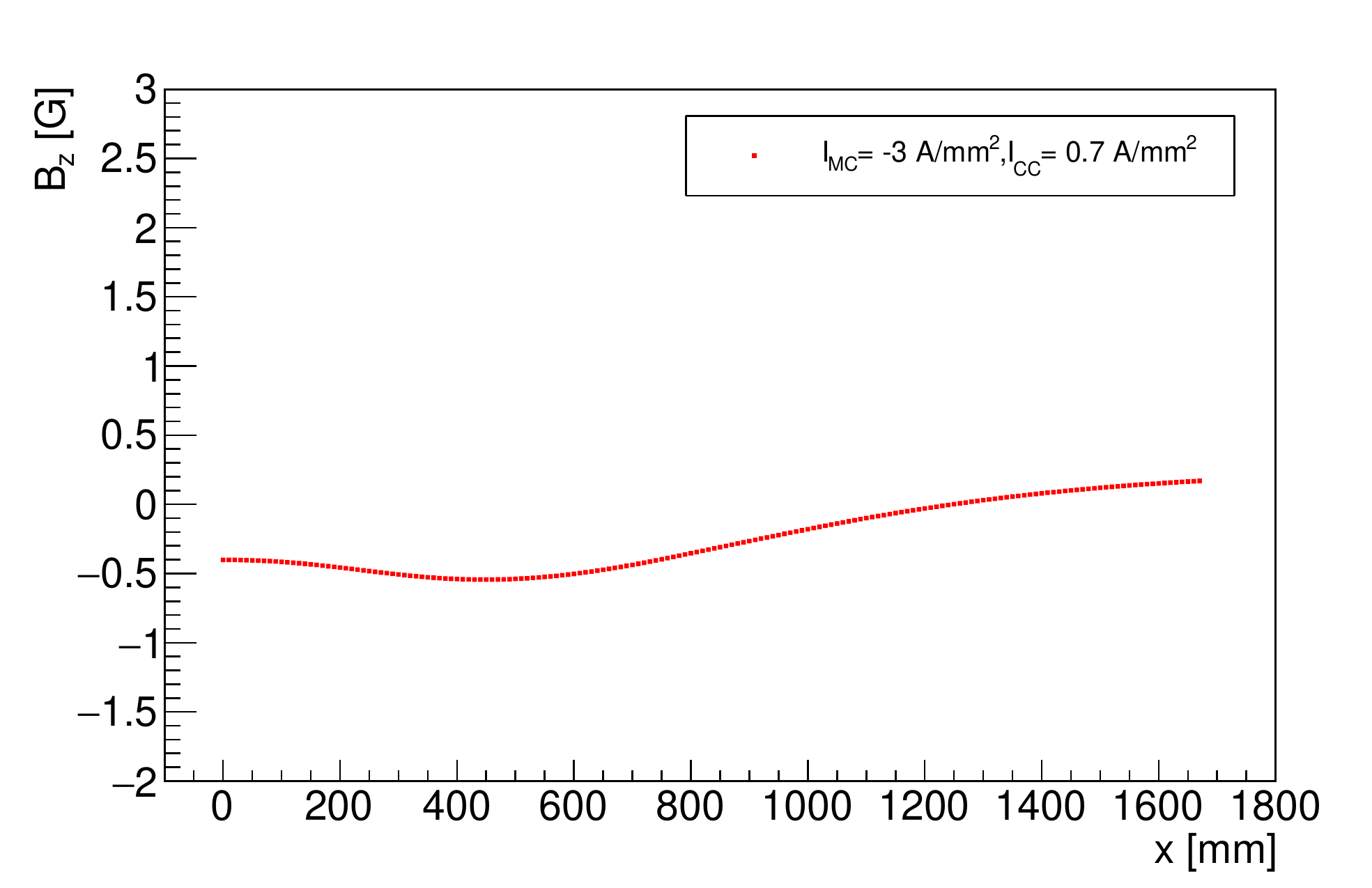}
		\caption{Model~3: the stray field $B_z$ along the EM-S beam section. The main coil (MC) works at a current density of 3~A/mm$^{2}$ and the compensation coils (CC) work at 0.7~A/mm$^{2}$ with opposite current direction}
		\label{FIG17} 
	\end{figure}

\begin{figure}[tbp]
		\centering
		\includegraphics[width=12cm]{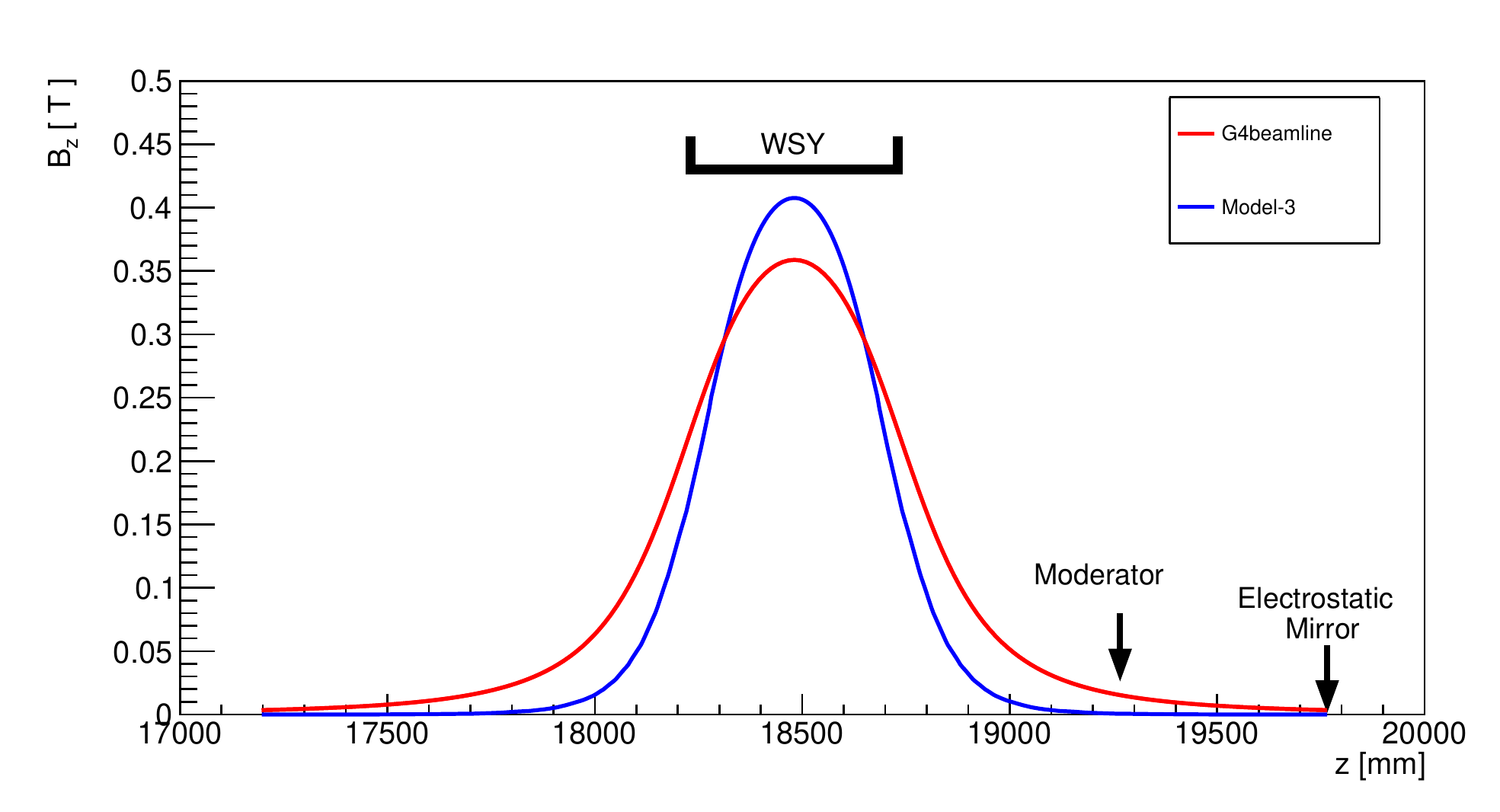}
		\\
		(a)
		\\
		\includegraphics[width=12cm]{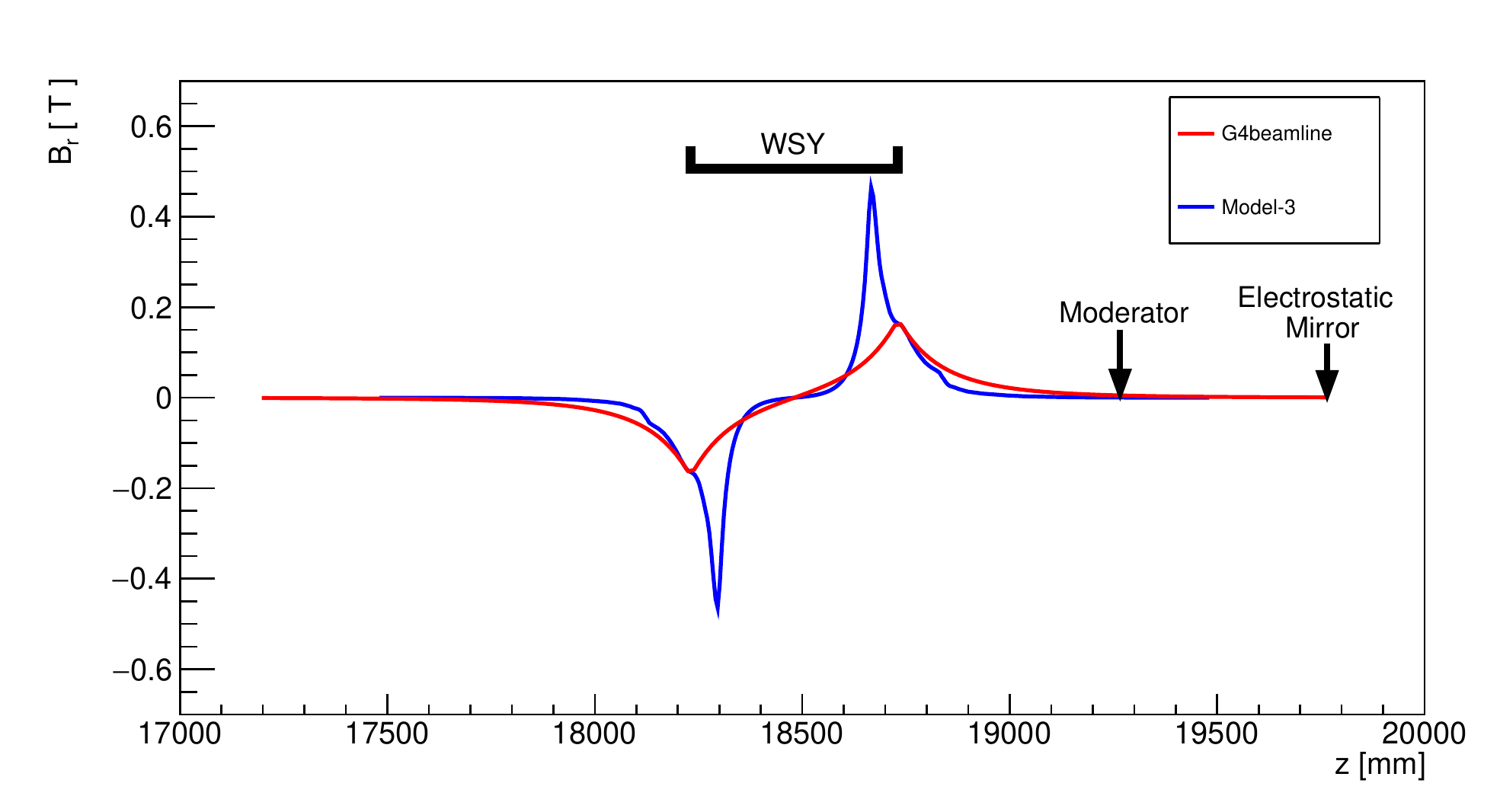}
		\\
		(b)
		\\
		\caption{Comparison of magnetic fields of the G4beamline solenoid and the Opera model-3: (a) $B_z$ along the solenoid central axis; (b) $B_r$ at the maximum aperture r = 20 cm. Both the fields of two solenoids have been normalized to have the same focusing power. }
		\label{FIG18} 
	\end{figure}

\begin{figure}[tbp]
		\centering
		\includegraphics[width=12cm]{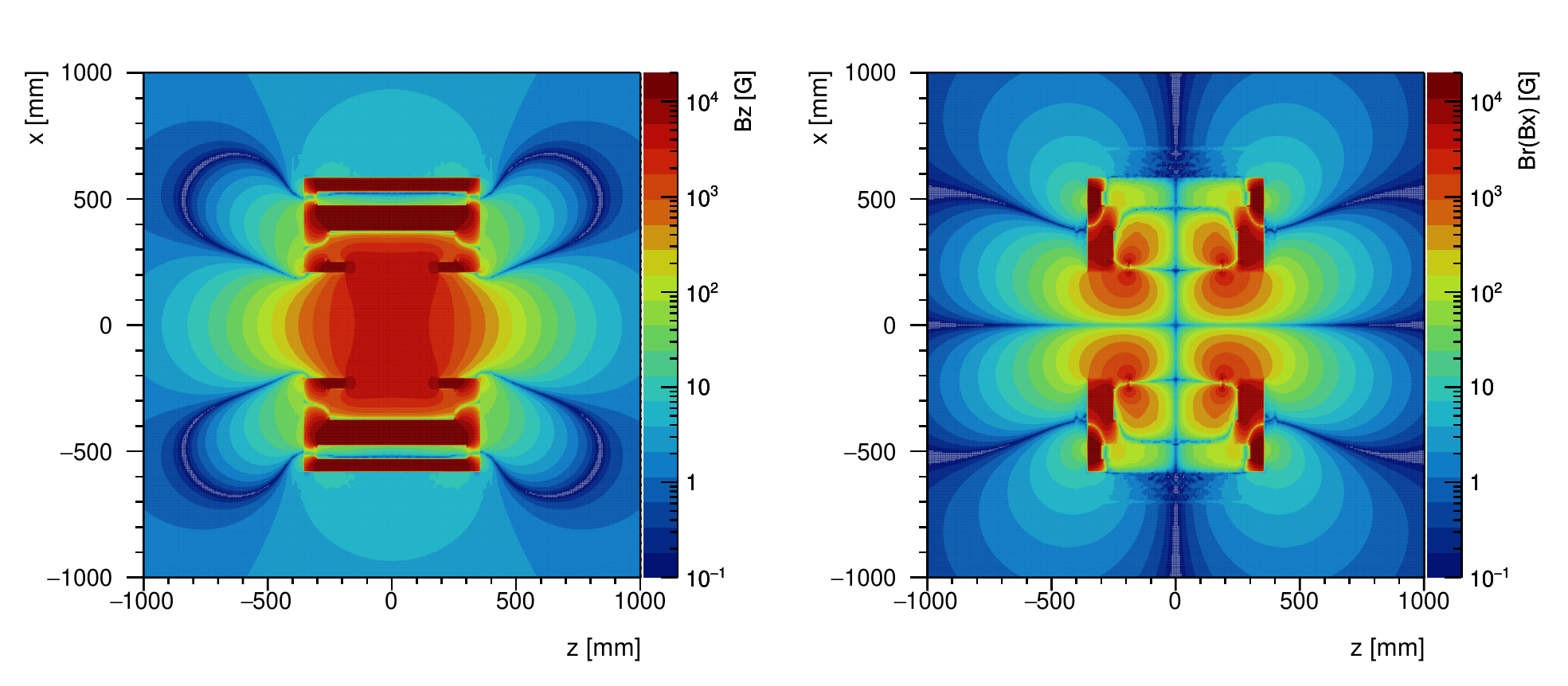}
		\caption{Contour map of $B_z/B_r$ on the XOZ plane of the model-3 WSY.}
		\label{FIG19} 
	\end{figure}

	At a ratio I${_2}$/I${_1}$ = -0.478, the stray field at the position of the Electrostatic Mirror can be compensated to almost 0. However, the stray field at the middle position of the EM-S beam section still can reach -5~G, and the beam will be lost in the EM-S beam section. Hence we conclude that the outer coil alone cannot compensate the stray fields to an acceptable low value along the entire EM-S beam section. Therefore, the third model which combines the first and the second model to further minimize the stray field along the EM-S beam section is investigated, as shown in Fig.~\ref{FIG16}. This model has been optimized to keep the iron housing unsaturated except the inner iron part located inside the main coil. In order to avoid the superconducting technology and considering the conductor area ratio in the water-cooled wire, all the current densities are kept below 3~A/mm${^2}$. The stray field of this model is shown in Fig.~\ref{FIG17} and its focusing power has been normalized as in the previous models. The center field becomes sharper compared to the solenoid center field used in G4beamline, as shown in Fig.~\ref{FIG18}. The stray field of model-3 along the EM-S beam section reaches the magnitude of the geomagnetic field, which is small enough for a nearly undisturbed beam transport, as shown below. A contour map of $B_z/B_r$ outside model-3 WSY is shown in Fig.~\ref{FIG19}.

	
\begin{figure}[tbp]
		\centering
		\includegraphics[width=12cm]{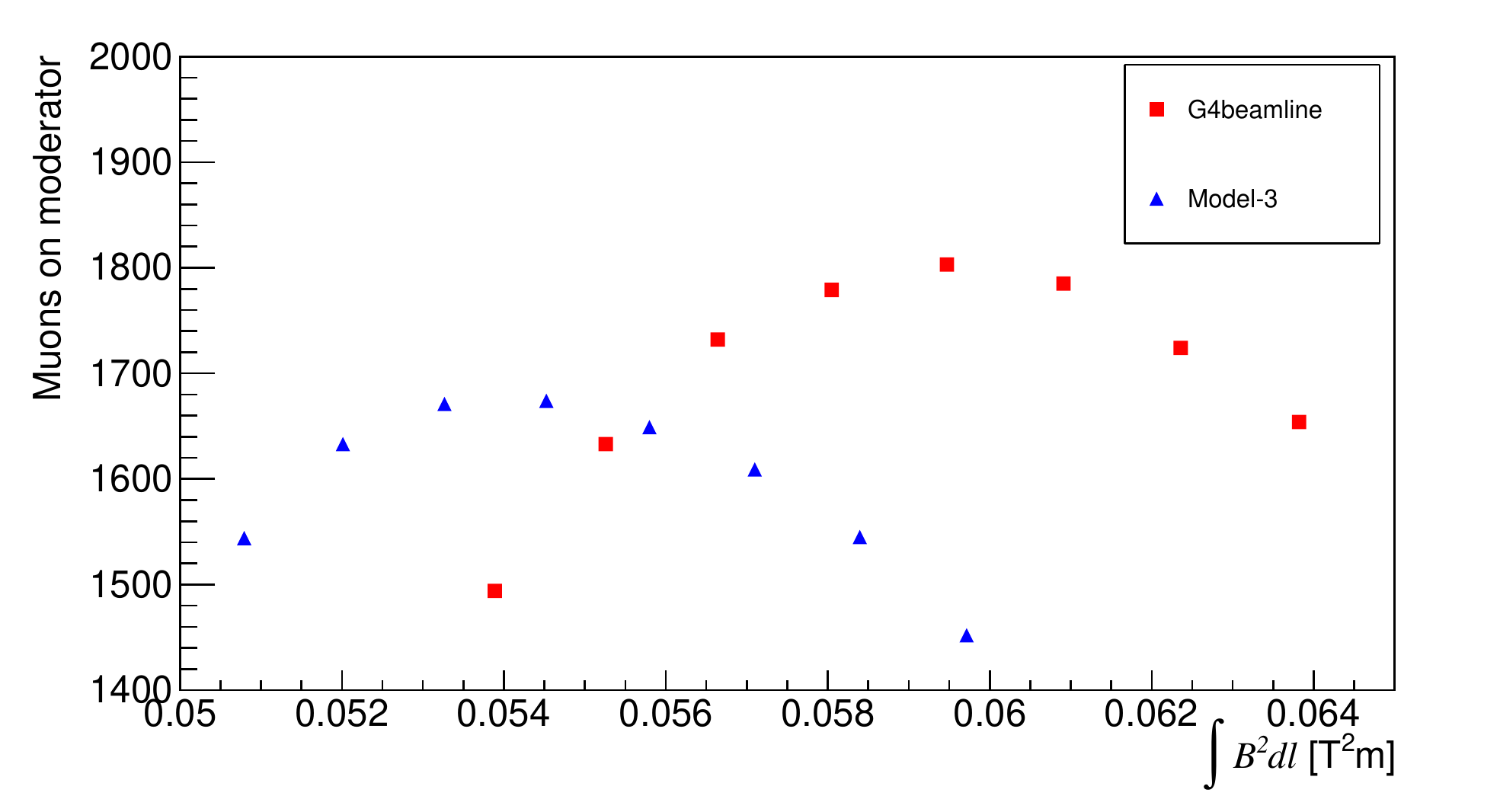}
		\caption{The comparison of focusing capabilities of model-3 WSY and G4beamline solenoid by using OTE in the simulation. The red squares show the number of muons on the moderator for the G4beamline solenoid as a function of focusing power (max = 1803~muons). The blue triangles show the result for model-3 WSY (max = 1685~muons).}
		\label{FIG20} 
	\end{figure}

	The model-3 WSY has different performance compared to the G4beamline solenoid due to the shielding of the stray field. Using the OTE in the simulation, the focusing ability of the model-3 solenoid WSY is weaker than that of the G4beamline solenoid by about 7\%, as shown in Fig.~\ref{FIG20}.

\begin{figure}[tbp]
		\centering
		\includegraphics[width=12cm]{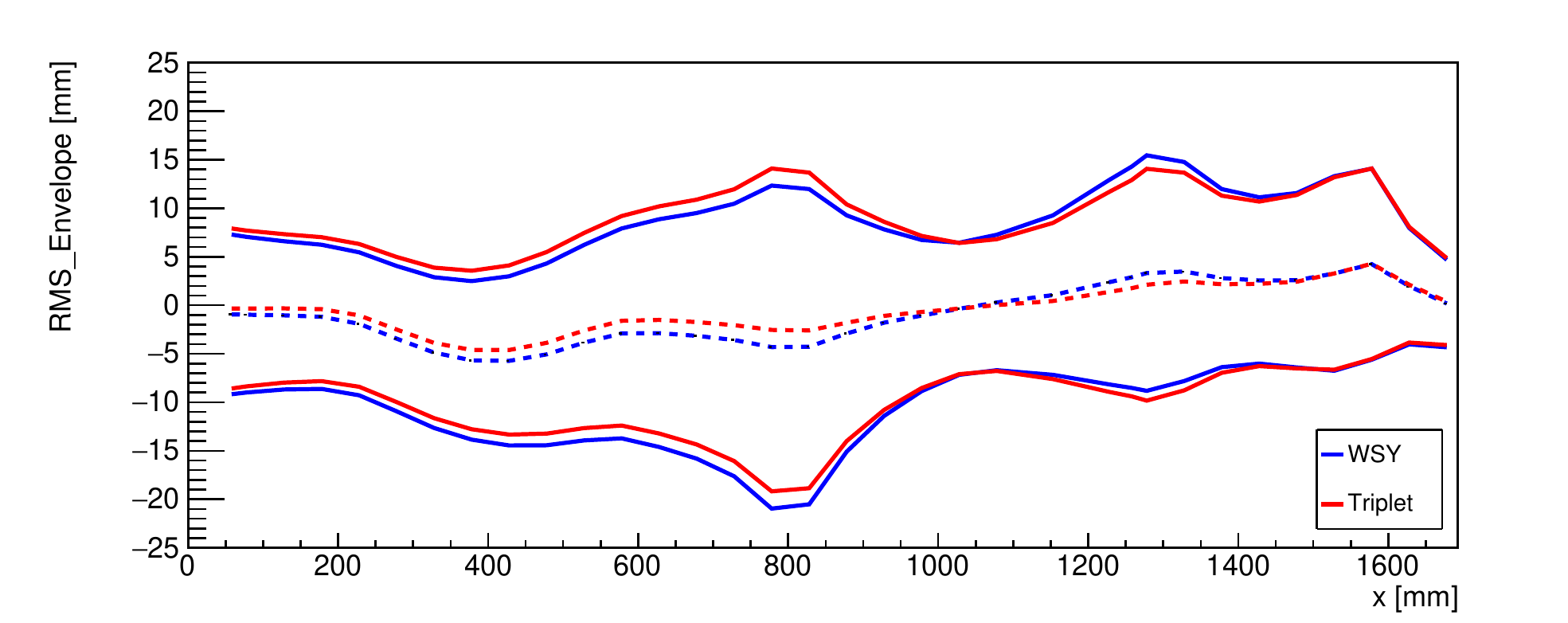}
		\\
		(a)
		\\
		\includegraphics[width=12cm]{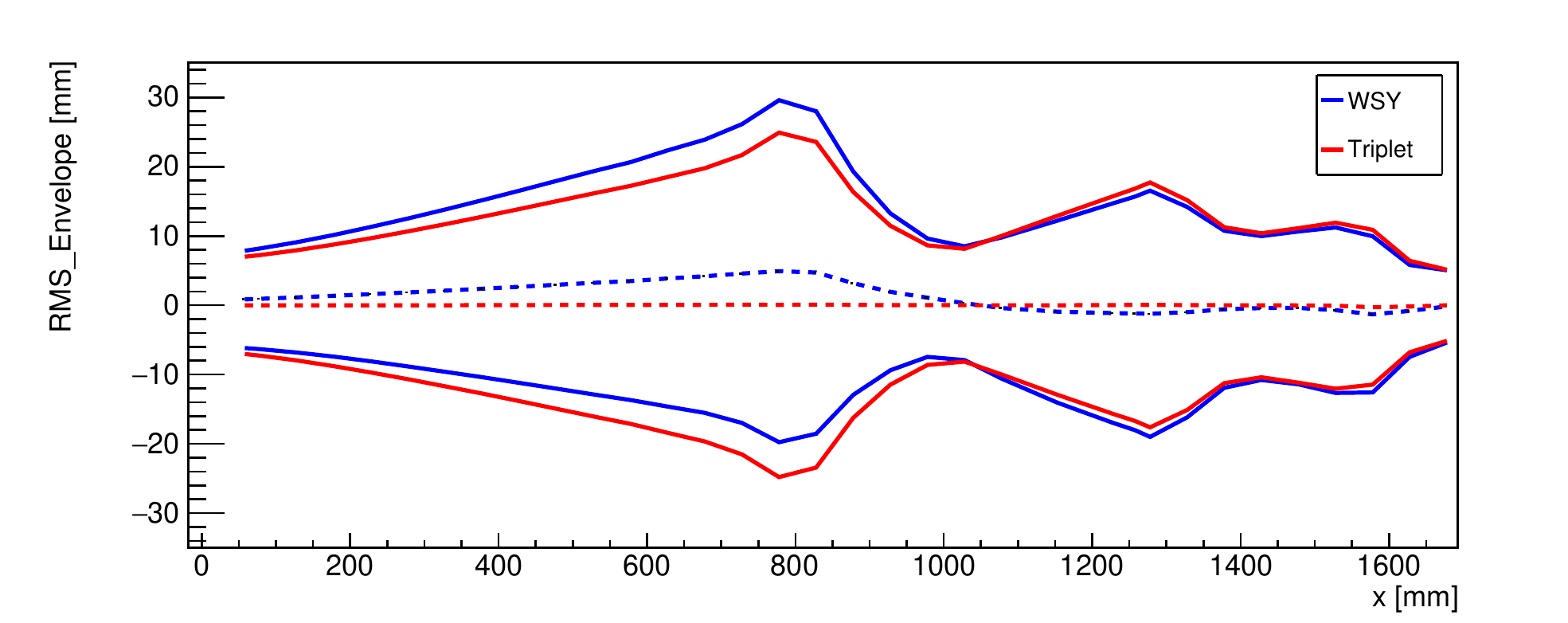}
		\\
		(b)
		\\
		\caption{The comparisons of (a) horizontal and (b) vertical RMS (solid lines) and center (dashed lines) of the beam for the 15~keV muon beam under the influence of different stray fields along the EM-S beam section.}
		\label{FIG21} 
	\end{figure}

\begin{figure}[tbp]
		\centering
		\includegraphics[width=6cm]{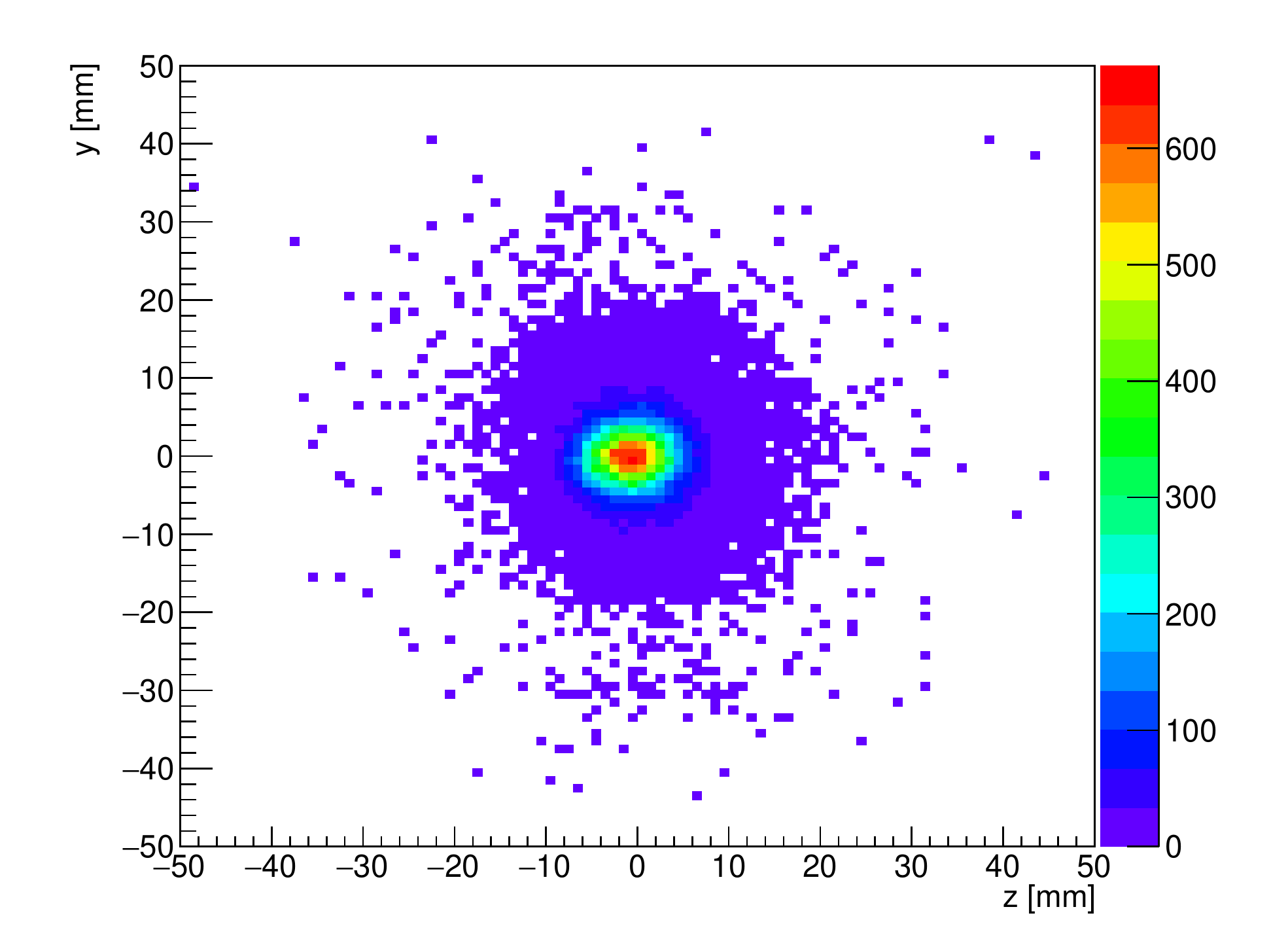}
		\\
		(a)
		\\
		\includegraphics[width=6cm]{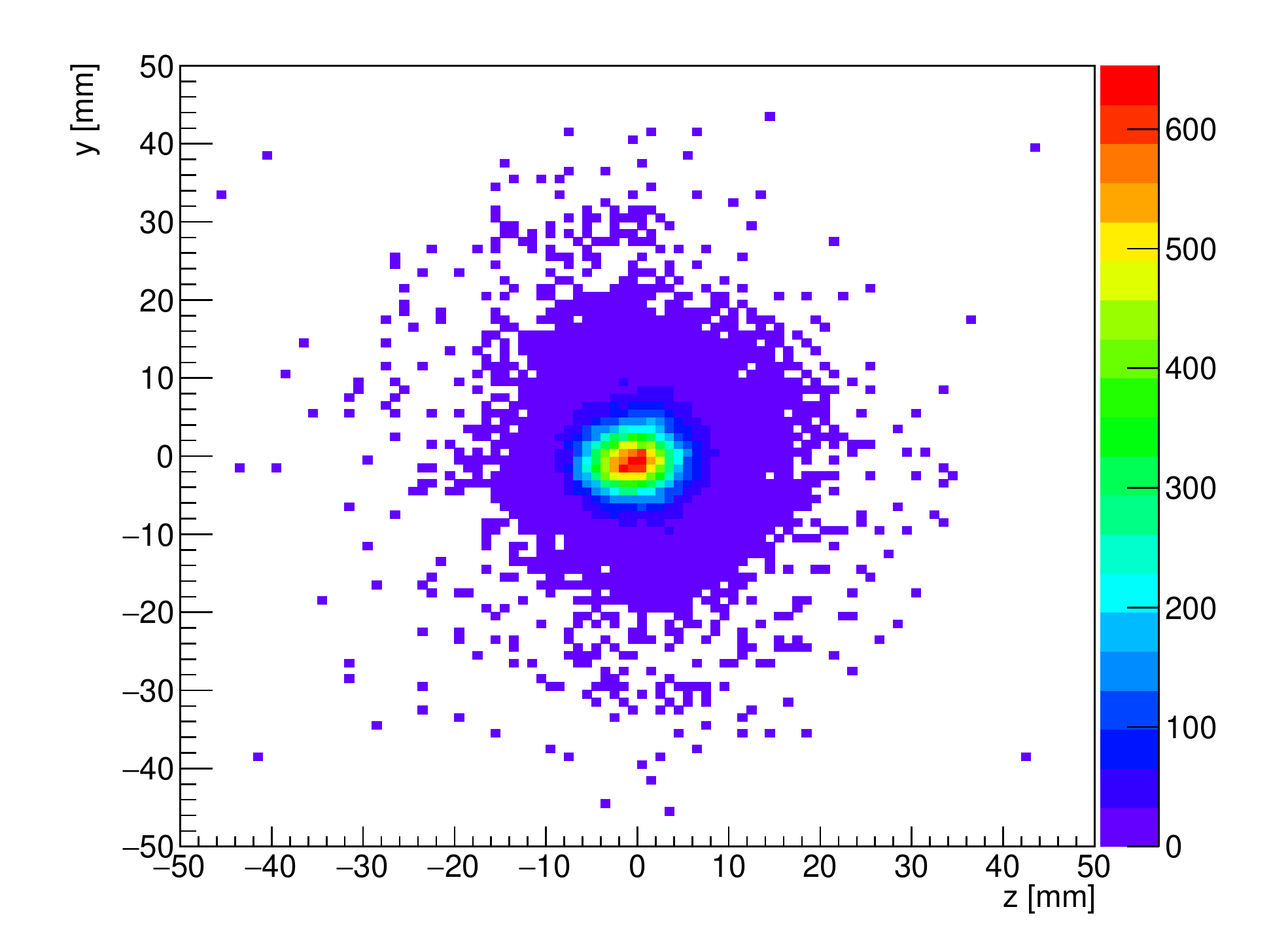}
		\\
		(b)
		\\
		\caption{Comparison of the 15~keV muon beam spots at the sample position under the influence of different stray fields: current $\mu$E4 with (a) quadrupole triplet; (b) solenoid WSY, model-3.}
		\label{FIG22} 
	\end{figure}

	For the beam transport of low-energy muons, the influence of the model-3 WSY stray fields along the EM-S beam section is simulated for a typical beam energy of 15~keV using the Geant4-based program {\tt musrSim} \cite{Sedlak_musrsim_2012}. The RMS beam envelopes and beam spots at the sample position of the  LE-$\mu$SR experiments are shown in Fig.~\ref{FIG21} and Fig.~\ref{FIG22}, respectively, calculated for the current LEM beam transport element settings. Table~\ref{Table1} shows the fraction of the beam in different areas at the sample position for the current $\mu$E4 with quadrupole triplet (stray field is zero) and model-3 WSY. Compared with the current beamline, the influence of stray fields of model-3 WSY on the low energy muon beam can be neglected.
    \\
    \\

	\begin{table}
		\caption{Comparison of beam spot fractions at the LEM sample position for different stray fields using 10000 simulated muons starting at the moderator.}\label{Table1}
        \begin{ruledtabular}
		\begin{tabular}{cccccccc}
			&\multicolumn{1}{c}{15 keV muons}&\multicolumn{4}{c}{fractions on area [mm$^{2}$] }\\
			&beam spot&5 $\times$ 5&10 $\times$ 10&20 $\times$ 20&25 $\times$ 25\\
			\hline
			triplet & 48176 & 27.6$\%$ & 65.2$\%$ & 92.4$\%$ & 95.9$\%$ \\
			WSY, model-3 & 47451 & 27.0$\%$ & 65.1$\%$ & 92.3$\%$ & 96.0$\%$ \\
		\end{tabular}
	    \end{ruledtabular}
	\end{table}

	\begin{center}{\section*{IV. The high intensity muon beamline “Super-$\mu$E4”}}\end{center}
	
\begin{figure}[tbp]
		\centering
		\includegraphics[width=16cm]{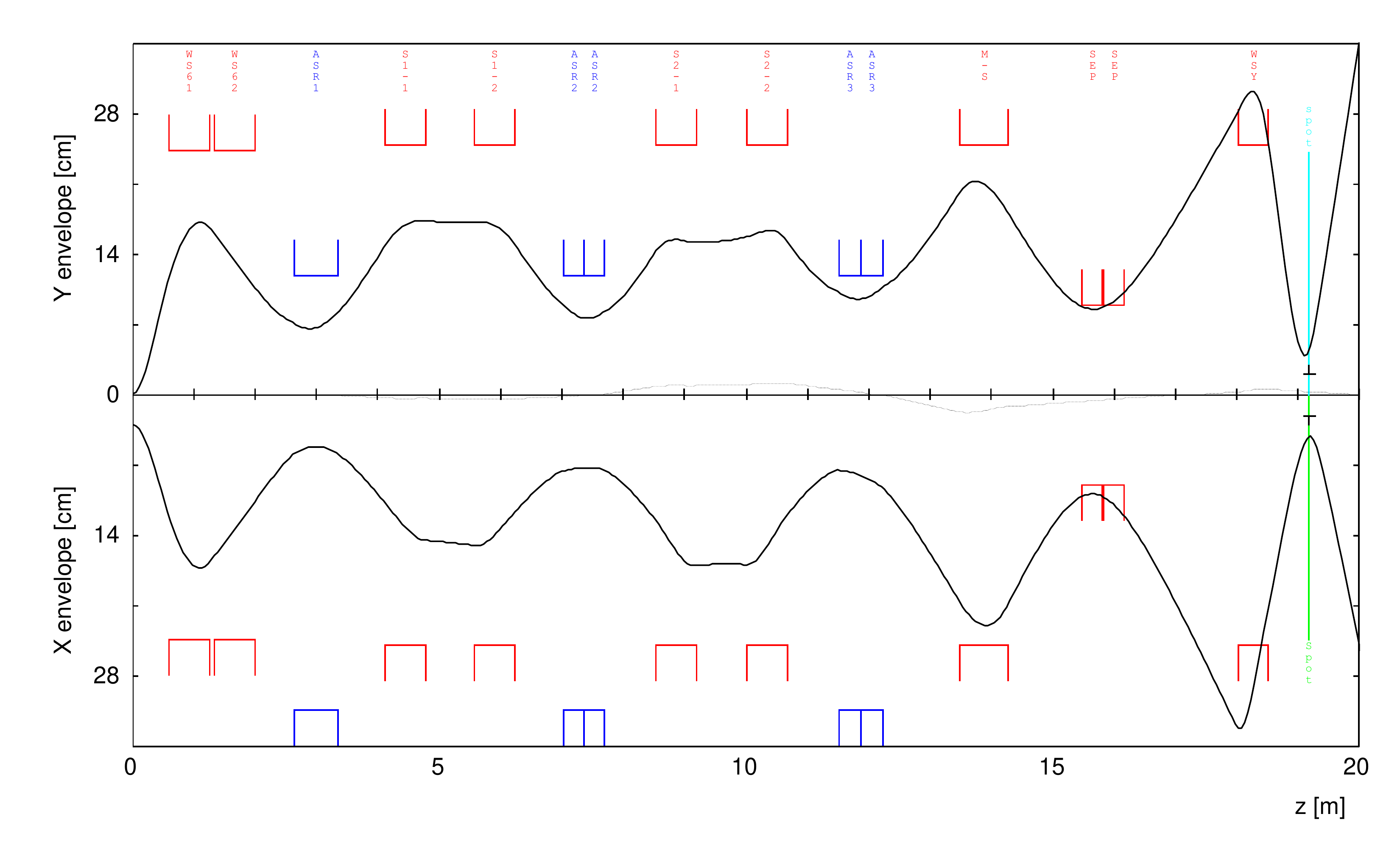}
		\\
		\vspace{-10mm}
		(a)
		\\
		\includegraphics[width=16cm]{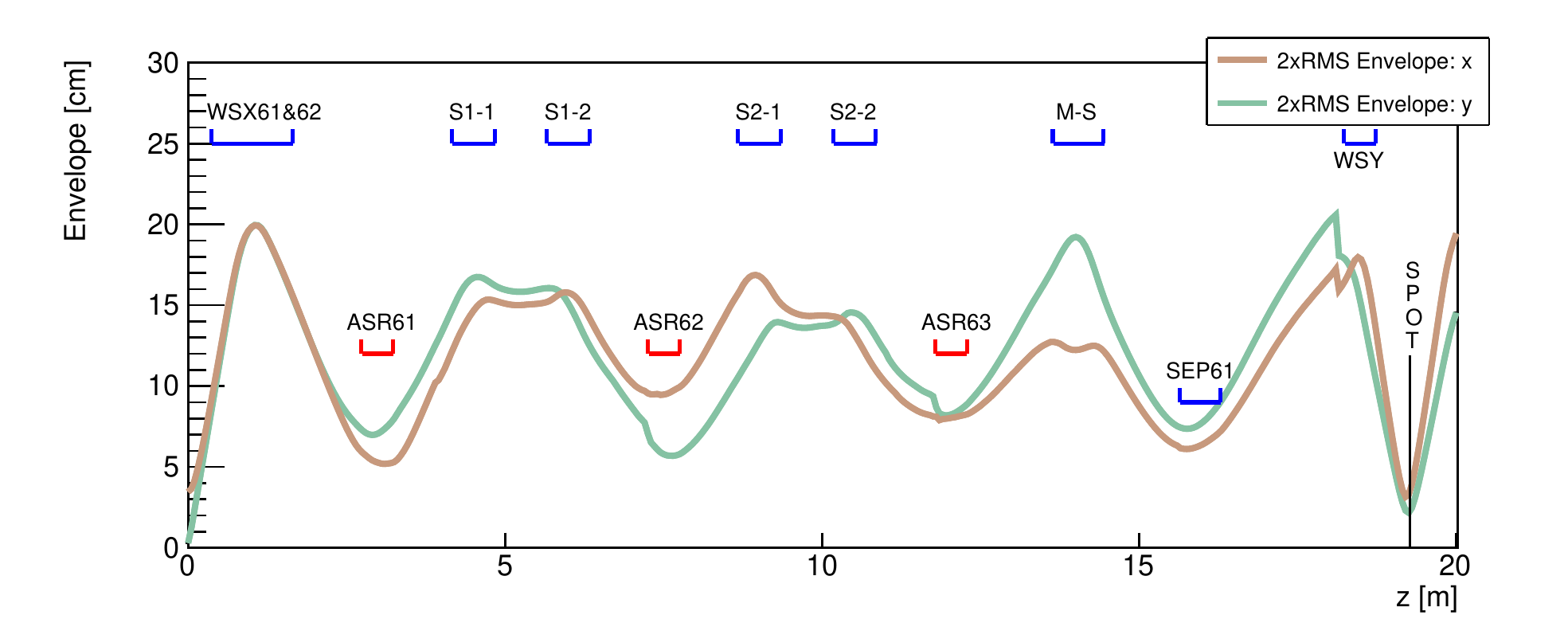}
		\\
		\vspace{-5mm}
		(b)
		\caption{(a) The 2nd order optical calculation for the Super-$\mu$E4 in {\tt TRANSPORT}. Initial parameters used in {\tt TRANSPORT} are (half-width): x =~30 mm, x’ = 200~mrad, y = 2.5~mm, y’ = 200~mrad, dp/p = 4.5$\%$. (b) Beam envelopes ($2\times$RMS) in multi-particle simulation after optimizing the magnetic fields using a homogeneous beam distribution at the source. Initial beam parameters as in (a).
		}
		\label{FIG23} 
	\end{figure}

	As an ultimate - although significantly more expensive - upgrade of the $\mu$E4 beamline, the so-called “Super-$\mu$E4” beamline has been explored, where all quadrupole triplets are replaced by normal-conducting solenoids with an aperture of 500~mm. The optimum beam transport can be achieved with short solenoids, as shown below, and there is no need to use superconductor technology. The “Super-$\mu$E4” upgrade is an example study how a beamline with practical constraints due to existing radiation shielding and already specified deflection angles can be further optimized with regard to beam transmission. This beamline is mainly designed for the large surface area of the NTE, therefore the acceptance is designed as 30~mm / 200~mrad and 2.5~mm / 200~mrad (half-widths) for horizontal and vertical directions, respectively. Two options are useful for controlling the beam profile in the solenoid transmission system when considering the first-order beam optics: (i) ensuring the phase space rotation angle by the system is close to $\theta$ = n$\cdot$($\pi$/2) (n=0,1,2,…); (ii) using solenoid pairs instead of a single solenoid, where the first solenoid ''couples'' transverse phase spaces and the second solenoid ''decouples'' the phase spaces by employing opposite magnetic fields \cite{Prokscha_NewMuE4_2008, kawamura_new_2018}. Figure~\ref{FIG23}(a) shows the optical calculation for the Super-$\mu$E4. The first two solenoids WSX61/WSX62 rotate the phase space by nearly 90 degrees. Stray fields of the following solenoid pairs S1-1/2 and S2-1/2 as well as the single solenoid M-S are shielded by iron housing as in our WSY model-1 to avoid the superposition of the fields between solenoids and the adjacent dipoles. The M-S solenoid controls the beam envelopes and divergences on entering the \textbf{E$\times$B} separator SEP61 and the focusing solenoid WSY, so that the beam envelopes in WSY are large enough to maximize the divergence of the beam focused on the moderator, thereby maximizing the beam intensity. The uniform initial distribution of the muon beam at target E, as used in {\tt TRANSPORT}, is used in G4beamline to help tuning and optimizing the fields of the solenoids in the multi-particle simulation. Figure~\ref{FIG23}(b) shows the RMS transverse beam envelopes in G4beamline after tuning the solenoids. The best beam transport has been found for S1-1/S1-2, S2-1/S2-2, and M-S all having same magnetic field direction, opposite to WSX61/62 and WSY.

\begin{figure}[tbp]
		\centering
		\includegraphics[width=16cm]{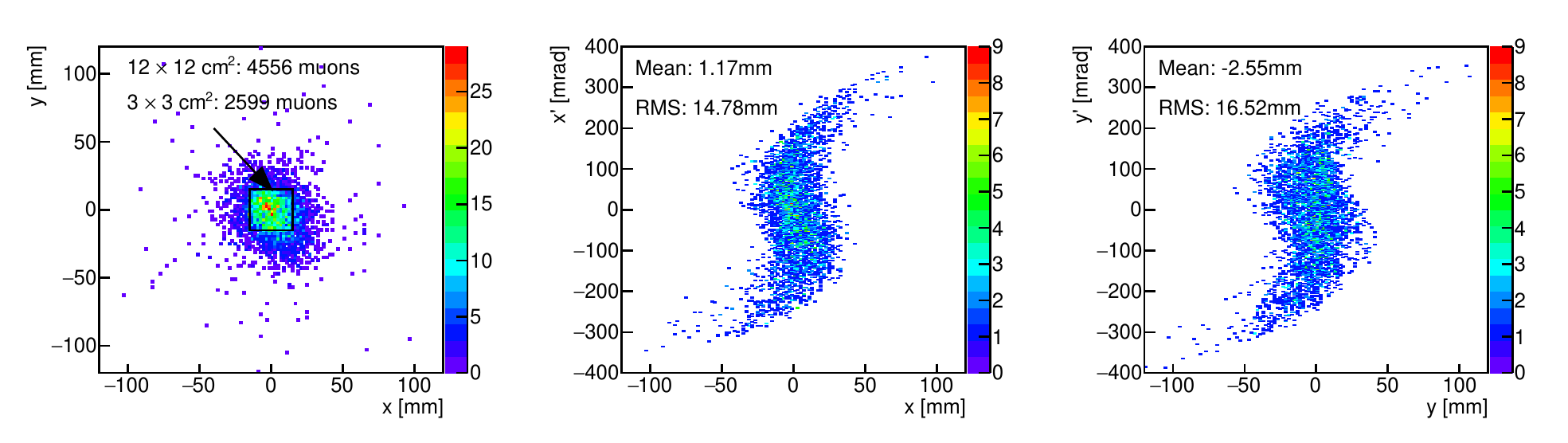}
		\\
		(a)
		\\
		\includegraphics[width=16cm]{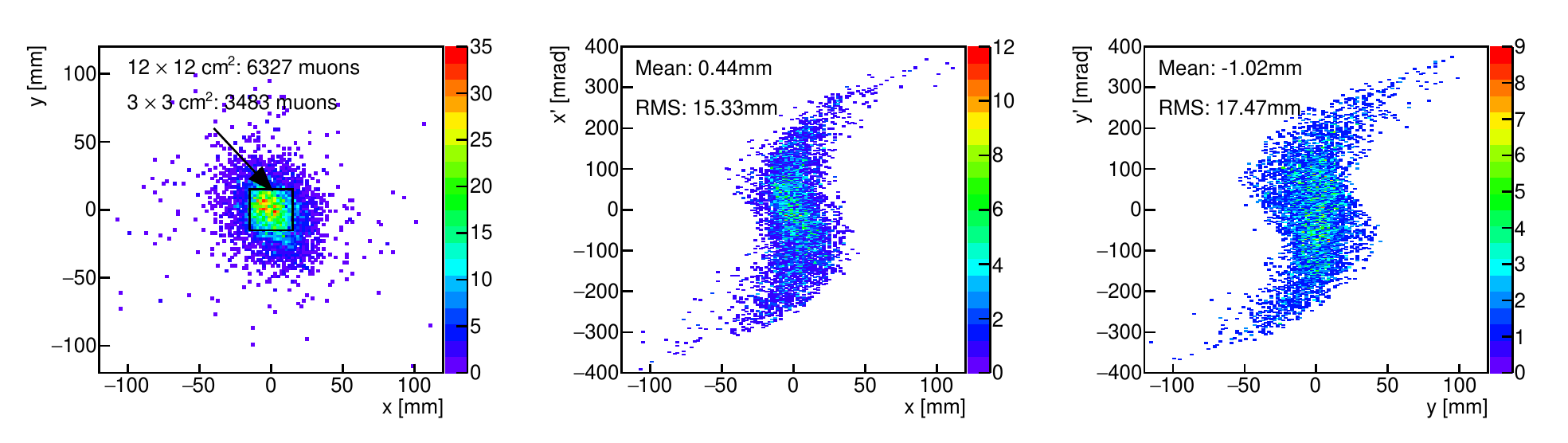}
		\\
		(b)
		\\
		\caption{The beam spots for the Super-$\mu$E4 beamline: (a) with OTE; (b) with NTE}
		\label{FIG24} 
	\end{figure}

\begin{figure}[tbp]
		\centering
		\includegraphics[width=12cm]{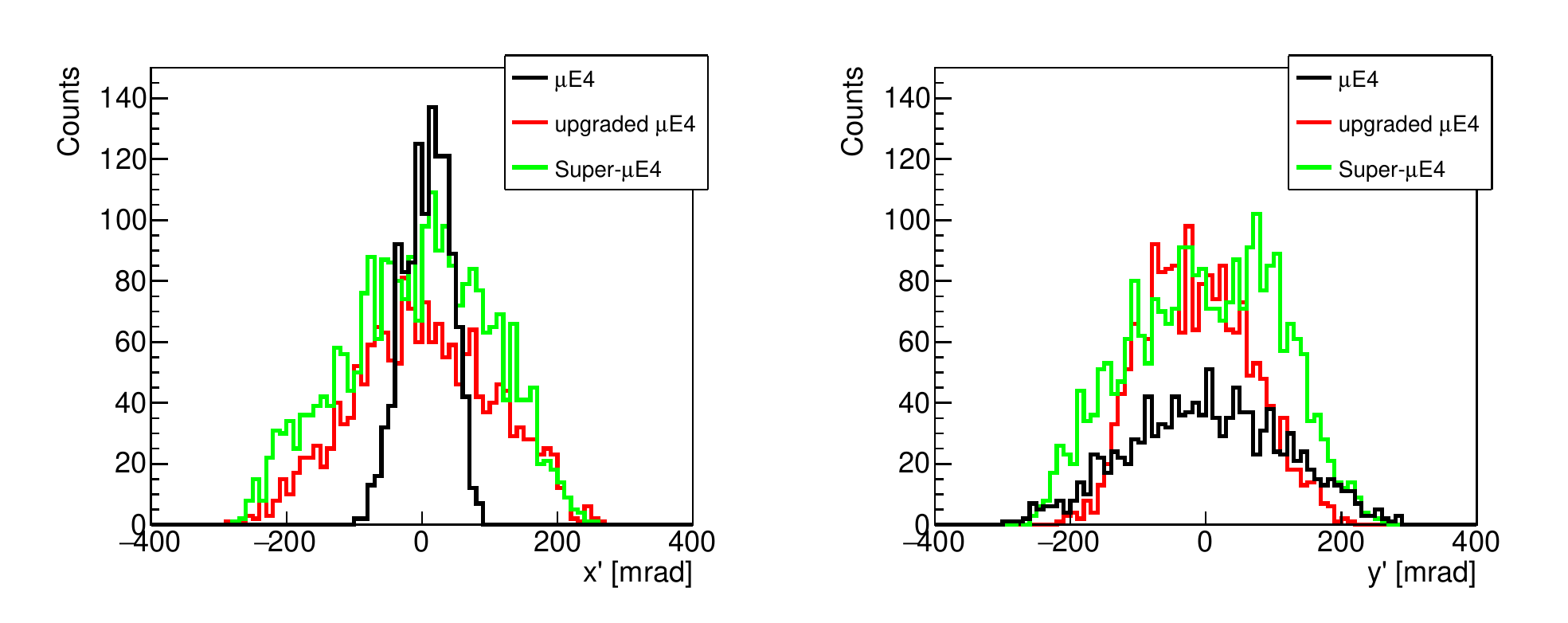}
		\caption{Comparison of divergences for different beamline settings with the OTE.}
		\label{FIG25} 
	\end{figure}

	Simulated beam spots and divergences of the beam focused on the moderator are shown in Fig.~\ref{FIG24} and Fig.~\ref{FIG25}, respectively. Table~\ref{Table2} summarizes the muon intensities on the total beam spot (12 $\times$ 12~cm$^{2}$) and the moderator(3 $\times$ 3~cm$^{2}$) for different target and beamline settings. While the upgraded $\mu$E4 (model-3 WSY) increases the muon rate on moderator by 42\%, Super-$\mu$E4 yields an additional increase of about 55\% for both, OTE and NTE. Overall, with the help of NTE, the Super-$\mu$E4 beamline can increase the muon beam rate on the moderator by a factor of 2.9 compared to the original intensity of the LEM facility with OTE.
	\\
	\\

	\begin{table}[tbp]
		\caption{Comparison of the simulated beam intensity of the $\mu$E4 beam at the moderator position for different target and beamline settings.}\label{Table2}
        \begin{ruledtabular}
		\begin{tabular}{cccccccc}
			&\multicolumn{2}{c}{$\mu$E4}&\multicolumn{2}{c}{upgraded $\mu$E4 (model-3 WSY)}&\multicolumn{2}{c}{Super-$\mu$E4}\\			
			&beam spot&moderator&beam spot&moderator&beam spot&moderator\\
			\hline
			OTE & 2949 & 1186 & 2593 & 1685 & 4556 & 2599 \\
			NTE & 4037 & 1577 & 4021 & 2235 & 6327 & 3483 \\
		\end{tabular}
		\end{ruledtabular}
	\end{table}

	\begin{center}{\section*{V.Conclusions}}\end{center}

	Different methods of upgrading the surface muon beamline $\mu$E4 have been investigated. The feasibility of replacing the last quadrupole triplet with a specially designed solenoid has been demonstrated where the influence of its minimized stray field on the LEM beam transport is negligible. In the simulation, this new solenoid (model-3 WSY) results in an increase of 42\% of surface muon beam rate on the moderator, which is 7\% less compared to the G4beamline solenoid for OTE due to the shielding changing the stray field. With the NTE, routinely used in operation since 2021, the current $\mu$E4 beamline gains 33\% in surface beam rate in the simulation, in good agreement with the measured increase in muon rate. The overall beam rate on moderator for model-3 WSY and NTE can be expected to increase by a factor of $\sim$~1.9 compared to the current $\mu$E4 and OTE. In general, the lower transmission efficiency of surface muon beamlines composed of quadrupole magnets can be overcome by the employment of solenoids. Also benefiting from larger aperture solenoids, the Super-$\mu$E4 can deliver about 55\% more muons to the end of the beamline, with a corresponding increase of muon rate on moderator. Compared to the current $\mu$E4 beamline the rate on moderator increases by a factor of 2.2, for either OTE or NTE. The gain in muon rate on moderator of Super-$\mu$E4/NTE compared to the original $\mu$E4/OTE is $\sim$~2.9.
	
    Our study demonstrates the feasibility of upgrading existing surface muon beams by replacing conventional quadrupole doublets or triplets by normal-conducting, large aperture solenoids to significantly improve the beam intensity at the end of the beamline. The last focusing solenoid can be critical for experiments due to the usually large stray magnetic fields. These stray fields can be minimized by a special solenoid design with iron housing and built-in compensation coils, while maintaining its large focusing power.  
    \\
    
	\begin{center}{\section*{Acknowledgement}}\end{center}
	
	This work was mainly performed at Paul Scherrer Institute in Switzerland. We are grateful to Andreas Knecht at PSI, and Nikolaos Vassilopoulos, Chang-Dong Deng and Zong-Tai Xie at Institute of High Energy Physics (IHEP, Beijing) for fruitful discussions. The first author gratefully acknowledges financial support from China Scholarship Council and Dr. Thomas Prokscha, Dr. Alex Amato and Ms. Isolde Fuchs for their warm hospitality, especially during the difficult pandemic time. The work of L.~P.~Zhou and J.~Y.~Tang was also supported by the National Natural Science Foundation of China (Projects No.~11527811).

    \begin{center}{\section*{Appendix: Comparison of simulated and measured muon rates}}\end{center}
    
    One can use the simulated numbers of Table~\ref{Table2} to check the reliability of the used QGSP\_BIC model for
    pion production, and
    the G4Beamline transport for predicting the observed muon rates in the entire beam spot and on
    moderator. The simulation used $10^{11}$ protons on target. In the experiment \cite{Prokscha_NewMuE4_2008}, muon rates are
    given normalized to proton beam current of 1 mA, i.e. 1/mAs. 
    A charge of 1~mAs corresponds to 
    $6.25\times 10^{15}$ protons. Thus, scaling the numbers of Table~\ref{Table2} by $6.25\times 10^4$ yields
    the simulated muon rates in Table~\ref{Table_APP}. Using the QGSP\_BIC model, the muon rates are underestimated
    by 13\%-14\% compared to the experimental results. This originates from an underestimation
    of the BIC pion production cross sections relative to the measured production cross sections \cite{Berg_SMtarget_2016}, where the enhanced measured pion production cross sections resulted
    in an even better agreement between simulated and measured beam rates. These results demonstrate
    the reliability of our beam transport simulation.
    \begin{table}[tbp]
		\caption{Comparison of muon rates, normalized to a proton beam current of
		1~mA for the current $\mu$E4, OTE. The experimental data are
		from ref.~\cite{Prokscha_NewMuE4_2008}. The simulation uses the BIC model for pion
		production.}\label{Table_APP}
        \begin{ruledtabular}
		\begin{tabular}{lcc}
		&muon rate on beam spot [1/mAs]& muon rate on moderator [1/mAs]\\
			\hline
			Experiment & $210\times 10^6$ & $86\times 10^6$\\
		    Simulation & $184\times 10^6$ & $74\times 10^6$\\
		\end{tabular}
		\end{ruledtabular}
	\end{table}
    \newpage
	\bibliography{reference}

\begin{thebibliography}{34}%
\makeatletter
\providecommand \@ifxundefined [1]{%
 \@ifx{#1\undefined}
}%
\providecommand \@ifnum [1]{%
 \ifnum #1\expandafter \@firstoftwo
 \else \expandafter \@secondoftwo
 \fi
}%
\providecommand \@ifx [1]{%
 \ifx #1\expandafter \@firstoftwo
 \else \expandafter \@secondoftwo
 \fi
}%
\providecommand \natexlab [1]{#1}%
\providecommand \enquote  [1]{``#1''}%
\providecommand \bibnamefont  [1]{#1}%
\providecommand \bibfnamefont [1]{#1}%
\providecommand \citenamefont [1]{#1}%
\providecommand \href@noop [0]{\@secondoftwo}%
\providecommand \href [0]{\begingroup \@sanitize@url \@href}%
\providecommand \@href[1]{\@@startlink{#1}\@@href}%
\providecommand \@@href[1]{\endgroup#1\@@endlink}%
\providecommand \@sanitize@url [0]{\catcode `\\12\catcode `\$12\catcode
  `\&12\catcode `\#12\catcode `\^12\catcode `\_12\catcode `\%12\relax}%
\providecommand \@@startlink[1]{}%
\providecommand \@@endlink[0]{}%
\providecommand \url  [0]{\begingroup\@sanitize@url \@url }%
\providecommand \@url [1]{\endgroup\@href {#1}{\urlprefix }}%
\providecommand \urlprefix  [0]{URL }%
\providecommand \Eprint [0]{\href }%
\providecommand \doibase [0]{http://dx.doi.org/}%
\providecommand \selectlanguage [0]{\@gobble}%
\providecommand \bibinfo  [0]{\@secondoftwo}%
\providecommand \bibfield  [0]{\@secondoftwo}%
\providecommand \translation [1]{[#1]}%
\providecommand \BibitemOpen [0]{}%
\providecommand \bibitemStop [0]{}%
\providecommand \bibitemNoStop [0]{.\EOS\space}%
\providecommand \EOS [0]{\spacefactor3000\relax}%
\providecommand \BibitemShut  [1]{\csname bibitem#1\endcsname}%
\let\auto@bib@innerbib\@empty
\bibitem [{\citenamefont {{Foroughi}}\ \emph {et~al.}(2001)\citenamefont
  {{Foroughi}}, \citenamefont {{Morenzoni}}, \citenamefont {{Prokscha}},
  \citenamefont {{Daum}}, \citenamefont {{Deiters}}, \citenamefont {{George}},
  \citenamefont {{Herlach}}, \citenamefont {{Petitjean}}, \citenamefont
  {{Renker}},\ and\ \citenamefont {{Vrankovic}}}]{Foroughi_Upgrading_2001}%
  \BibitemOpen
  \bibfield  {author} {\bibinfo {author} {\bibfnamefont {F.}~\bibnamefont
  {{Foroughi}}}, \bibinfo {author} {\bibfnamefont {E.}~\bibnamefont
  {{Morenzoni}}}, \bibinfo {author} {\bibfnamefont {T.}~\bibnamefont
  {{Prokscha}}}, \bibinfo {author} {\bibfnamefont {M.}~\bibnamefont {{Daum}}},
  \bibinfo {author} {\bibfnamefont {K.}~\bibnamefont {{Deiters}}}, \bibinfo
  {author} {\bibfnamefont {D.}~\bibnamefont {{George}}}, \bibinfo {author}
  {\bibfnamefont {D.}~\bibnamefont {{Herlach}}}, \bibinfo {author}
  {\bibfnamefont {C.}~\bibnamefont {{Petitjean}}}, \bibinfo {author}
  {\bibfnamefont {D.}~\bibnamefont {{Renker}}}, \ and\ \bibinfo {author}
  {\bibfnamefont {V.}~\bibnamefont {{Vrankovic}}},\ }\href {\doibase
  10.1023/A:1020830830050} {\bibfield  {journal} {\bibinfo  {journal}
  {Hyperfine Interact.}\ }\textbf {\bibinfo {volume} {138}},\ \bibinfo {pages}
  {483} (\bibinfo {year} {2001})}\BibitemShut {NoStop}%
\bibitem [{\citenamefont {{Prokscha}}\ \emph
  {et~al.}(2004{\natexlab{a}})\citenamefont {{Prokscha}}, \citenamefont
  {{Morenzoni}}, \citenamefont {{Deiters}}, \citenamefont {{Foroughi}},
  \citenamefont {{George}}, \citenamefont {{Kobler}},\ and\ \citenamefont
  {{Vrankovic}}}]{Prokscha_NewMB_2004}%
  \BibitemOpen
  \bibfield  {author} {\bibinfo {author} {\bibfnamefont {T.}~\bibnamefont
  {{Prokscha}}}, \bibinfo {author} {\bibfnamefont {E.}~\bibnamefont
  {{Morenzoni}}}, \bibinfo {author} {\bibfnamefont {K.}~\bibnamefont
  {{Deiters}}}, \bibinfo {author} {\bibfnamefont {F.}~\bibnamefont
  {{Foroughi}}}, \bibinfo {author} {\bibfnamefont {D.}~\bibnamefont
  {{George}}}, \bibinfo {author} {\bibfnamefont {R.}~\bibnamefont {{Kobler}}},
  \ and\ \bibinfo {author} {\bibfnamefont {V.}~\bibnamefont {{Vrankovic}}},\
  }\href {\doibase 10.1007/s10751-005-9129-9} {\bibfield  {journal} {\bibinfo
  {journal} {Hyperfine Interact.}\ }\textbf {\bibinfo {volume} {159}},\
  \bibinfo {pages} {385} (\bibinfo {year} {2004}{\natexlab{a}})}\BibitemShut
  {NoStop}%
\bibitem [{\citenamefont {{Prokscha}}\ \emph {et~al.}(2008)\citenamefont
  {{Prokscha}}, \citenamefont {{Morenzoni}}, \citenamefont {{Deiters}},
  \citenamefont {{Foroughi}}, \citenamefont {{George}}, \citenamefont
  {{Kobler}}, \citenamefont {{Suter}},\ and\ \citenamefont
  {{Vrankovic}}}]{Prokscha_NewMuE4_2008}%
  \BibitemOpen
  \bibfield  {author} {\bibinfo {author} {\bibfnamefont {T.}~\bibnamefont
  {{Prokscha}}}, \bibinfo {author} {\bibfnamefont {E.}~\bibnamefont
  {{Morenzoni}}}, \bibinfo {author} {\bibfnamefont {K.}~\bibnamefont
  {{Deiters}}}, \bibinfo {author} {\bibfnamefont {F.}~\bibnamefont
  {{Foroughi}}}, \bibinfo {author} {\bibfnamefont {D.}~\bibnamefont
  {{George}}}, \bibinfo {author} {\bibfnamefont {R.}~\bibnamefont {{Kobler}}},
  \bibinfo {author} {\bibfnamefont {A.}~\bibnamefont {{Suter}}}, \ and\
  \bibinfo {author} {\bibfnamefont {V.}~\bibnamefont {{Vrankovic}}},\ }\href
  {\doibase 10.1016/j.nima.2008.07.081} {\bibfield  {journal} {\bibinfo
  {journal} {Nucl. Instrum. Methods Phys. Res. Sect. A}\ }\textbf {\bibinfo
  {volume} {595}},\ \bibinfo {pages} {317} (\bibinfo {year}
  {2008})}\BibitemShut {NoStop}%
\bibitem [{\citenamefont {{Morenzoni}}\ \emph {et~al.}(2000)\citenamefont
  {{Morenzoni}}, \citenamefont {{Gl{\"u}ckler}}, \citenamefont {{Prokscha}},
  \citenamefont {{Weber}}, \citenamefont {{Forgan}}, \citenamefont {{Jackson}},
  \citenamefont {{Luetkens}}, \citenamefont {{Niedermayer}}, \citenamefont
  {{Pleines}}, \citenamefont {{Birke}}, \citenamefont {{Hofer}}, \citenamefont
  {{Litterst}}, \citenamefont {{Riseman}},\ and\ \citenamefont
  {{Schatz}}}]{Morenzoni_LEmuPSI_2000}%
  \BibitemOpen
  \bibfield  {author} {\bibinfo {author} {\bibfnamefont {E.}~\bibnamefont
  {{Morenzoni}}}, \bibinfo {author} {\bibfnamefont {H.}~\bibnamefont
  {{Gl{\"u}ckler}}}, \bibinfo {author} {\bibfnamefont {T.}~\bibnamefont
  {{Prokscha}}}, \bibinfo {author} {\bibfnamefont {H.~P.}\ \bibnamefont
  {{Weber}}}, \bibinfo {author} {\bibfnamefont {E.~M.}\ \bibnamefont
  {{Forgan}}}, \bibinfo {author} {\bibfnamefont {T.~J.}\ \bibnamefont
  {{Jackson}}}, \bibinfo {author} {\bibfnamefont {H.}~\bibnamefont
  {{Luetkens}}}, \bibinfo {author} {\bibfnamefont {C.}~\bibnamefont
  {{Niedermayer}}}, \bibinfo {author} {\bibfnamefont {M.}~\bibnamefont
  {{Pleines}}}, \bibinfo {author} {\bibfnamefont {M.}~\bibnamefont {{Birke}}},
  \bibinfo {author} {\bibfnamefont {A.}~\bibnamefont {{Hofer}}}, \bibinfo
  {author} {\bibfnamefont {J.}~\bibnamefont {{Litterst}}}, \bibinfo {author}
  {\bibfnamefont {T.}~\bibnamefont {{Riseman}}}, \ and\ \bibinfo {author}
  {\bibfnamefont {G.}~\bibnamefont {{Schatz}}},\ }\href {\doibase
  10.1016/S0921-4526(00)00303-3} {\bibfield  {journal} {\bibinfo  {journal}
  {Physica B}\ }\textbf {\bibinfo {volume} {289}},\ \bibinfo {pages} {653}
  (\bibinfo {year} {2000})}\BibitemShut {NoStop}%
\bibitem [{\citenamefont {{Morenzoni}}\ \emph {et~al.}(2003)\citenamefont
  {{Morenzoni}}, \citenamefont {{Khasanov}}, \citenamefont {{Luetkens}},
  \citenamefont {{Prokscha}}, \citenamefont {{Suter}}, \citenamefont
  {{Garifianov}}, \citenamefont {{Gl{\"u}ckler}}, \citenamefont {{Birke}},
  \citenamefont {{Forgan}}, \citenamefont {{Keller}}, \citenamefont
  {{Litterst}}, \citenamefont {{Niedermayer}},\ and\ \citenamefont
  {{Nieuwenhuys}}}]{Morenzoni_LEmu_2003}%
  \BibitemOpen
  \bibfield  {author} {\bibinfo {author} {\bibfnamefont {E.}~\bibnamefont
  {{Morenzoni}}}, \bibinfo {author} {\bibfnamefont {R.}~\bibnamefont
  {{Khasanov}}}, \bibinfo {author} {\bibfnamefont {H.}~\bibnamefont
  {{Luetkens}}}, \bibinfo {author} {\bibfnamefont {T.}~\bibnamefont
  {{Prokscha}}}, \bibinfo {author} {\bibfnamefont {A.}~\bibnamefont {{Suter}}},
  \bibinfo {author} {\bibfnamefont {N.}~\bibnamefont {{Garifianov}}}, \bibinfo
  {author} {\bibfnamefont {H.}~\bibnamefont {{Gl{\"u}ckler}}}, \bibinfo
  {author} {\bibfnamefont {M.}~\bibnamefont {{Birke}}}, \bibinfo {author}
  {\bibfnamefont {E.}~\bibnamefont {{Forgan}}}, \bibinfo {author}
  {\bibfnamefont {H.}~\bibnamefont {{Keller}}}, \bibinfo {author}
  {\bibfnamefont {J.}~\bibnamefont {{Litterst}}}, \bibinfo {author}
  {\bibfnamefont {C.}~\bibnamefont {{Niedermayer}}}, \ and\ \bibinfo {author}
  {\bibfnamefont {G.}~\bibnamefont {{Nieuwenhuys}}},\ }\href {\doibase
  10.1016/S0921-4526(02)01601-0} {\bibfield  {journal} {\bibinfo  {journal}
  {Physica B}\ }\textbf {\bibinfo {volume} {326}},\ \bibinfo {pages} {196}
  (\bibinfo {year} {2003})}\BibitemShut {NoStop}%
\bibitem [{\citenamefont {{Prokscha}}\ \emph
  {et~al.}(2004{\natexlab{b}})\citenamefont {{Prokscha}}, \citenamefont
  {{Morenzoni}}, \citenamefont {{Suter}}, \citenamefont {{Khasanov}},
  \citenamefont {{Luetkens}}, \citenamefont {{Eshchenko}}, \citenamefont
  {{Garifianov}}, \citenamefont {{Forgan}}, \citenamefont {{Keller}},
  \citenamefont {{Litterst}}, \citenamefont {{Niedermayer}},\ and\
  \citenamefont {{Nieuwenhuys}}}]{Prokscha_Thinfilm_2004}%
  \BibitemOpen
  \bibfield  {author} {\bibinfo {author} {\bibfnamefont {T.}~\bibnamefont
  {{Prokscha}}}, \bibinfo {author} {\bibfnamefont {E.}~\bibnamefont
  {{Morenzoni}}}, \bibinfo {author} {\bibfnamefont {A.}~\bibnamefont
  {{Suter}}}, \bibinfo {author} {\bibfnamefont {R.}~\bibnamefont {{Khasanov}}},
  \bibinfo {author} {\bibfnamefont {H.}~\bibnamefont {{Luetkens}}}, \bibinfo
  {author} {\bibfnamefont {D.}~\bibnamefont {{Eshchenko}}}, \bibinfo {author}
  {\bibfnamefont {N.}~\bibnamefont {{Garifianov}}}, \bibinfo {author}
  {\bibfnamefont {E.~M.}\ \bibnamefont {{Forgan}}}, \bibinfo {author}
  {\bibfnamefont {H.}~\bibnamefont {{Keller}}}, \bibinfo {author}
  {\bibfnamefont {J.}~\bibnamefont {{Litterst}}}, \bibinfo {author}
  {\bibfnamefont {C.}~\bibnamefont {{Niedermayer}}}, \ and\ \bibinfo {author}
  {\bibfnamefont {G.}~\bibnamefont {{Nieuwenhuys}}},\ }\href {\doibase
  10.1007/s10751-005-9104-5} {\bibfield  {journal} {\bibinfo  {journal}
  {Hyperfine Interact.}\ }\textbf {\bibinfo {volume} {159}},\ \bibinfo {pages}
  {227} (\bibinfo {year} {2004}{\natexlab{b}})}\BibitemShut {NoStop}%
\bibitem [{\citenamefont {{Morenzoni}}\ \emph {et~al.}(2004)\citenamefont
  {{Morenzoni}}, \citenamefont {{Prokscha}}, \citenamefont {{Suter}},
  \citenamefont {{Luetkens}},\ and\ \citenamefont
  {{Khasanov}}}]{Morenzoni_NanoFilm_2003}%
  \BibitemOpen
  \bibfield  {author} {\bibinfo {author} {\bibfnamefont {E.}~\bibnamefont
  {{Morenzoni}}}, \bibinfo {author} {\bibfnamefont {T.}~\bibnamefont
  {{Prokscha}}}, \bibinfo {author} {\bibfnamefont {A.}~\bibnamefont {{Suter}}},
  \bibinfo {author} {\bibfnamefont {H.}~\bibnamefont {{Luetkens}}}, \ and\
  \bibinfo {author} {\bibfnamefont {R.}~\bibnamefont {{Khasanov}}},\ }\href
  {\doibase 10.1088/0953-8984/16/40/010} {\bibfield  {journal} {\bibinfo
  {journal} {J. Phys.-Condes. Matter}\ }\textbf {\bibinfo {volume} {16}},\
  \bibinfo {pages} {S4583} (\bibinfo {year} {2004})}\BibitemShut {NoStop}%
\bibitem [{\citenamefont {{Prokscha}}\ \emph {et~al.}(2001)\citenamefont
  {{Prokscha}}, \citenamefont {{Morenzoni}}, \citenamefont {{David}},
  \citenamefont {{Hofer}}, \citenamefont {{Gl{\"u}ckler}},\ and\ \citenamefont
  {{Scandella}}}]{Prokscha_ModforMup_2001}%
  \BibitemOpen
  \bibfield  {author} {\bibinfo {author} {\bibfnamefont {T.}~\bibnamefont
  {{Prokscha}}}, \bibinfo {author} {\bibfnamefont {E.}~\bibnamefont
  {{Morenzoni}}}, \bibinfo {author} {\bibfnamefont {C.}~\bibnamefont
  {{David}}}, \bibinfo {author} {\bibfnamefont {A.}~\bibnamefont {{Hofer}}},
  \bibinfo {author} {\bibfnamefont {H.}~\bibnamefont {{Gl{\"u}ckler}}}, \ and\
  \bibinfo {author} {\bibfnamefont {L.}~\bibnamefont {{Scandella}}},\ }\href
  {\doibase 10.1016/S0169-4332(00)00857-6} {\bibfield  {journal} {\bibinfo
  {journal} {Appl. Surf. Sci.}\ }\textbf {\bibinfo {volume} {172}},\ \bibinfo
  {pages} {235} (\bibinfo {year} {2001})}\BibitemShut {NoStop}%
\bibitem [{\citenamefont {Xiao}\ \emph {et~al.}(2017)\citenamefont {Xiao},
  \citenamefont {Morenzoni}, \citenamefont {Salman}, \citenamefont {Ye},\ and\
  \citenamefont {Prokscha}}]{Xiao_Segmented_2017}%
  \BibitemOpen
  \bibfield  {author} {\bibinfo {author} {\bibfnamefont {R.}~\bibnamefont
  {Xiao}}, \bibinfo {author} {\bibfnamefont {E.}~\bibnamefont {Morenzoni}},
  \bibinfo {author} {\bibfnamefont {Z.}~\bibnamefont {Salman}}, \bibinfo
  {author} {\bibfnamefont {B.~J.}\ \bibnamefont {Ye}}, \ and\ \bibinfo {author}
  {\bibfnamefont {T.}~\bibnamefont {Prokscha}},\ }\href {\doibase
  10.1007/s41365-017-0190-2} {\bibfield  {journal} {\bibinfo  {journal} {Nucl.
  Sci. Tech.}\ }\textbf {\bibinfo {volume} {28}} (\bibinfo {year} {2017}),\
  10.1007/s41365-017-0190-2}\BibitemShut {NoStop}%
\bibitem [{\citenamefont {{Khaw}}\ \emph {et~al.}(2015)\citenamefont {{Khaw}},
  \citenamefont {{Antognini}}, \citenamefont {{Crivelli}}, \citenamefont
  {{Kirch}}, \citenamefont {{Morenzoni}}, \citenamefont {{Salman}},
  \citenamefont {{Suter}},\ and\ \citenamefont
  {{Prokscha}}}]{Khaw_SimuLEM_2015}%
  \BibitemOpen
  \bibfield  {author} {\bibinfo {author} {\bibfnamefont {K.~S.}\ \bibnamefont
  {{Khaw}}}, \bibinfo {author} {\bibfnamefont {A.}~\bibnamefont {{Antognini}}},
  \bibinfo {author} {\bibfnamefont {P.}~\bibnamefont {{Crivelli}}}, \bibinfo
  {author} {\bibfnamefont {K.}~\bibnamefont {{Kirch}}}, \bibinfo {author}
  {\bibfnamefont {E.}~\bibnamefont {{Morenzoni}}}, \bibinfo {author}
  {\bibfnamefont {Z.}~\bibnamefont {{Salman}}}, \bibinfo {author}
  {\bibfnamefont {A.}~\bibnamefont {{Suter}}}, \ and\ \bibinfo {author}
  {\bibfnamefont {T.}~\bibnamefont {{Prokscha}}},\ }\href {\doibase
  10.1088/1748-0221/10/10/P10025} {\bibfield  {journal} {\bibinfo  {journal}
  {J. Instrum.}\ }\textbf {\bibinfo {volume} {10}},\ \bibinfo {eid} {P10025}
  (\bibinfo {year} {2015})},\ \Eprint {http://arxiv.org/abs/1506.01779}
  {arXiv:1506.01779 [physics.ins-det]} \BibitemShut {NoStop}%
\bibitem [{\citenamefont {{Hillier}}\ \emph {et~al.}(2014)\citenamefont
  {{Hillier}}, \citenamefont {{Adams}}, \citenamefont {{Baker}}, \citenamefont
  {{Bekasovs}}, \citenamefont {{Coomer}}, \citenamefont {{Cottrell}},
  \citenamefont {{Higgins}}, \citenamefont {{Jago}}, \citenamefont {{Jones}},
  \citenamefont {{Lord}}, \citenamefont {{Markvardsen}}, \citenamefont
  {{Parker}}, \citenamefont {{Peck}}, \citenamefont {{Pratt}}, \citenamefont
  {{Telling}},\ and\ \citenamefont {{Williamson}}}]{HillierAD_ISISmuon_2014}%
  \BibitemOpen
  \bibfield  {author} {\bibinfo {author} {\bibfnamefont {A.~D.}\ \bibnamefont
  {{Hillier}}}, \bibinfo {author} {\bibfnamefont {D.~J.}\ \bibnamefont
  {{Adams}}}, \bibinfo {author} {\bibfnamefont {P.~J.}\ \bibnamefont
  {{Baker}}}, \bibinfo {author} {\bibfnamefont {A.}~\bibnamefont {{Bekasovs}}},
  \bibinfo {author} {\bibfnamefont {F.~C.}\ \bibnamefont {{Coomer}}}, \bibinfo
  {author} {\bibfnamefont {S.~P.}\ \bibnamefont {{Cottrell}}}, \bibinfo
  {author} {\bibfnamefont {S.~D.}\ \bibnamefont {{Higgins}}}, \bibinfo {author}
  {\bibfnamefont {S.~J.~S.}\ \bibnamefont {{Jago}}}, \bibinfo {author}
  {\bibfnamefont {K.~G.}\ \bibnamefont {{Jones}}}, \bibinfo {author}
  {\bibfnamefont {J.~S.}\ \bibnamefont {{Lord}}}, \bibinfo {author}
  {\bibfnamefont {A.}~\bibnamefont {{Markvardsen}}}, \bibinfo {author}
  {\bibfnamefont {P.~G.}\ \bibnamefont {{Parker}}}, \bibinfo {author}
  {\bibfnamefont {J.~N.~T.}\ \bibnamefont {{Peck}}}, \bibinfo {author}
  {\bibfnamefont {F.~L.}\ \bibnamefont {{Pratt}}}, \bibinfo {author}
  {\bibfnamefont {M.~T.~F.}\ \bibnamefont {{Telling}}}, \ and\ \bibinfo
  {author} {\bibfnamefont {R.~E.}\ \bibnamefont {{Williamson}}},\ }in\ \href
  {\doibase 10.1088/1742-6596/551/1/012067} {\emph {\bibinfo {booktitle}
  {Journal of Physics Conference Series}}},\ \bibinfo {series} {Journal of
  Physics Conference Series}, Vol.\ \bibinfo {volume} {551}\ (\bibinfo {year}
  {2014})\ p.\ \bibinfo {pages} {012067}\BibitemShut {NoStop}%
\bibitem [{\citenamefont {{Hillier}}\ \emph {et~al.}(2019)\citenamefont
  {{Hillier}}, \citenamefont {{Lord}}, \citenamefont {{Ishida}},\ and\
  \citenamefont {{Rogers}}}]{HillierAD_ISISmuon_2019}%
  \BibitemOpen
  \bibfield  {author} {\bibinfo {author} {\bibfnamefont {A.~D.}\ \bibnamefont
  {{Hillier}}}, \bibinfo {author} {\bibfnamefont {J.~S.}\ \bibnamefont
  {{Lord}}}, \bibinfo {author} {\bibfnamefont {K.}~\bibnamefont {{Ishida}}}, \
  and\ \bibinfo {author} {\bibfnamefont {C.}~\bibnamefont {{Rogers}}},\ }\href
  {\doibase 10.1098/rsta.2018.0064} {\bibfield  {journal} {\bibinfo  {journal}
  {Philos. Trans. R. Soc. Lond. Ser. A-Math. Phys. Eng. Sci.}\ }\textbf
  {\bibinfo {volume} {377}},\ \bibinfo {eid} {20180064} (\bibinfo {year}
  {2019})}\BibitemShut {NoStop}%
\bibitem [{\citenamefont {{The COMET Collaboration}}\ \emph
  {et~al.}(2020)\citenamefont {{The COMET Collaboration}}, \citenamefont
  {Abramishvili} \emph {et~al.}}]{Comet_collaboration_comet_2020}%
  \BibitemOpen
  \bibfield  {author} {\bibinfo {author} {\bibnamefont {{The COMET
  Collaboration}}}, \bibinfo {author} {\bibfnamefont {R.}~\bibnamefont
  {Abramishvili}},  \emph {et~al.},\ }\href {\doibase 10.1093/ptep/ptz125}
  {\bibfield  {journal} {\bibinfo  {journal} {Progress of Theoretical and
  Experimental Physics}\ }\textbf {\bibinfo {volume} {2020}},\ \bibinfo {pages}
  {033C01} (\bibinfo {year} {2020})}\BibitemShut {NoStop}%
\bibitem [{\citenamefont {Cook}\ \emph {et~al.}(2017)\citenamefont {Cook},
  \citenamefont {D'Arcy}, \citenamefont {Edmonds}, \citenamefont {Fukuda},
  \citenamefont {Hatanaka}, \citenamefont {Hino}, \citenamefont {Kuno},
  \citenamefont {Lancaster}, \citenamefont {Mori}, \citenamefont {Ogitsu},
  \citenamefont {Sakamoto}, \citenamefont {Sato}, \citenamefont {Tran},
  \citenamefont {Truong}, \citenamefont {Wing}, \citenamefont {Yamamoto},\ and\
  \citenamefont {Yoshida}}]{Cook_MuSIC_2017}%
  \BibitemOpen
  \bibfield  {author} {\bibinfo {author} {\bibfnamefont {S.}~\bibnamefont
  {Cook}}, \bibinfo {author} {\bibfnamefont {R.}~\bibnamefont {D'Arcy}},
  \bibinfo {author} {\bibfnamefont {A.}~\bibnamefont {Edmonds}}, \bibinfo
  {author} {\bibfnamefont {M.}~\bibnamefont {Fukuda}}, \bibinfo {author}
  {\bibfnamefont {K.}~\bibnamefont {Hatanaka}}, \bibinfo {author}
  {\bibfnamefont {Y.}~\bibnamefont {Hino}}, \bibinfo {author} {\bibfnamefont
  {Y.}~\bibnamefont {Kuno}}, \bibinfo {author} {\bibfnamefont {M.}~\bibnamefont
  {Lancaster}}, \bibinfo {author} {\bibfnamefont {Y.}~\bibnamefont {Mori}},
  \bibinfo {author} {\bibfnamefont {T.}~\bibnamefont {Ogitsu}}, \bibinfo
  {author} {\bibfnamefont {H.}~\bibnamefont {Sakamoto}}, \bibinfo {author}
  {\bibfnamefont {A.}~\bibnamefont {Sato}}, \bibinfo {author} {\bibfnamefont
  {N.~H.}\ \bibnamefont {Tran}}, \bibinfo {author} {\bibfnamefont {N.~M.}\
  \bibnamefont {Truong}}, \bibinfo {author} {\bibfnamefont {M.}~\bibnamefont
  {Wing}}, \bibinfo {author} {\bibfnamefont {A.}~\bibnamefont {Yamamoto}}, \
  and\ \bibinfo {author} {\bibfnamefont {M.}~\bibnamefont {Yoshida}},\ }\href
  {\doibase 10.1103/PhysRevAccelBeams.20.030101} {\bibfield  {journal}
  {\bibinfo  {journal} {Phys. Rev. Accel. Beams}\ }\textbf {\bibinfo {volume}
  {20}},\ \bibinfo {pages} {030101} (\bibinfo {year} {2017})}\BibitemShut
  {NoStop}%
\bibitem [{\citenamefont {{Tomono}}\ \emph {et~al.}(2018)\citenamefont
  {{Tomono}}, \citenamefont {{Fukuda}}, \citenamefont {{Hatanaka}},
  \citenamefont {{Higemoto}}, \citenamefont {{Kawashima}}, \citenamefont
  {{Kojima}}, \citenamefont {{Kuno}}, \citenamefont {{Matsuda}}, \citenamefont
  {{Matsuzaki}}, \citenamefont {{Miyake}}, \citenamefont {{Miyamoto}},
  \citenamefont {{Morita}}, \citenamefont {{Motoishi}}, \citenamefont
  {{Nakazawa}}, \citenamefont {{Ninomiya}}, \citenamefont {{Nishikawa}},
  \citenamefont {{Ohta}}, \citenamefont {{Sato}}, \citenamefont {{Shimomura}},
  \citenamefont {{Takahisa}}, \citenamefont {{Weichao}},\ and\ \citenamefont
  {{Wong}}}]{Tomono_MuSIC_2018}%
  \BibitemOpen
  \bibfield  {author} {\bibinfo {author} {\bibfnamefont {D.}~\bibnamefont
  {{Tomono}}}, \bibinfo {author} {\bibfnamefont {M.}~\bibnamefont {{Fukuda}}},
  \bibinfo {author} {\bibfnamefont {K.}~\bibnamefont {{Hatanaka}}}, \bibinfo
  {author} {\bibfnamefont {W.}~\bibnamefont {{Higemoto}}}, \bibinfo {author}
  {\bibfnamefont {Y.}~\bibnamefont {{Kawashima}}}, \bibinfo {author}
  {\bibfnamefont {K.~M.}\ \bibnamefont {{Kojima}}}, \bibinfo {author}
  {\bibfnamefont {Y.}~\bibnamefont {{Kuno}}}, \bibinfo {author} {\bibfnamefont
  {Y.}~\bibnamefont {{Matsuda}}}, \bibinfo {author} {\bibfnamefont
  {T.}~\bibnamefont {{Matsuzaki}}}, \bibinfo {author} {\bibfnamefont
  {Y.}~\bibnamefont {{Miyake}}}, \bibinfo {author} {\bibfnamefont
  {K.}~\bibnamefont {{Miyamoto}}}, \bibinfo {author} {\bibfnamefont
  {Y.}~\bibnamefont {{Morita}}}, \bibinfo {author} {\bibfnamefont
  {T.}~\bibnamefont {{Motoishi}}}, \bibinfo {author} {\bibfnamefont
  {Y.}~\bibnamefont {{Nakazawa}}}, \bibinfo {author} {\bibfnamefont
  {K.}~\bibnamefont {{Ninomiya}}}, \bibinfo {author} {\bibfnamefont
  {R.}~\bibnamefont {{Nishikawa}}}, \bibinfo {author} {\bibfnamefont
  {S.}~\bibnamefont {{Ohta}}}, \bibinfo {author} {\bibfnamefont
  {A.}~\bibnamefont {{Sato}}}, \bibinfo {author} {\bibfnamefont
  {K.}~\bibnamefont {{Shimomura}}}, \bibinfo {author} {\bibfnamefont
  {K.}~\bibnamefont {{Takahisa}}}, \bibinfo {author} {\bibfnamefont
  {Y.}~\bibnamefont {{Weichao}}}, \ and\ \bibinfo {author} {\bibfnamefont
  {M.~L.}\ \bibnamefont {{Wong}}},\ }in\ \href {\doibase
  10.7566/JPSCP.21.011057} {\emph {\bibinfo {booktitle} {Proceedings of the
  14th International Conference on Muon Spin Rotation}}}\ (\bibinfo {year}
  {2018})\ p.\ \bibinfo {pages} {011057}\BibitemShut {NoStop}%
\bibitem [{\citenamefont {{Strasser}}\ \emph {et~al.}(2014)\citenamefont
  {{Strasser}}, \citenamefont {{Ikedo}}, \citenamefont {{Makimura}},
  \citenamefont {{Nakamura}}, \citenamefont {{Nishiyama}}, \citenamefont
  {{Shimomura}}, \citenamefont {{Fujimori}}, \citenamefont {{Adachi}},
  \citenamefont {{Koda}}, \citenamefont {{Kawamura}}, \citenamefont
  {{Kobayashi}}, \citenamefont {{Higemoto}}, \citenamefont {{Ito}},
  \citenamefont {{Nagatomo}}, \citenamefont {{Torikai}}, \citenamefont
  {{Kadono}},\ and\ \citenamefont {{Miyake}}}]{Strasser_ULM_2014}%
  \BibitemOpen
  \bibfield  {author} {\bibinfo {author} {\bibfnamefont {P.}~\bibnamefont
  {{Strasser}}}, \bibinfo {author} {\bibfnamefont {Y.}~\bibnamefont {{Ikedo}}},
  \bibinfo {author} {\bibfnamefont {S.}~\bibnamefont {{Makimura}}}, \bibinfo
  {author} {\bibfnamefont {J.}~\bibnamefont {{Nakamura}}}, \bibinfo {author}
  {\bibfnamefont {K.}~\bibnamefont {{Nishiyama}}}, \bibinfo {author}
  {\bibfnamefont {K.}~\bibnamefont {{Shimomura}}}, \bibinfo {author}
  {\bibfnamefont {H.}~\bibnamefont {{Fujimori}}}, \bibinfo {author}
  {\bibfnamefont {T.}~\bibnamefont {{Adachi}}}, \bibinfo {author}
  {\bibfnamefont {A.}~\bibnamefont {{Koda}}}, \bibinfo {author} {\bibfnamefont
  {N.}~\bibnamefont {{Kawamura}}}, \bibinfo {author} {\bibfnamefont
  {Y.}~\bibnamefont {{Kobayashi}}}, \bibinfo {author} {\bibfnamefont
  {W.}~\bibnamefont {{Higemoto}}}, \bibinfo {author} {\bibfnamefont {T.~U.}\
  \bibnamefont {{Ito}}}, \bibinfo {author} {\bibfnamefont {T.}~\bibnamefont
  {{Nagatomo}}}, \bibinfo {author} {\bibfnamefont {E.}~\bibnamefont
  {{Torikai}}}, \bibinfo {author} {\bibfnamefont {R.}~\bibnamefont {{Kadono}}},
  \ and\ \bibinfo {author} {\bibfnamefont {Y.}~\bibnamefont {{Miyake}}},\ }in\
  \href {\doibase 10.1088/1742-6596/551/1/012065} {\emph {\bibinfo {booktitle}
  {Journal of Physics Conference Series}}},\ \bibinfo {series} {Journal of
  Physics Conference Series}, Vol.\ \bibinfo {volume} {551}\ (\bibinfo {year}
  {2014})\ p.\ \bibinfo {pages} {012065}\BibitemShut {NoStop}%
\bibitem [{\citenamefont {{Nakamura}}\ \emph {et~al.}(2014)\citenamefont
  {{Nakamura}}, \citenamefont {{Nagatomo}}, \citenamefont {{Oishi}},
  \citenamefont {{Ikedo}}, \citenamefont {{Strasser}}, \citenamefont {{Saito}},
  \citenamefont {{Miyazaki}}, \citenamefont {{Yokoyama}}, \citenamefont
  {{Okamura}}, \citenamefont {{Miyake}}, \citenamefont {{Makimura}},
  \citenamefont {{Nishiyama}}, \citenamefont {{Shimomura}}, \citenamefont
  {{Kawamura}}, \citenamefont {{Koda}}, \citenamefont {{Higemoto}},
  \citenamefont {{Wada}}, \citenamefont {{Iwasaki}},\ and\ \citenamefont
  {{Torikai}}}]{Nakamura_ULM_2014}%
  \BibitemOpen
  \bibfield  {author} {\bibinfo {author} {\bibfnamefont {J.}~\bibnamefont
  {{Nakamura}}}, \bibinfo {author} {\bibfnamefont {T.}~\bibnamefont
  {{Nagatomo}}}, \bibinfo {author} {\bibfnamefont {Y.}~\bibnamefont {{Oishi}}},
  \bibinfo {author} {\bibfnamefont {Y.}~\bibnamefont {{Ikedo}}}, \bibinfo
  {author} {\bibfnamefont {P.}~\bibnamefont {{Strasser}}}, \bibinfo {author}
  {\bibfnamefont {N.}~\bibnamefont {{Saito}}}, \bibinfo {author} {\bibfnamefont
  {K.}~\bibnamefont {{Miyazaki}}}, \bibinfo {author} {\bibfnamefont
  {K.}~\bibnamefont {{Yokoyama}}}, \bibinfo {author} {\bibfnamefont
  {K.}~\bibnamefont {{Okamura}}}, \bibinfo {author} {\bibfnamefont
  {Y.}~\bibnamefont {{Miyake}}}, \bibinfo {author} {\bibfnamefont
  {S.}~\bibnamefont {{Makimura}}}, \bibinfo {author} {\bibfnamefont
  {K.}~\bibnamefont {{Nishiyama}}}, \bibinfo {author} {\bibfnamefont
  {K.}~\bibnamefont {{Shimomura}}}, \bibinfo {author} {\bibfnamefont
  {N.}~\bibnamefont {{Kawamura}}}, \bibinfo {author} {\bibfnamefont
  {A.}~\bibnamefont {{Koda}}}, \bibinfo {author} {\bibfnamefont
  {W.}~\bibnamefont {{Higemoto}}}, \bibinfo {author} {\bibfnamefont
  {S.}~\bibnamefont {{Wada}}}, \bibinfo {author} {\bibfnamefont
  {M.}~\bibnamefont {{Iwasaki}}}, \ and\ \bibinfo {author} {\bibfnamefont
  {E.}~\bibnamefont {{Torikai}}},\ }in\ \href {\doibase
  10.1088/1742-6596/502/1/012042} {\emph {\bibinfo {booktitle} {Journal of
  Physics Conference Series}}},\ \bibinfo {series} {Journal of Physics
  Conference Series}, Vol.\ \bibinfo {volume} {502}\ (\bibinfo {year} {2014})\
  p.\ \bibinfo {pages} {012042}\BibitemShut {NoStop}%
\bibitem [{\citenamefont {{Ikedo}}\ \emph {et~al.}(2011)\citenamefont
  {{Ikedo}}, \citenamefont {{Miyake}}, \citenamefont {{Shimomura}},
  \citenamefont {{Strasser}}, \citenamefont {{Nishiyama}}, \citenamefont
  {{Kawamura}}, \citenamefont {{Fujimori}}, \citenamefont {{Makimura}},
  \citenamefont {{Koda}}, \citenamefont {{Nakahara}}, \citenamefont {{Ogitsu}},
  \citenamefont {{Makida}}, \citenamefont {{Adachi}}, \citenamefont
  {{Yoshida}}, \citenamefont {{Yamamoto}}, \citenamefont {{Nakamoto}},
  \citenamefont {{Sasaki}}, \citenamefont {{Tanaka}}, \citenamefont {{Kimura}},
  \citenamefont {{Higemoto}}, \citenamefont {{Ajima}}, \citenamefont
  {{Ishida}}, \citenamefont {{Matsuda}},\ and\ \citenamefont
  {{Sato}}}]{Ikedo_SuperOmega_2011}%
  \BibitemOpen
  \bibfield  {author} {\bibinfo {author} {\bibfnamefont {Y.}~\bibnamefont
  {{Ikedo}}}, \bibinfo {author} {\bibfnamefont {Y.}~\bibnamefont {{Miyake}}},
  \bibinfo {author} {\bibfnamefont {K.}~\bibnamefont {{Shimomura}}}, \bibinfo
  {author} {\bibfnamefont {P.}~\bibnamefont {{Strasser}}}, \bibinfo {author}
  {\bibfnamefont {K.}~\bibnamefont {{Nishiyama}}}, \bibinfo {author}
  {\bibfnamefont {N.}~\bibnamefont {{Kawamura}}}, \bibinfo {author}
  {\bibfnamefont {H.}~\bibnamefont {{Fujimori}}}, \bibinfo {author}
  {\bibfnamefont {S.}~\bibnamefont {{Makimura}}}, \bibinfo {author}
  {\bibfnamefont {A.}~\bibnamefont {{Koda}}}, \bibinfo {author} {\bibfnamefont
  {K.}~\bibnamefont {{Nakahara}}}, \bibinfo {author} {\bibfnamefont
  {T.}~\bibnamefont {{Ogitsu}}}, \bibinfo {author} {\bibfnamefont
  {Y.}~\bibnamefont {{Makida}}}, \bibinfo {author} {\bibfnamefont
  {T.}~\bibnamefont {{Adachi}}}, \bibinfo {author} {\bibfnamefont
  {M.}~\bibnamefont {{Yoshida}}}, \bibinfo {author} {\bibfnamefont
  {A.}~\bibnamefont {{Yamamoto}}}, \bibinfo {author} {\bibfnamefont
  {T.}~\bibnamefont {{Nakamoto}}}, \bibinfo {author} {\bibfnamefont
  {K.}~\bibnamefont {{Sasaki}}}, \bibinfo {author} {\bibfnamefont
  {K.}~\bibnamefont {{Tanaka}}}, \bibinfo {author} {\bibfnamefont
  {N.}~\bibnamefont {{Kimura}}}, \bibinfo {author} {\bibfnamefont
  {W.}~\bibnamefont {{Higemoto}}}, \bibinfo {author} {\bibfnamefont
  {Y.}~\bibnamefont {{Ajima}}}, \bibinfo {author} {\bibfnamefont
  {K.}~\bibnamefont {{Ishida}}}, \bibinfo {author} {\bibfnamefont
  {Y.}~\bibnamefont {{Matsuda}}}, \ and\ \bibinfo {author} {\bibfnamefont
  {A.}~\bibnamefont {{Sato}}},\ }in\ \href {\doibase 10.1063/1.3644317} {\emph
  {\bibinfo {booktitle} {12th International Workshop on Neutrino Factories,
  Superbeams, and Betabeams: NuFact10}}},\ \bibinfo {series} {American
  Institute of Physics Conference Series}, Vol.\ \bibinfo {volume} {1382},\
  \bibinfo {editor} {edited by\ \bibinfo {editor} {\bibfnamefont {B.~S.}\
  \bibnamefont {{Acharya}}}, \bibinfo {editor} {\bibfnamefont {M.}~\bibnamefont
  {{Goodman}}}, \ and\ \bibinfo {editor} {\bibfnamefont {N.~K.}\ \bibnamefont
  {{Mondal}}}}\ (\bibinfo {year} {2011})\ pp.\ \bibinfo {pages}
  {220--222}\BibitemShut {NoStop}%
\bibitem [{\citenamefont {{Nakahara}}\ \emph {et~al.}(2009)\citenamefont
  {{Nakahara}}, \citenamefont {{Miyake}}, \citenamefont {{Shimomura}},
  \citenamefont {{Strasser}}, \citenamefont {{Nishiyama}}, \citenamefont
  {{Kawamura}}, \citenamefont {{Fujimori}}, \citenamefont {{Makimura}},
  \citenamefont {{Koda}}, \citenamefont {{Nagamine}}, \citenamefont {{Ogitsu}},
  \citenamefont {{Yamamoto}}, \citenamefont {{Adachi}}, \citenamefont
  {{Sasaki}}, \citenamefont {{Tanaka}}, \citenamefont {{Kimura}}, \citenamefont
  {{Makida}}, \citenamefont {{Ajima}}, \citenamefont {{Ishida}},\ and\
  \citenamefont {{Matsuda}}}]{Nakahara_SuperOmega_2009}%
  \BibitemOpen
  \bibfield  {author} {\bibinfo {author} {\bibfnamefont {K.}~\bibnamefont
  {{Nakahara}}}, \bibinfo {author} {\bibfnamefont {Y.}~\bibnamefont
  {{Miyake}}}, \bibinfo {author} {\bibfnamefont {K.}~\bibnamefont
  {{Shimomura}}}, \bibinfo {author} {\bibfnamefont {P.}~\bibnamefont
  {{Strasser}}}, \bibinfo {author} {\bibfnamefont {K.}~\bibnamefont
  {{Nishiyama}}}, \bibinfo {author} {\bibfnamefont {N.}~\bibnamefont
  {{Kawamura}}}, \bibinfo {author} {\bibfnamefont {H.}~\bibnamefont
  {{Fujimori}}}, \bibinfo {author} {\bibfnamefont {S.}~\bibnamefont
  {{Makimura}}}, \bibinfo {author} {\bibfnamefont {A.}~\bibnamefont {{Koda}}},
  \bibinfo {author} {\bibfnamefont {K.}~\bibnamefont {{Nagamine}}}, \bibinfo
  {author} {\bibfnamefont {T.}~\bibnamefont {{Ogitsu}}}, \bibinfo {author}
  {\bibfnamefont {A.}~\bibnamefont {{Yamamoto}}}, \bibinfo {author}
  {\bibfnamefont {T.}~\bibnamefont {{Adachi}}}, \bibinfo {author}
  {\bibfnamefont {K.}~\bibnamefont {{Sasaki}}}, \bibinfo {author}
  {\bibfnamefont {K.}~\bibnamefont {{Tanaka}}}, \bibinfo {author}
  {\bibfnamefont {N.}~\bibnamefont {{Kimura}}}, \bibinfo {author}
  {\bibfnamefont {Y.}~\bibnamefont {{Makida}}}, \bibinfo {author}
  {\bibfnamefont {Y.}~\bibnamefont {{Ajima}}}, \bibinfo {author} {\bibfnamefont
  {K.}~\bibnamefont {{Ishida}}}, \ and\ \bibinfo {author} {\bibfnamefont
  {Y.}~\bibnamefont {{Matsuda}}},\ }\href {\doibase 10.1016/j.nima.2008.11.106}
  {\bibfield  {journal} {\bibinfo  {journal} {Nucl. Instrum. Methods Phys. Res.
  Sect. A-Accel. Spectrom. Dect. Assoc. Equip.}\ }\textbf {\bibinfo {volume}
  {600}},\ \bibinfo {pages} {132} (\bibinfo {year} {2009})}\BibitemShut
  {NoStop}%
\bibitem [{\citenamefont {{Miyake}}\ \emph {et~al.}(2018)\citenamefont
  {{Miyake}}, \citenamefont {{Shimomura}}, \citenamefont {{Kawamura}},
  \citenamefont {{Koda}}, \citenamefont {{Strasser}}, \citenamefont {{Kojima}},
  \citenamefont {{Fujimori}}, \citenamefont {{Makimura}}, \citenamefont
  {{Ikedo}}, \citenamefont {{Kobayashi}}, \citenamefont {{Nakamura}},
  \citenamefont {{Oishi}}, \citenamefont {{Takeshita}}, \citenamefont
  {{Adachi}}, \citenamefont {{Datt Pant}}, \citenamefont {{Okabe}},
  \citenamefont {{Matoba}}, \citenamefont {{Tampo}}, \citenamefont
  {{Hiraishi}}, \citenamefont {{Hamada}}, \citenamefont {{Doiuchi}},
  \citenamefont {{Higemoto}}, \citenamefont {{Ito}},\ and\ \citenamefont
  {{Kadono}}}]{Miyake_MuSE_2018}%
  \BibitemOpen
  \bibfield  {author} {\bibinfo {author} {\bibfnamefont {Y.}~\bibnamefont
  {{Miyake}}}, \bibinfo {author} {\bibfnamefont {K.}~\bibnamefont
  {{Shimomura}}}, \bibinfo {author} {\bibfnamefont {N.}~\bibnamefont
  {{Kawamura}}}, \bibinfo {author} {\bibfnamefont {A.}~\bibnamefont {{Koda}}},
  \bibinfo {author} {\bibfnamefont {P.}~\bibnamefont {{Strasser}}}, \bibinfo
  {author} {\bibfnamefont {K.~M.}\ \bibnamefont {{Kojima}}}, \bibinfo {author}
  {\bibfnamefont {H.}~\bibnamefont {{Fujimori}}}, \bibinfo {author}
  {\bibfnamefont {S.}~\bibnamefont {{Makimura}}}, \bibinfo {author}
  {\bibfnamefont {Y.}~\bibnamefont {{Ikedo}}}, \bibinfo {author} {\bibfnamefont
  {Y.}~\bibnamefont {{Kobayashi}}}, \bibinfo {author} {\bibfnamefont
  {J.}~\bibnamefont {{Nakamura}}}, \bibinfo {author} {\bibfnamefont
  {Y.}~\bibnamefont {{Oishi}}}, \bibinfo {author} {\bibfnamefont
  {S.}~\bibnamefont {{Takeshita}}}, \bibinfo {author} {\bibfnamefont
  {T.}~\bibnamefont {{Adachi}}}, \bibinfo {author} {\bibfnamefont
  {A.}~\bibnamefont {{Datt Pant}}}, \bibinfo {author} {\bibfnamefont
  {H.}~\bibnamefont {{Okabe}}}, \bibinfo {author} {\bibfnamefont
  {S.}~\bibnamefont {{Matoba}}}, \bibinfo {author} {\bibfnamefont
  {M.}~\bibnamefont {{Tampo}}}, \bibinfo {author} {\bibfnamefont
  {M.}~\bibnamefont {{Hiraishi}}}, \bibinfo {author} {\bibfnamefont
  {K.}~\bibnamefont {{Hamada}}}, \bibinfo {author} {\bibfnamefont
  {S.}~\bibnamefont {{Doiuchi}}}, \bibinfo {author} {\bibfnamefont
  {W.}~\bibnamefont {{Higemoto}}}, \bibinfo {author} {\bibfnamefont {T.~U.}\
  \bibnamefont {{Ito}}}, \ and\ \bibinfo {author} {\bibfnamefont
  {R.}~\bibnamefont {{Kadono}}},\ }in\ \href {\doibase 10.7566/JPSCP.21.011054}
  {\emph {\bibinfo {booktitle} {Proceedings of the 14th International
  Conference on Muon Spin Rotation}}}\ (\bibinfo {year} {2018})\ p.\ \bibinfo
  {pages} {011054}\BibitemShut {NoStop}%
\bibitem [{\citenamefont {Kawamura}\ \emph {et~al.}(2018)\citenamefont
  {Kawamura}, \citenamefont {Aoki}, \citenamefont {Doornbos}, \citenamefont
  {Mibe}, \citenamefont {Miyake}, \citenamefont {Morimoto}, \citenamefont
  {Nakatsugawa}, \citenamefont {Otani}, \citenamefont {Saito}, \citenamefont
  {Seiya}, \citenamefont {Shimomura}, \citenamefont {Toyoda},\ and\
  \citenamefont {Yamazaki}}]{kawamura_new_2018}%
  \BibitemOpen
  \bibfield  {author} {\bibinfo {author} {\bibfnamefont {N.}~\bibnamefont
  {Kawamura}}, \bibinfo {author} {\bibfnamefont {M.}~\bibnamefont {Aoki}},
  \bibinfo {author} {\bibfnamefont {J.}~\bibnamefont {Doornbos}}, \bibinfo
  {author} {\bibfnamefont {T.}~\bibnamefont {Mibe}}, \bibinfo {author}
  {\bibfnamefont {Y.}~\bibnamefont {Miyake}}, \bibinfo {author} {\bibfnamefont
  {F.}~\bibnamefont {Morimoto}}, \bibinfo {author} {\bibfnamefont
  {Y.}~\bibnamefont {Nakatsugawa}}, \bibinfo {author} {\bibfnamefont
  {M.}~\bibnamefont {Otani}}, \bibinfo {author} {\bibfnamefont
  {N.}~\bibnamefont {Saito}}, \bibinfo {author} {\bibfnamefont
  {Y.}~\bibnamefont {Seiya}}, \bibinfo {author} {\bibfnamefont
  {K.}~\bibnamefont {Shimomura}}, \bibinfo {author} {\bibfnamefont
  {A.}~\bibnamefont {Toyoda}}, \ and\ \bibinfo {author} {\bibfnamefont
  {T.}~\bibnamefont {Yamazaki}},\ }\href {\doibase 10.1093/ptep/pty116}
  {\bibfield  {journal} {\bibinfo  {journal} {Progress of Theoretical and
  Experimental Physics}\ }\textbf {\bibinfo {volume} {2018}},\ \bibinfo {pages}
  {113G01} (\bibinfo {year} {2018})}\BibitemShut {NoStop}%
\bibitem [{\citenamefont {Papa}(2019)}]{Apapa_HiMB_2019}%
  \BibitemOpen
  \bibfield  {author} {\bibinfo {author} {\bibfnamefont {A.}~\bibnamefont
  {Papa}},\ }\href@noop {} {\enquote {\bibinfo {title} {Towards an high
  intensity muon beam {(HiMB)} at {PSI}},}\ } (\bibinfo {year} {2019}),\
  \bibinfo {note}
  {\url{https://indico.cern.ch/event/577856/contributions/3420391/attachments/1879682/3097488/Papa_HiMB_EPS2019.pdf}}\BibitemShut
  {NoStop}%
\bibitem [{\citenamefont {Kirch}(2020)}]{KSKirch_HiMB_2020}%
  \BibitemOpen
  \bibfield  {author} {\bibinfo {author} {\bibfnamefont {K.~S.}\ \bibnamefont
  {Kirch}},\ }\href@noop {} {\enquote {\bibinfo {title} {Welcome report at
  {Open User Meeting BV51}},}\ } (\bibinfo {year} {2020}),\ \bibinfo {note}
  {\url{https://indico.psi.ch/event/8337/contributions/22169/attachments/16011/22593/KK-BVR2020.pdf}}\BibitemShut
  {NoStop}%
\bibitem [{\citenamefont {Aiba}\ \emph {et~al.}(2021)\citenamefont {Aiba} \emph
  {et~al.}}]{Aiba:2021bxe}%
  \BibitemOpen
  \bibfield  {author} {\bibinfo {author} {\bibfnamefont {M.}~\bibnamefont
  {Aiba}} \emph {et~al.},\ }\href@noop {} {\enquote {\bibinfo {title} {{Science
  Case for the new High-Intensity Muon Beams HIMB at PSI}},}\ } (\bibinfo
  {year} {2021}),\ \Eprint {http://arxiv.org/abs/2111.05788} {arXiv:2111.05788
  [hep-ex]} \BibitemShut {NoStop}%
\bibitem [{\citenamefont {{Tang}}\ \emph {et~al.}(2018)\citenamefont {{Tang}},
  \citenamefont {{Ni}}, \citenamefont {{Ma}}, \citenamefont {{Luo}},
  \citenamefont {{Bao}}, \citenamefont {{Yuan}}, \citenamefont {{Chen}},
  \citenamefont {{Chen}}, \citenamefont {{Deng}}, \citenamefont {{Dong}},
  \citenamefont {{Hou}}, \citenamefont {{Hu}}, \citenamefont {{Jing}},
  \citenamefont {{Liang}}, \citenamefont {{Mu}}, \citenamefont {{Ning}},
  \citenamefont {{Pan}}, \citenamefont {{Song}}, \citenamefont {{Tang}},
  \citenamefont {{Vassilopoulos}}, \citenamefont {{Wang}}, \citenamefont
  {{Xie}}, \citenamefont {{Ye}}, \citenamefont {{Zhang}}, \citenamefont
  {{Zhang}}, \citenamefont {{Zhao}}, \citenamefont {{Zhao}}, \citenamefont
  {{Zhou}}, \citenamefont {{Zhu}}, \citenamefont {{Zhu}},\ and\ \citenamefont
  {{Zhuang}}}]{Tang_EMuS_2018}%
  \BibitemOpen
  \bibfield  {author} {\bibinfo {author} {\bibfnamefont {J.}~\bibnamefont
  {{Tang}}}, \bibinfo {author} {\bibfnamefont {X.}~\bibnamefont {{Ni}}},
  \bibinfo {author} {\bibfnamefont {X.}~\bibnamefont {{Ma}}}, \bibinfo {author}
  {\bibfnamefont {H.}~\bibnamefont {{Luo}}}, \bibinfo {author} {\bibfnamefont
  {Y.}~\bibnamefont {{Bao}}}, \bibinfo {author} {\bibfnamefont
  {Y.}~\bibnamefont {{Yuan}}}, \bibinfo {author} {\bibfnamefont
  {Y.}~\bibnamefont {{Chen}}}, \bibinfo {author} {\bibfnamefont
  {Y.}~\bibnamefont {{Chen}}}, \bibinfo {author} {\bibfnamefont
  {F.}~\bibnamefont {{Deng}}}, \bibinfo {author} {\bibfnamefont
  {J.}~\bibnamefont {{Dong}}}, \bibinfo {author} {\bibfnamefont
  {Z.}~\bibnamefont {{Hou}}}, \bibinfo {author} {\bibfnamefont
  {C.}~\bibnamefont {{Hu}}}, \bibinfo {author} {\bibfnamefont {H.}~\bibnamefont
  {{Jing}}}, \bibinfo {author} {\bibfnamefont {H.}~\bibnamefont {{Liang}}},
  \bibinfo {author} {\bibfnamefont {Q.}~\bibnamefont {{Mu}}}, \bibinfo {author}
  {\bibfnamefont {C.}~\bibnamefont {{Ning}}}, \bibinfo {author} {\bibfnamefont
  {Z.}~\bibnamefont {{Pan}}}, \bibinfo {author} {\bibfnamefont
  {Y.}~\bibnamefont {{Song}}}, \bibinfo {author} {\bibfnamefont
  {J.}~\bibnamefont {{Tang}}}, \bibinfo {author} {\bibfnamefont
  {N.}~\bibnamefont {{Vassilopoulos}}}, \bibinfo {author} {\bibfnamefont
  {H.}~\bibnamefont {{Wang}}}, \bibinfo {author} {\bibfnamefont
  {Z.}~\bibnamefont {{Xie}}}, \bibinfo {author} {\bibfnamefont
  {B.}~\bibnamefont {{Ye}}}, \bibinfo {author} {\bibfnamefont {G.}~\bibnamefont
  {{Zhang}}}, \bibinfo {author} {\bibfnamefont {Y.}~\bibnamefont {{Zhang}}},
  \bibinfo {author} {\bibfnamefont {G.}~\bibnamefont {{Zhao}}}, \bibinfo
  {author} {\bibfnamefont {W.}~\bibnamefont {{Zhao}}}, \bibinfo {author}
  {\bibfnamefont {L.}~\bibnamefont {{Zhou}}}, \bibinfo {author} {\bibfnamefont
  {D.}~\bibnamefont {{Zhu}}}, \bibinfo {author} {\bibfnamefont
  {Z.}~\bibnamefont {{Zhu}}}, \ and\ \bibinfo {author} {\bibfnamefont
  {M.}~\bibnamefont {{Zhuang}}},\ }\href {\doibase 10.3390/qubs2040023}
  {\bibfield  {journal} {\bibinfo  {journal} {Quantum Beam Sci.}\ }\textbf
  {\bibinfo {volume} {2}},\ \bibinfo {pages} {23} (\bibinfo {year}
  {2018})}\BibitemShut {NoStop}%
\bibitem [{\citenamefont {{Berg}}\ \emph {et~al.}(2016)\citenamefont {{Berg}},
  \citenamefont {{Desorgher}}, \citenamefont {{Fuchs}}, \citenamefont
  {{Hajdas}}, \citenamefont {{Hodge}}, \citenamefont {{Kettle}}, \citenamefont
  {{Knecht}}, \citenamefont {{L{\"u}scher}}, \citenamefont {{Papa}},
  \citenamefont {{Rutar}},\ and\ \citenamefont
  {{Wohlmuther}}}]{Berg_SMtarget_2016}%
  \BibitemOpen
  \bibfield  {author} {\bibinfo {author} {\bibfnamefont {F.}~\bibnamefont
  {{Berg}}}, \bibinfo {author} {\bibfnamefont {L.}~\bibnamefont {{Desorgher}}},
  \bibinfo {author} {\bibfnamefont {A.}~\bibnamefont {{Fuchs}}}, \bibinfo
  {author} {\bibfnamefont {W.}~\bibnamefont {{Hajdas}}}, \bibinfo {author}
  {\bibfnamefont {Z.}~\bibnamefont {{Hodge}}}, \bibinfo {author} {\bibfnamefont
  {P.~R.}\ \bibnamefont {{Kettle}}}, \bibinfo {author} {\bibfnamefont
  {A.}~\bibnamefont {{Knecht}}}, \bibinfo {author} {\bibfnamefont
  {R.}~\bibnamefont {{L{\"u}scher}}}, \bibinfo {author} {\bibfnamefont
  {A.}~\bibnamefont {{Papa}}}, \bibinfo {author} {\bibfnamefont
  {G.}~\bibnamefont {{Rutar}}}, \ and\ \bibinfo {author} {\bibfnamefont
  {M.}~\bibnamefont {{Wohlmuther}}},\ }\href {\doibase
  10.1103/PhysRevAccelBeams.19.024701} {\bibfield  {journal} {\bibinfo
  {journal} {Phys. Rev. Accel. Beams}\ }\textbf {\bibinfo {volume} {19}},\
  \bibinfo {eid} {024701} (\bibinfo {year} {2016})},\ \Eprint
  {http://arxiv.org/abs/1511.01288} {arXiv:1511.01288 [physics.ins-det]}
  \BibitemShut {NoStop}%
\bibitem [{\citenamefont {{Kaplan}}\ and\ \citenamefont
  {{Roberts}}(2007)}]{Kaplan_G4beamline_2007}%
  \BibitemOpen
  \bibfield  {author} {\bibinfo {author} {\bibfnamefont {D.~M.}\ \bibnamefont
  {{Kaplan}}}\ and\ \bibinfo {author} {\bibfnamefont {T.~J.}\ \bibnamefont
  {{Roberts}}},\ }in\ \href@noop {} {\emph {\bibinfo {booktitle} {22nd Particle
  Accelerator Conference (PAC 07)}}}\ (\bibinfo {year} {2007})\ p.\ \bibinfo
  {pages} {3468}\BibitemShut {NoStop}%
\bibitem [{\citenamefont {{Eshraqi}}\ \emph {et~al.}(2009)\citenamefont
  {{Eshraqi}}, \citenamefont {{Franchetti}},\ and\ \citenamefont
  {{Lombardi}}}]{Eshraqi_Solenoid_2009}%
  \BibitemOpen
  \bibfield  {author} {\bibinfo {author} {\bibfnamefont {M.}~\bibnamefont
  {{Eshraqi}}}, \bibinfo {author} {\bibfnamefont {G.}~\bibnamefont
  {{Franchetti}}}, \ and\ \bibinfo {author} {\bibfnamefont {A.~M.}\
  \bibnamefont {{Lombardi}}},\ }\href {\doibase 10.1103/PhysRevSTAB.12.024201}
  {\bibfield  {journal} {\bibinfo  {journal} {Physical Review Accelerators and
  Beams}\ }\textbf {\bibinfo {volume} {12}},\ \bibinfo {eid} {024201} (\bibinfo
  {year} {2009})}\BibitemShut {NoStop}%
\bibitem [{Roh()}]{Rohrer_TRANSPORT}%
  \BibitemOpen
  \href@noop {} {\enquote {\bibinfo {title} {{PSI Graphic Transport Framework},
  by {U. Rohrer} based on a {CERN-SLAC-FERMILAB} version by {K. L. Brown} et
  al},}\ }\BibitemShut {NoStop}%
\bibitem [{OPE()}]{OPERA-3D_web}%
  \BibitemOpen
  \href@noop {} {\enquote {\bibinfo {title} {Opera: A finite element analysis
  software suite developed by dassault systèmes.
  https://www.3ds.com/products-services/simulia/products/opera/},}\
  }\BibitemShut {NoStop}%
\bibitem [{\citenamefont {{Kumar}}(2009)}]{Kumar_Solenoid_2009}%
  \BibitemOpen
  \bibfield  {author} {\bibinfo {author} {\bibfnamefont {V.}~\bibnamefont
  {{Kumar}}},\ }\href {\doibase 10.1119/1.3129242} {\bibfield  {journal}
  {\bibinfo  {journal} {Am. J. Phys.}\ }\textbf {\bibinfo {volume} {77}},\
  \bibinfo {pages} {737} (\bibinfo {year} {2009})}\BibitemShut {NoStop}%
\bibitem [{\citenamefont {Laxdal}(2013)}]{Laxdal_Solshielding_2013}%
  \BibitemOpen
  \bibfield  {author} {\bibinfo {author} {\bibfnamefont {R.~E.}\ \bibnamefont
  {Laxdal}},\ }in\ \href@noop {} {\emph {\bibinfo {booktitle} {Proceedings of
  the 16th International Conference on RF Superconductivity (SRF 2013)}}}\
  (\bibinfo {year} {2013})\BibitemShut {NoStop}%
\bibitem [{\citenamefont {{Takeuchi}}(1997)}]{Takeuchi_Dualsol_1997}%
  \BibitemOpen
  \bibfield  {author} {\bibinfo {author} {\bibfnamefont {T.}~\bibnamefont
  {{Takeuchi}}},\ }\href {\doibase 10.1119/1.2344695} {\bibfield  {journal}
  {\bibinfo  {journal} {Phys. Teach.}\ }\textbf {\bibinfo {volume} {35}},\
  \bibinfo {pages} {306} (\bibinfo {year} {1997})}\BibitemShut {NoStop}%
\bibitem [{\citenamefont {Sedlak}\ \emph {et~al.}(2012)\citenamefont {Sedlak},
  \citenamefont {Scheuermann}, \citenamefont {Shiroka}, \citenamefont
  {Stoykov}, \citenamefont {Raselli},\ and\ \citenamefont
  {Amato}}]{Sedlak_musrsim_2012}%
  \BibitemOpen
  \bibfield  {author} {\bibinfo {author} {\bibfnamefont {K.}~\bibnamefont
  {Sedlak}}, \bibinfo {author} {\bibfnamefont {R.}~\bibnamefont {Scheuermann}},
  \bibinfo {author} {\bibfnamefont {T.}~\bibnamefont {Shiroka}}, \bibinfo
  {author} {\bibfnamefont {A.}~\bibnamefont {Stoykov}}, \bibinfo {author}
  {\bibfnamefont {A.}~\bibnamefont {Raselli}}, \ and\ \bibinfo {author}
  {\bibfnamefont {A.}~\bibnamefont {Amato}},\ }\href {\doibase
  10.1016/j.phpro.2012.04.040} {\bibfield  {journal} {\bibinfo  {journal}
  {Phys. Procedia}\ }\textbf {\bibinfo {volume} {30}},\ \bibinfo {pages} {61}
  (\bibinfo {year} {2012})}\BibitemShut {NoStop}%
\end{thebibliography}%

\end{document}